\documentclass[11pt]{article}
\usepackage[authoryear,round]{natbib}
\usepackage{econometrics}
\usepackage{amssymb}
\usepackage{siunitx,booktabs,xcolor,colortbl}
\usepackage{graphicx}
\usepackage{amsmath}
\usepackage{longtable}
\usepackage{array,multirow,makecell}
\usepackage{float}
\usepackage{arydshln}
\usepackage{subcaption}

\DeclareGraphicsExtensions{.pdf,.png,.jpg}
\usepackage{xcolor}
\definecolor{orange-red}{rgb}{1.0, 0.27, 0.0}
\definecolor{myred}{RGB}{16, 141, 193}
\definecolor{ballblue}{rgb}{0.13, 0.67, 0.8}
\definecolor{bluencs}{rgb}{0.0, 0.53, 0.74}
\definecolor{brandeisblue}{rgb}{0.0, 0.44, 1.0}
\definecolor{deepskyblue}{rgb}{0.0, 0.75, 1.0}
\definecolor{dodgerblue}{rgb}{0.12, 0.56, 1.0}
\definecolor{frenchblue}{rgb}{0.0, 0.45, 0.73}
\definecolor{mediumtealblue}{rgb}{0.0, 0.33, 0.71}
\definecolor{oceanboatblue}{rgb}{0.0, 0.47, 0.75}
\definecolor{richelectricblue}{rgb}{0.03, 0.57, 0.82}
\definecolor{tuftsblue}{rgb}{0.28, 0.57, 0.81}
\definecolor{yaleblue}{rgb}{0.06, 0.3, 0.57}
\definecolor{burntorange}{rgb}{0.8, 0.33, 0.0}
\definecolor{cadmiumorange}{rgb}{0.93, 0.53, 0.18}
\definecolor{chocolate}{rgb}{0.82, 0.41, 0.12}

\usepackage[colorlinks=true,citecolor=blue,unicode=true,  
            urlcolor  = mediumtealblue, linkcolor=orange-red]{hyperref}%

\usepackage{color}
\definecolor{Gray}{gray}{0.9}
\usepackage{hhline}
\usepackage{dcolumn}
\usepackage{mathtools}
\usepackage{pdflscape}
\usepackage{rotating}
\usepackage[onehalfspacing]{setspace}
\usepackage{changepage}
\usepackage{dirtytalk}
\usepackage{comment}

\usepackage[para]{threeparttable}
\usepackage[toc,page]{appendix}
\usepackage{fancyhdr}

\usepackage{XCharter}

\usepackage[left=25mm, right=25mm, top=22.5mm, bottom=25mm, footskip=29pt]{geometry} 
\usepackage{setspace} 
\usepackage[indent=1.5em]{parskip}
\usepackage[overload]{empheq}

\title{\textbf{Monetary Policy in the Media Spotlight: Sentiments, Signals, and Economic Impact}\thanks{Corresponding Author: Dalibor Stavanovic (\href{dstevanovic.econ@gmail.com}{dstevanovic.econ@gmail.com}). Abdoul Massaoudou and Gedeon Gbedonou provided excellent research assistance. This research was partly enabled by support provided by Calcul Quebec, the Digital Research Alliance of Canada and Social Sciences and Humanities Research Council. The views expressed herein are those of the authors and should not be attributed to the IMF, its Executive Board, or its management.}}

\author{Firmin Ayivodji \\
\textit{IMF}  %
\and Etienne Briand \\
\textit{UQAM}
\and Kevin Moran \\
\textit{Université Laval} \\ \textit{CIRANO} 
\and Dalibor Stevanovic  \\
\textit{UQAM} \\ \textit{CIRANO} 
}
\date{This version: \textcolor{blue}{\today}}

\begin{document}

\pagenumbering{gobble}

\maketitle

\vspace{-0.5cm}
\begin{abstract}
\noindent
News media coverage of monetary policy is not a passive transcript of central-bank communication: it filters announcements, macroeconomic news, and editorial choices into narratives that move expectations and policy decisions. We embed media sentiment into a behavioral New-Keynesian model in which the central bank reacts to sentiment and sentiment follows an explicit law of motion. We construct monetary-policy sentiment indicators from more than 50,000 Canadian newspaper articles using dictionary methods, transformer models, and a generative-AI framework. Media sentiment shifts household inflation and wage expectations, improves out-of-sample forecasts of GDP growth and inflation, and loads positively on the Bank of Canada's estimated Taylor rule once treated as endogenous. A Bayesian SVAR identifies anticipated and unanticipated monetary-policy shocks together with a narrative shock; the narrative shock contributes a non-trivial share of medium-horizon macroeconomic variance, and a counterfactual that shuts down the dynamic feedback from media sentiment attenuates the propagation of monetary policy to output and prices. 

\vspace{0.1 cm} 

\vspace{0.1 cm}

\end{abstract}

\newpage
\pagenumbering{arabic}

\begin{flushright}
\say{\textit{Monetary policy is $98\%$ talk and $2\%$ action, and communication is a big part.}}\\
-- \textbf{Ben Bernanke (2015)}
\end{flushright}

\vspace{-1cm}
\section{Introduction}

Central banks devote a large share of their work to communicating with the public. The Bernanke quote is a deliberate exaggeration, but it captures a now-mainstream view: explaining current decisions, signaling future intentions, and shaping public expectations are integral to the conduct of monetary policy. A growing empirical literature documents systematic effects of central-bank language on interest rates, asset prices, and survey expectations.\footnote{See \cite{doh2022deciphering, silva2025text, sekkel2025money, aruoba2024identifying, alexopoulos2024more, gorodnichenko2023voice, cieslak2023policymakers, gardner2022words, shapiro2022taking, hansen2018transparency}.} 
This work builds on a long-standing tradition emphasizing the role of narratives and communication in monetary policy \citep{fortinthesis, romer1989does}.

\begin{figure}[H]
\caption{How the public learns about monetary policy: euro-area evidence\label{fig:ecb_sources}}
\vspace{-1cm}
    \centering
        \includegraphics[trim=0 320 0 0,clip,width=0.85\textwidth,height=0.4\textheight]{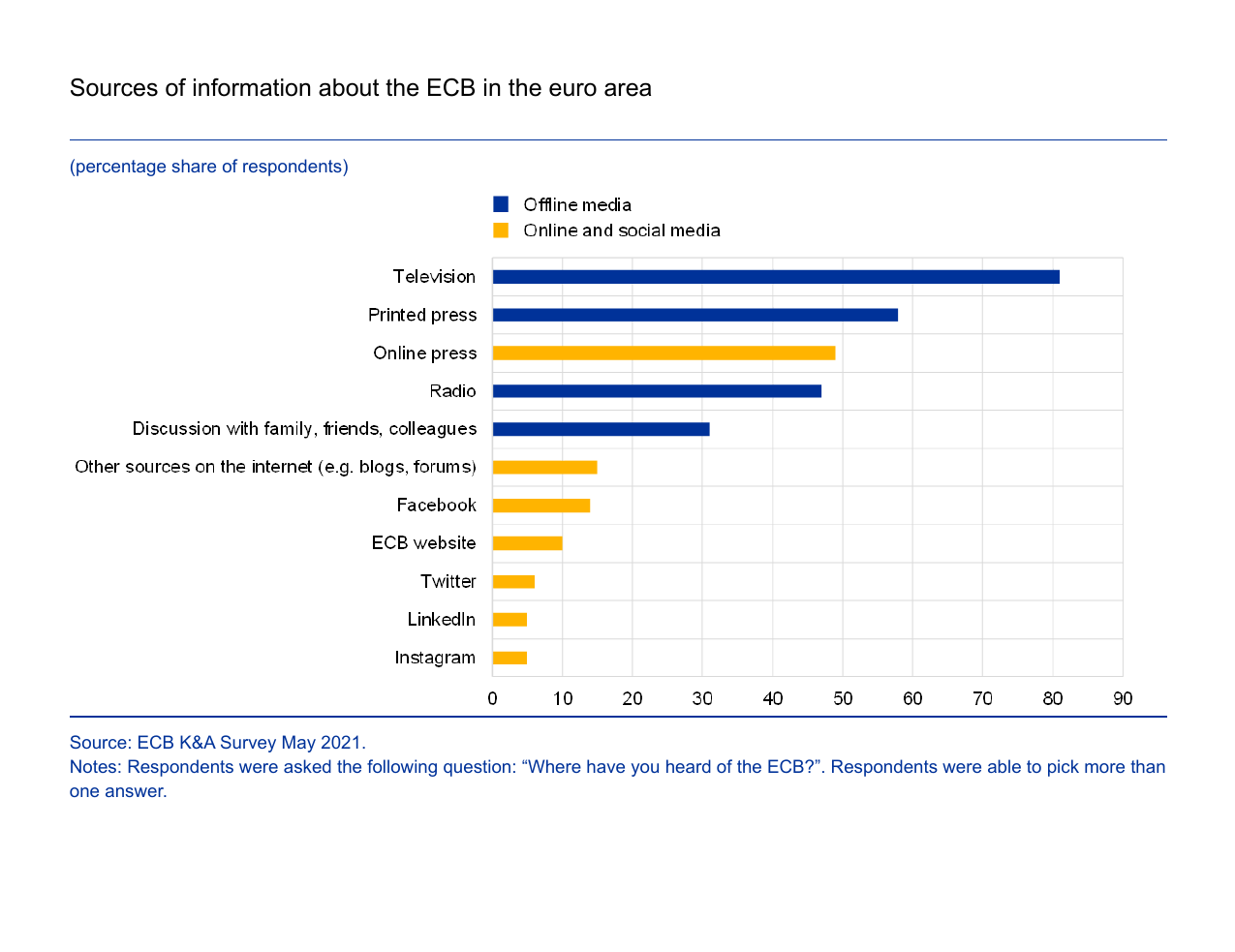}
\vspace{-2cm}
\end{figure}

Central-bank messages do not, however, reach the public directly. Survey evidence on both sides of the Atlantic shows that well over half of households learn about monetary policy through traditional and online news media, while only a small minority reads central-bank communications first-hand \citep{blinder2024communication}. Figure~\ref{fig:ecb_sources} illustrates this pattern for the euro area. The media therefore stand between the central bank and the public as an intermediate producer of signals.\footnote{\citet{coibion2022mp} provide experimental evidence that this indirect news-media route is a measurable but partial channel of monetary-policy communication: households exposed to a USA Today summary of an FOMC meeting revise their inflation expectations, but only about half as strongly as those reading the FOMC statement directly.}

The media coverage contains pieces of the central bank's own communication, but also factual reporting of the macroeconomic events that the central bank is also reacting to, and editorial choices that depend on news values, journalistic conventions, and competition for attention \citep{nimark2014man, chahrour2021sectoral}. As \cite{carroll2003macroeconomic} shows, household expectations are formed in part by reading what the news contains, so a media coverage that selectively emphasizes some aspects of monetary policy and downplays others can shift the public's view of the policy stance even when the underlying actions are unchanged.

\begin{figure}[H]
\caption{Media narratives as an endogenous outcome of monetary policy\label{sec:fig1}}
\vspace{-0.2cm}
    \centering
        \includegraphics[width=0.8\textwidth,height=0.38\textheight]{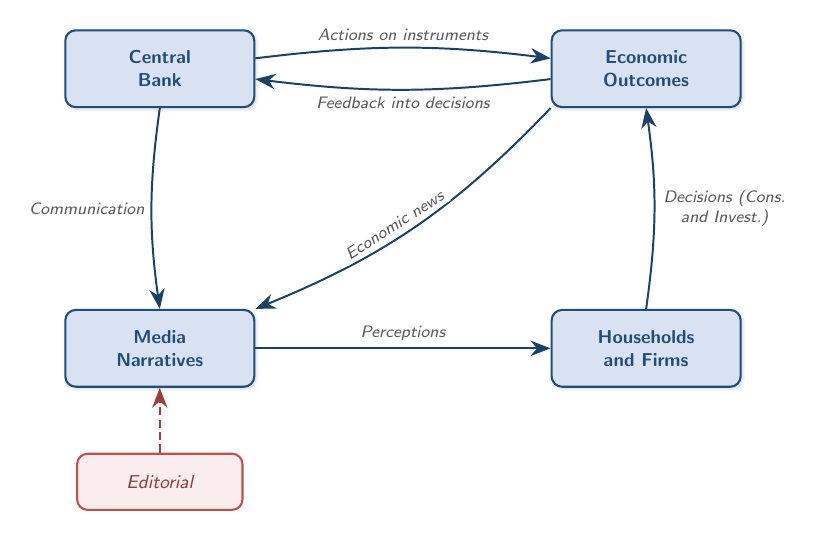} 
\end{figure}

This composite nature makes media coverage an endogenous outcome of the same macroeconomic system that monetary policy operates on. Figure~\ref{sec:fig1} depicts the resulting structure. The central bank acts on its instruments, economic outcomes feed back into the next policy decision, and the central bank's own communication reaches the media alongside the flow of economic news. The media reshape these inputs through editorial choices into the narratives that households and firms read; agents form perceptions and adjust their consumption and investment decisions, which in turn affect economic outcomes. The diagram has two key implications. First, the media-narrative node has three labeled inputs so any indicator of media sentiment is, by construction, a composite of central-bank communication (\textit{Communication}), current fundamentals (\textit{Economic News}), and an editorial residual (\textit{Editorial}). Second that residual is plausibly orthogonal to policy decisions and 
identifying how monetary policy responds to media sentiment requires an empirical strategy that respects this endogeneity, rather than treating sentiment as a passive informational signal.

To formalize the endogenous role of media narratives, we adopt the behavioral New-Keynesian framework of \cite{gabaix2020behavioral}, in which a fixed attention parameter $M^h$ down-weights the response of current decisions to expected future variables, and we augment it along two dimensions. The central-bank reaction function includes a feedback to a media-sentiment index, and that index follows a law of motion that decomposes it explicitly into four blocks: a slow-moving autoregressive component, a fundamentals component, a communication component built from anticipated policy shocks, and a residual narrative innovation that captures editorial choices. Sentiment is therefore an endogenous equilibrium object: it does not enter the IS or Phillips-curve equations directly, but it interacts with monetary policy decisions through the central bank's reaction function and feeds back into the expectations-relevant macroeconomic dynamics through the equilibrium response of the policy rate. The model produces sharp implications for impulse responses to forward guidance.

We then turn to measurement. Using natural-language-processing tools applied to a corpus of more than 50,000 articles from major Canadian newspapers between 1977 and 2024, we construct media indicators of monetary-policy sentiment along four dimensions: topic, tone, temporal orientation, and uncertainty. We use dictionary methods, transformer-based encoder models (FinBERT and ModernFinBERT), and a structured generative-LLM framework (CBILA). We find that media sentiments affect consumer inflation and wage expectations, and improve the out-of-sample forecast precision of GDP growth and CPI inflation.  

We then estimate forward-looking Taylor rules for the Bank of Canada, treating media sentiment as endogenous as the model dictates. Estimated by GMM with appropriate instruments, the dictionary, FinBERT and ModernFinBERT sentiment measures all load positively and significantly on the policy rate; the gap between OLS and GMM estimates confirms that simultaneity between sentiment and policy is a first-order concern. The CBILA generative-LLM tones fail standard IV validity tests because their pre-training data extend into the estimation sample, which contaminates the lagged-instrument exclusion restriction.

We then move from the partial-equilibrium evidence on the Taylor rule to the full dynamic transmission of monetary policy in a Bayesian structural VAR. Sign restrictions derived from the model identify three structural shocks: an anticipated monetary policy shock, an unanticipated monetary policy shock, and a narrative shock that captures autonomous variation in media coverage orthogonal to fundamentals and to central-bank communication. The unanticipated easing cuts the policy rate on impact and is followed by a hump-shaped expansion of output and prices; the anticipated easing produces a smaller impact response of the rate, a slow build-up of output that peaks near the announced implementation date, and a positive price response, consistent with the cognitive-discounting mechanism. The narrative shock contributes a non-trivial share of medium-horizon macroeconomic variance. A counterfactual that shuts down the dynamic feedback from media sentiment to the rest of the system attenuates the response of output and prices to the anticipated monetary policy shock, confirming that media sentiment operates as a quantitatively relevant propagation channel. Media coverage therefore matters both as a separate source of macroeconomic fluctuation and as a transmission mechanism for conventional monetary policy.

Two recent papers study the role of narratives in monetary policy. \citet{andre2026narratives} measure the narratives US households use to interpret macroeconomic phenomena, show in randomized experiments that these narratives shape inflation expectations, and embed them in an otherwise standard New Keynesian model. \citet{kaminski2026narratives} extract narratives from FOMC transcripts via an LLM-based directed-acyclic-graph approach and find that the transmission of monetary policy is strongly narrative dependent. Our focus is the media that mediate between the central bank and households: the news coverage that filters central-bank communication and economic events into the narratives households read and the central bank observes. The three papers therefore cover complementary nodes of monetary-policy transmission --- households, central bank, and the media in between.

The paper is structured as follows. Section~\ref{sec:model} introduces the behavioral New-Keynesian model with media as a narrative friction. Section~\ref{sec:measure} constructs measures of media coverage about monetary policy, including the dictionary, FinBERT, and CBILA generative-AI frameworks. Section~\ref{sec:empirical} relates these measures to household expectations and evaluates the predictive content of the indicators for realized macroeconomic outcomes. Section~\ref{sec:taylor} estimates text-augmented Taylor rules. Section~\ref{sec:svar} identifies the dynamic transmission of monetary policy and media sentiment in a Bayesian SVAR. Section~\ref{sec:conclusion} concludes.

\section{A Behavioral New-Keynesian Model with Media Narratives}\label{sec:model}

We embed the media into a stylized New-Keynesian model with cognitive discounting in order to (i) discipline how narratives interact with monetary policy, and (ii) guide the empirical strategy of Sections~\ref{sec:expectations}, \ref{sec:forecasting}, \ref{sec:taylor} and \ref{sec:svar}. Two features of the model are central. First, following \cite{gabaix2020behavioral}, both the Euler equation and the Phillips curve carry an attention parameter $M$ that downweights distant-future events, so that forward guidance does not pass through one-for-one. Second, the public's view of monetary policy is filtered by a media-sentiment variable $s_t$ that aggregates four objects: the central bank's own communication, current fundamentals, an autoregressive component, and a residual narrative innovation. The first three components are by construction correlated with the policy decision; only the last is plausibly independent. The model is therefore the structural counterpart of Figure~\ref{sec:fig1}: the three labeled inputs of the media-narrative node (Communication, Economic News, Editorial) correspond to three of the four blocks of the sentiment law of motion introduced in Section~\ref{sec:model_narratives}, and the central-bank reaction function closes the feedback loop in the diagram. 

\subsection{Households and firms}\label{sec:model_hh_firms}

A representative household consumes a final good $C_t$, supplies labor $N_t$, and invests in nominal bonds $B_t$ that pay gross return $R_{t-1}$ between $t-1$ and $t$. Lifetime utility is
\begin{equation}
    E_0 \sum_{t=0}^\infty \beta^t \bigg[ \frac{C_t^{1-\sigma} - 1}{1-\sigma} - \chi\, \frac{N_t^{1+\psi}}{1+\psi} \bigg],
\end{equation}
with period budget constraint $P_t C_t + B_t = W_t N_t + R_{t-1} B_{t-1} + \int \Pi_t(j)\, dj$. Here $\sigma$ and $\psi$ are the inverse intertemporal and Frisch elasticities, $\chi > 0$ is the labor-disutility weight, $\beta$ is the discount factor, $W_t$ is the nominal wage, and $\Pi_t(j)$ are firm $j$'s profits. A continuum of price-setting firms $j \in [0,1]$ produces differentiated varieties with linear technology $Y_t(j) = N_t(j)$ and faces Calvo pricing rigidities with reset probability $1-\theta$. The final good is a CES aggregator with elasticity $\epsilon$, yielding the standard demand schedules and the aggregate price index $P_t = \big( \int_0^1 P_t(j)^{1-\epsilon} dj \big)^{1/(1-\epsilon)}$.

\subsection{Central bank with feedback to sentiment}\label{sec:model_cb}

The central bank sets the gross nominal rate $R_t$ following an inertial Taylor rule that, compared to the textbook specification, includes a feedback to media sentiment $S_t$:
\begin{equation}\label{eq:taylor_struct}
    \frac{R_t}{R} = \bigg( \frac{R_{t-1}}{R} \bigg)^{\rho_r} \bigg[ \bigg( \frac{\Pi_t}{\Pi} \bigg)^{\phi_\pi} \bigg( \frac{\tilde{Y}_t}{\tilde{Y}} \bigg)^{\phi_x} \bigg(\frac{S_t}{S} \bigg)^{\phi_s} \bigg]^{1-\rho_r} \exp(\varepsilon^u_t)\, \exp(\varepsilon^a_{t-\tau}).
\end{equation}
The smoothing parameter is $\rho_r$. The elasticities $\phi_\pi$, $\phi_x$, and $\phi_s$ govern the response to inflation $\Pi_t$, the output gap $\tilde Y_t \equiv Y_t/Y_t^n$, and the sentiment index $S_t$, respectively. Variables without subscripts denote steady-state values. Two innovations drive the policy rate. The conventional surprise $\varepsilon^u_t$ is i.i.d. The anticipated shock $\varepsilon^a_{t-\tau}$ is announced $\tau$ periods in advance and is meant to capture explicit central-bank communication about a future change in the policy stance, that is, forward guidance.\footnote{The Bank of Canada has progressively expanded its set of communication tools and now publishes a summary of policy deliberations after each decision, starting in February 2023; see \cite{jain2023summaries} for the institutional context.} The cognitive-discounting block below determines how strongly agents react to the announced component.

\subsection{Cognitive discounting and equilibrium}\label{sec:model_eq}

For the news media of Figure~\ref{sec:fig1} to play a substantive role in the framework, agents cannot be fully attentive to every signal that the central bank and the macroeconomy generate. If they were, fully rational agents would extract the same information from the underlying communication and fundamentals as journalists do, and the media node would be mechanically redundant. We follow \cite{gabaix2020behavioral} and add a friction on agents' attention to distant signals; the next subsection then interprets the news media as the institution that produces the summaries on which the inattentive public relies.

We adopt the reduced-form behavioral specification of \citet{gabaix2020behavioral}, in which agents' perceived law of motion of the state vector $\mathbf{X}_t$. This is operationalized by downweighted future variables with an attention parameter $M^h \in [0,1]$ for households (with $M^h = 1$ recovering full rationality), as well as $M^f$ for firms making pricing decisions. We do not re-derive the linearization here; following Gabaix's derivation, the equilibrium around the zero-inflation steady state takes the cognitively discounted New-Keynesian form
\begin{align}
    x_t &= M^h\, E_t[x_{t+1}] - \sigma^{-1}(r_t - E_t[\pi_{t+1}] - r_t^n), \label{eq:IS}\\
    \pi_t &= \beta M^f\, E_t[\pi_{t+1}] + \kappa\, x_t, \label{eq:NKPC}\\
    r_t &= \rho_r\, r_{t-1} + (1-\rho_r)(\phi_\pi \pi_t + \phi_x x_t + \phi_s s_t) + \varepsilon^u_t + \varepsilon^a_{t-\tau}, \label{eq:taylor_lin}
\end{align}
where lowercase letters denote log deviations and $\kappa$ takes its standard New-Keynesian value. We treat $M^h$ and $M^f$ as exogenous attention parameters, in line with \cite{gabaix2020behavioral}. The Taylor rule (\ref{eq:taylor_lin}) is itself a reduced-form structural specification: the loading $\phi_s$ on sentiment is taken as a parameter of the rule rather than derived from a central-bank optimisation problem.

\subsection{Media as a narrative friction}\label{sec:model_narratives}

Cognitive discounting reduces the response of current decisions to expected future variables, but it does not by itself say who summarizes those distant signals for an inattentive public. The news media are the natural candidate, and the link between the two frictions is tight: limited attention is what gives media narratives traction, and media narratives are what an inattentive public actually reads \citep{briand2024inflation}. The channel through which the central bank reaches agents is therefore not direct but is mediated by journalists, who can amplify, distort, or filter the original signal \citep{nimark2014man,chahrour2021sectoral,carroll2003macroeconomic}. Sentiment $s_t$ is an intermediate variable, not a primitive: it summarizes how the public reads both the current state of the economy and the central bank's stance toward the future. Closely related, \cite{andre2026narratives} embed household narratives as subjective causal models of inflation in an otherwise standard New-Keynesian framework and show that they affect aggregate equilibrium outcomes. Our framework can be read as a media-side counterpart: the narratives are produced by an intermediate institution (journalists) rather than held directly by agents, and they enter both the law of motion of $s_t$ and the central bank's reaction function.

\begin{figure}[H]
\caption{Examples of monetary-policy media coverage in Canada\label{fig:articles}}
\vspace{0.1cm}
\centering
\begin{minipage}{0.32\textwidth}\centering\includegraphics[width=\linewidth]{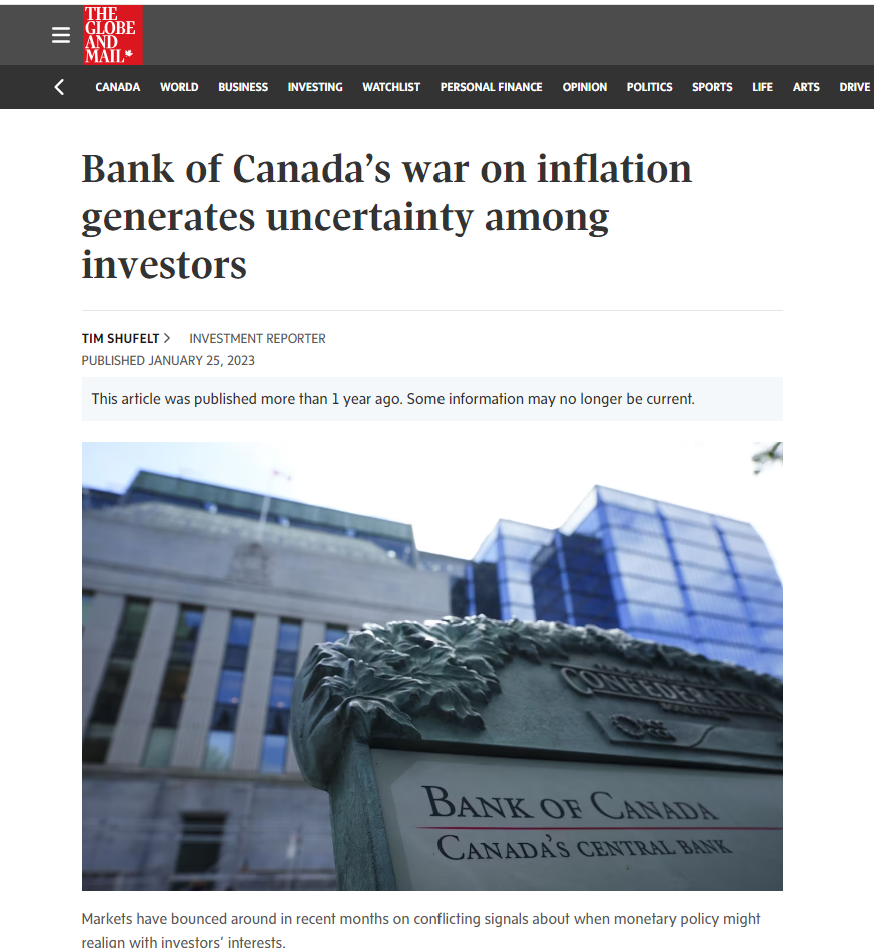}\end{minipage}\hfill
\begin{minipage}{0.32\textwidth}\centering\includegraphics[width=\linewidth]{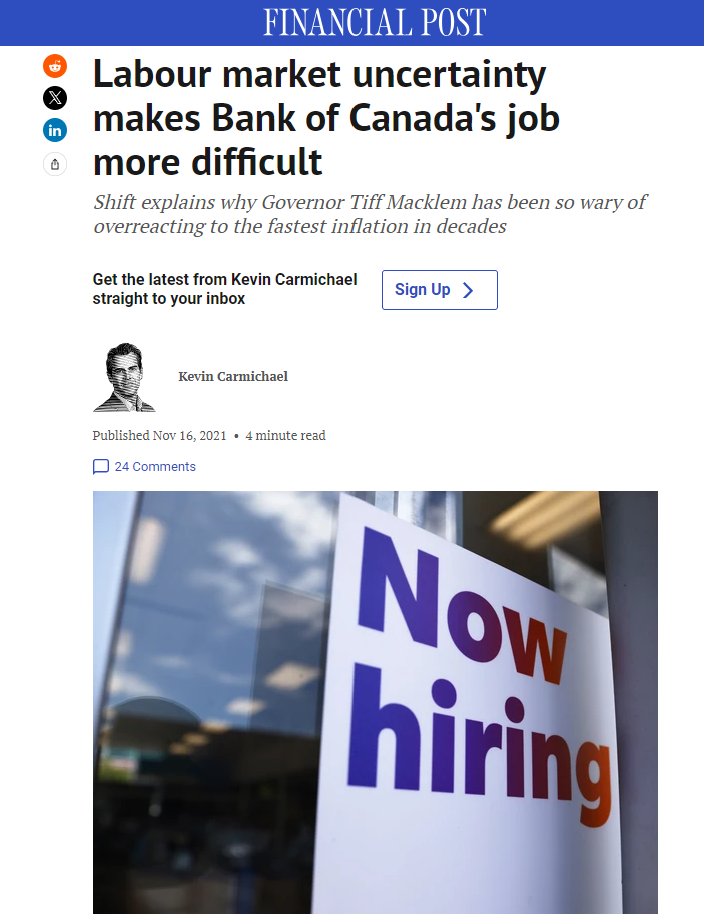}\end{minipage}\hfill
\begin{minipage}{0.32\textwidth}\centering\includegraphics[width=\linewidth]{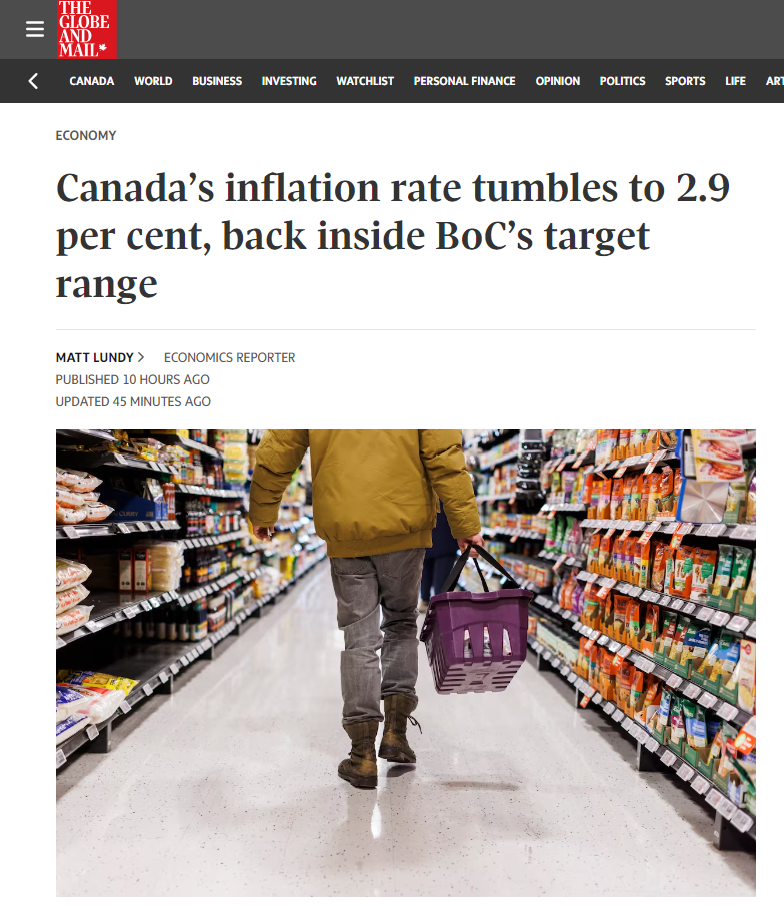}\end{minipage}
\end{figure}

Figure~\ref{fig:articles} illustrates the editorial dimension on three Canadian articles published around inflation-policy turning points. The same underlying policy environment is rendered as a war on inflation that generates uncertainty, as an obstacle that complicates the central bank's task, or as a return to the target range, depending on the journalistic angle. The factual content of each article overlaps substantially with the macroeconomic news the central bank itself reads, but the headline tone, framing, and word choice are not pinned down by that content alone. The variation that is left after conditioning on observable fundamentals and on the central bank's communication is what we call the editorial component, and it is the empirical counterpart of the residual narrative innovation $\varepsilon^s_t$ in the law of motion below.

We model the law of motion of media sentiment as
\begin{equation}\label{eq:sentiment_lom}
    s_t = \rho_s\, s_{t-1} + \sum_{k=0}^\tau M^{\tau-k} \big[ \lambda_x x_{t-\tau+k} + \lambda_\pi \pi_{t-\tau+k} + \lambda_a \varepsilon^a_{t-\tau+k} \big] + \varepsilon^s_t,
\end{equation}
treating $s_t$ as an aggregate of the press tone read by households and firms. This decomposes $s_t$ into four blocks: an autoregressive component capturing slow-moving narrative momentum, a fundamentals component (output gap and inflation up to horizon $\tau$), a communication component built from the announced policy shocks $\{\varepsilon^a_{t-\tau+k}\}_{k=0}^\tau$, and a residual term $\varepsilon^s_t$ that the model treats as orthogonal to the previous three blocks. Strictly speaking, $\varepsilon^s_t$ is the noise term of the LoM in the sense of the news and noise literature \citep{lorenzoni2009theory, blanchard2013news, angeletos2013sentiments}; we label it a \emph{narrative shock} as a substantive interpretation, since the orthogonal residual of a sentiment LoM is most naturally attributed to editorial choices and framing \citep{nimark2014man, chahrour2021sectoral}. The orthogonality of $\varepsilon^s_t$ to fundamentals and communication is an identifying assumption, not an empirical claim about the data: it is the assumption that, jointly with the sign restrictions of Section~\ref{sec:svar_signs}, lets the SVAR isolate the narrative shock as a separate structural object. The discount factor $M \in (0,1]$ in (\ref{eq:sentiment_lom}) has the same interpretation as $M^h$ in (\ref{eq:IS})--(\ref{eq:NKPC}), reflecting that agents read distant news with declining intensity. The fundamentals loadings $\lambda_x, \lambda_\pi > 0$ encode the assumption that the press reads good fundamentals (high $x_t, \pi_t$) as positive sentiment.\footnote{There are only demand-side shocks in the model.} The sign of the loading $\lambda_a$ on the announced policy shock is more subtle, because an announcement of a future easing ($\varepsilon^a < 0$) admits two readings.

Under $\lambda_a < 0$, the press reads the announcement at face value like households: an anticipated easing is read as positive news because the rate is going to fall. \cite{grigoli2026monetary} provide direct survey evidence consistent with this reading. Using a  randomized-information experiment they show that U.S. households interpret a higher federal funds rate as raising inflation and worsening economic conditions, a pattern that fits a cost-channel reading of monetary policy rather than the central-bank information effect of \cite{romer2000federal}. Therefore, households read an anticipated rate cut as good news for the outlook, which is exactly the face-value reading that $\lambda_a < 0$ encodes on the media side. Both the fundamentals block and the communication block of (\ref{eq:sentiment_lom}) then contribute positively to $s_0$.

Under $\lambda_a > 0$, the press attaches to the announcement a sign that is opposite to its substantive policy content, through a central bank information effect on the media side: journalists read through the literal cut to the implicit signal of incipient weakness that motivates it, so the announced easing is framed as bad news for the outlook \citep{romer2000federal, nakamura2018high}. Under this case, the impact response of sentiment to an anticipated easing is positive, $s_0 > 0$, only when the fundamentals block of the LoM dominates the direct loading of the announcement, $\lambda_x x_0 + \lambda_\pi \pi_0 > \lambda_a |\varepsilon^a_0|$. We verify numerically that the inequality holds under our calibration. Even when it fails, the contemporaneous lean against the wind on $r_t$ can still hold under a weaker condition: the policy response to inflation outweighs the negative sentiment contribution to the rule, $\phi_\pi \pi_0 + \phi_x x_0 > \phi_s |s_0|$. With $\phi_\pi = 1.5$ and $\phi_s = 0.5$, the policy weight on $\pi_0$ is three times the weight on $s_0$, so this weaker condition is satisfied for substantially smaller values of $\pi_0$ than the dominance for $s_t$.

This decomposition is the structural counterpart of the diagram in Figure~\ref{sec:fig1}, and it has direct implications for empirical work. Three of the four blocks of $s_t$ are mechanically correlated with the policy innovations in equation~(\ref{eq:taylor_lin}): the autoregressive component is predetermined and feeds into $r_{t-1}$; the fundamentals component covaries with $r_t$ through the central bank's response to $\pi_t$ and $x_t$; the communication component is a function of the same announced shocks $\varepsilon^a$ that already enter the policy rule. Only the residual $\varepsilon^s_t$ is plausibly orthogonal to $\varepsilon^u_t$. Section~\ref{sec:taylor} returns to this decomposition when discussing the identification of $\phi_s$.

\subsection{Impulse responses}\label{sec:model_irf}

We solve the linearized system (\ref{eq:IS})-(\ref{eq:taylor_lin}) together with the sentiment law of motion (\ref{eq:sentiment_lom}) under the calibration in Table~\ref{tab:calibration}, and trace out impulse responses to a one-standard-deviation \emph{anticipated} expansionary shock to the policy rate, $\varepsilon^a_{t-4}$, under different combinations of the attention parameter $M^h$ and the sentiment loading $\phi_s$.

\begin{table}[H]
\centering
\caption{Calibration of the behavioral NK model.\label{tab:calibration}}
\begin{tabular}{llc@{\hskip 1em}llc}
\toprule
Param & Description & Value & Param & Description & Value \\
\midrule
$\beta$    & Discount factor              & $0.99$         & $\tau$        & Announcement horizon  & $4$            \\
$\theta$   & Calvo reset prob.            & $0.75$         & $M^h$         & Household attention   & $\{0.8, 1.0\}$ \\
$\sigma$   & Inv. intertemporal elasticity & $1$           & $M^f$         & Firm attention        & $0.85$         \\
$\kappa$   & NKPC slope                   & $0.15$         & $M$           & Sentiment discount    & $0.85$         \\
$\rho_r$   & Taylor smoothing             & $0.9$          & $\rho_s$      & Sentiment persistence & $0$            \\
$\phi_\pi$ & Inflation response           & $1.5$          & $\lambda_x$   & Output-gap loading    & $1$            \\
$\phi_x$   & Output-gap response          & $0$            & $\lambda_\pi$ & Inflation loading     & $1$            \\
$\phi_s$   & Sentiment response           & $\{0, 0.5\}$   & $\lambda_a$   & Communication loading & $1$            \\
\bottomrule
\end{tabular}
\end{table}

\begin{figure}[ht]
\centering
\caption{Impulse responses to an anticipated expansionary monetary policy shock ($\tau=4$)\label{fig:irf}}
\includegraphics[width=0.95\linewidth]{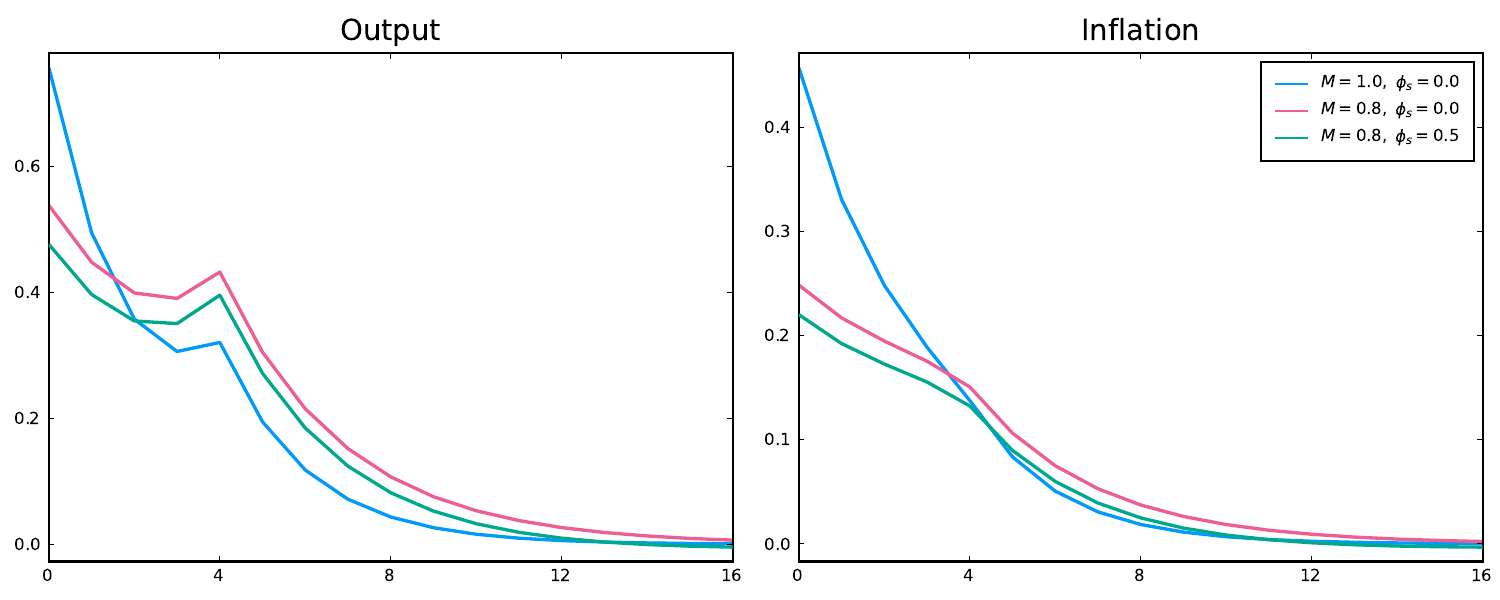}
\end{figure}

Figure~\ref{fig:irf} traces impulse responses of output and inflation to the anticipated expansionary shock under three calibrations: $M^h = 1.0, \phi_s = 0$ (full attention, no sentiment feedback); $M^h = 0.8, \phi_s = 0$ (cognitive discounting, no sentiment feedback); and $M^h = 0.8, \phi_s = 0.5$ (cognitive discounting with active sentiment feedback). Three patterns are worth noting.

First, lowering household attention from $M^h = 1$ to $M^h = 0.8$ flattens and lengthens the responses of output and inflation, with the peak shifted from impact toward the announced implementation date $t = 4$: cognitive discounting is a propagation mechanism in itself, since forward-looking agents do not fully internalize the announced future cut at the moment it is revealed.

Second, when $\phi_s > 0$ the policy rate \emph{rises} on impact, despite the announcement being for a future cut. The mechanism is a lean against the wind that runs through the sentiment law of motion and the reaction function. At $t = 0$ the announced cut has not yet materialized, so $r_0$ moves only through $(1-\rho_r)(\phi_\pi \pi_0 + \phi_x x_0 + \phi_s s_0)$. Forward-looking households and firms anticipate the future easing and bid up current $\pi_0$ and $x_0$ through the cognitive-discounting Euler equation and Phillips curve. The press loads on this incipient optimism through the fundamentals block of (\ref{eq:sentiment_lom}), and $s_0$ rises under the dominance condition $\lambda_x x_0 + \lambda_\pi \pi_0 > \lambda_a |\varepsilon^a_0|$ discussed in Section~\ref{sec:model_narratives}, which holds in our $M^h = 0.8$ calibration. The central bank reads this rise in $s_0$ as a signal of incipient overheating and tightens contemporaneously via $\phi_s s_0 > 0$. The net effect of the lean is to dampen the expansionary impact of the announcement: under $\phi_s = 0.5$ the responses of output and inflation lie \emph{inside} the corresponding $\phi_s = 0$ benchmark, closer to the zero-line. 

Third, were an independent narrative shock $\varepsilon^s_t$ to realize alongside the announcement, the policy rate would move in the same direction as $\varepsilon^s_t$: a positive narrative shock would reinforce the contractionary lean already at work, leading to a stronger rate hike on impact and a more muted real response; a negative narrative shock would produce the opposite. The model therefore predicts that media-driven optimism or pessimism, conditional on the same fundamental announcement, can amplify or dampen the macroeconomic effects of monetary policy.

The model has three empirical implications. The sentiment law of motion gives $s_t$ a role in shifting household and firm expectations beyond their own persistence; the Taylor rule puts $\phi_s$ inside the policy rule as a structural loading whose consistent estimation has to handle the simultaneous determination of $s_t$ and $r_t$ through (\ref{eq:taylor_lin})--(\ref{eq:sentiment_lom}); and the impact responses derived above deliver sign restrictions that identify the three structural shocks $\{\varepsilon^u, \varepsilon^a, \varepsilon^s\}$ in a Bayesian SVAR.

\section{Measuring sentiments and narratives about monetary policy}\label{sec:measure}

We extract signals about monetary policy narratives from news media using four features of natural language processing (NLP) techniques : (i) topic or narrative with Latent Dirichlet Allocation (LDA) developed by  \cite{blei2003latent} ; (ii) tone or sentiment with the dictionary method and large language models (LLMs), particularly FinBERT approach;  (iii) time dimension with temporal tagging and (iv) uncertainty-adjusted topic attention leveraging word embedding (word2vec)  as described by  \cite{mikolov2013distributed}.\footnote{See \cite{ash2023text} for an overview of the usage of text in economic research.} We measure aggregated and topic-adjusted sentiment/uncertainty/tense from over 50,000 news articles published in major Canadian newspapers including the \textit{National Post, Calgary Herald, Edmonton Journal, Montreal Gazette, Ottawa Citizen, Regina Leader-Post, The Globe and Mail, Vancouver Sun, and the Victoria Times-Colonist} from January 1977 to January 2024.\footnote{A parallel corpus of US press coverage of Canadian monetary policy from January 1980 onward is also available, totalling approximately 6,000 articles. The \textit{Wall Street Journal} is the largest contributor with about 4,000 articles, followed by the \textit{New York Times}, \textit{Washington Post}, \textit{Los Angeles Times}, and \textit{Chicago Tribune}. We use it as a robustness sample given its substantially smaller volume than the Canadian corpus.}  Our data, sourced from the ProQuest Database, is available at daily frequency and in real-time.

Sentiment and confidence have long been measured using qualitative data that draw on surveys asking people whether they think the economy is improving, staying the same, or getting worse. Such sentiment measures have been shown to have predictive power for macroeconomic outcomes, even when controlling for other factors. We consider model-free and model-based approaches to measure the tone or sentiment about the monetary policy sentences. 

The model-free methods use dictionaries from \cite{gonzalez2021monetary}. Model-based estimation of the tone of each document uses BERT  (bidirectional encoder representations from transformers. BERT is a deep learning model introduced by \citep{devlin2018bert} for natural language processing. It focuses on sequences of words rather than simply counting particular words in isolation, as in the economic policy uncertainty index. Pretrained on a corpus of more than 3.3 million words, with over 110 million parameters \citep{devlin2018bert}, BERT avoids the subjective use of judgment in choosing the dictionary used to define, in our context, sentiment. Importantly, by using surrounding text, rather than simply reading from left to right, BERT aims to establish the context of the text and thus infer the meaning of language that might otherwise be ambiguous. To this day transformer models are considered the state-of-the-art NLP technique, having superseded all other language
models.\footnote{Indicatively, GPT models, such as OpenAI’s ChatGPT, LLaMA, PaLM, and Claude are also based on the transformer paradigm. Also, see \citealp{chu2024history} for an overview.}

\subsection{Dictionary and model-based sentiment}\label{sec:measure_dict}
We first measure the tone of the monetary-policy news corpus in a model-free manner, following the bag-of-words methodology of \cite{tetlock2007giving} and \cite{loughran2011liability}. We count the frequency of words appearing in a tone lexicon calibrated to central-bank vocabulary by \cite{gonzalez2021monetary} and aggregate over the corpus at time $t$ as
$$
\operatorname{Tone}_t^{\mathrm{Dictionary}}=\frac{n_t^{\text{Positive}}-n_t^{\text{Negative}}}{n_t^{\text{Words}}},
$$
where $n_t^{\text{Positive}}$ and $n_t^{\text{Negative}}$ are the counts of positive and negative tonal words at $t$, and $n_t^{\text{Words}}$ is the total word count. A higher value indicates a less negative sentiment in the news.

To reduce the subjective judgment involved in choosing a dictionary, we construct an alternative model-based measure of tone using FinBERT \citep{yang2020finbert}, a variant of BERT pre-trained on financial texts.\footnote{The FinBERT weights are released by \citet{yang2020finbert} and were estimated on a corpus of corporate filings, analyst reports and financial news running from $1990$ through $2014$, with the underlying BERT pre-training cut-off in October 2018. Both endpoints predate the bulk of our 1977--2024 sample on the right tail and predate or coincide with it elsewhere, so FinBERT scoring of articles in our corpus does not embed knowledge of post-publication macroeconomic outcomes. The look-ahead concern that we discuss for CBILA in Section~\ref{sec:cbila} therefore does not apply to FinBERT.} We apply FinBERT to the same corpus at the sentence level: each sentence is classified as positive, negative, or neutral, and we aggregate at time $t$ using the same ratio as in Section~\ref{sec:measure_dict} with sentences in place of words,
$$
\text{Tone}_t^{\text{FinBERT}}=\frac{n_t^{\text{Positive}}-n_t^{\text{Negative}}}{n_t^{\text{Sentences}}}.
$$

\begin{figure}[ht]
\vspace{-0.2cm}
    \centering
    \caption{Monetary Policy Sentiment: Dictionary vs. FinBERT (3-Month Moving Average) \label{gr_sent}}
    \includegraphics[width=0.9\linewidth,height=0.4\textheight]{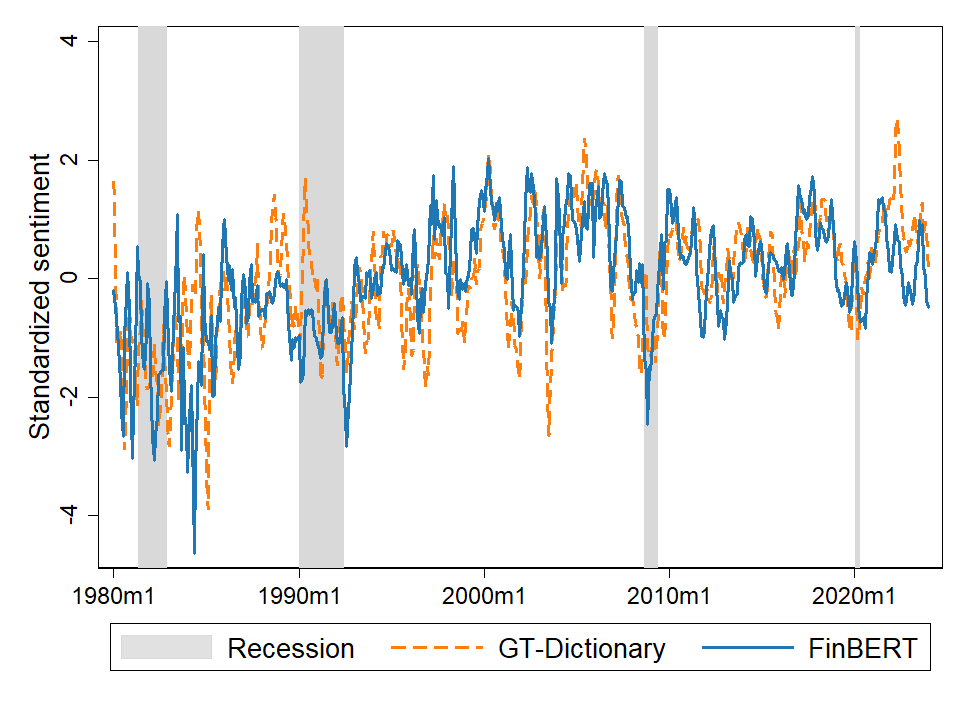}  
\end{figure}

We plot the two standardized sentiments in Figure \ref{gr_sent} and note how the BERT-based estimates of tone appear to exhibit greater concordance with C.D. Howe recessions than those using the GT dictionary. As expected, media monetary policy sentiment
falls in recessions.

However, Figure \ref{gr_sent} indicates that sentiment did not decline as sharply during the COVID-19 recession or the global financial crisis as it did following the 1980 recession when interest rates were significantly increased to combat high inflation.

\subsection{Generative-AI sentiment: the CBILA framework}\label{sec:cbila}

The dictionary and FinBERT measures share a structural limitation: their scoring rule is fixed once and for all, either by the lexicon or by the pre-trained transformer weights. They do not exploit context beyond the document being scored. Recent generative large language models can, in principle, do better. They can read a sentence about monetary policy with attention to qualifiers, conditional statements, implicit forward guidance, and credibility cues, and they can return both a classification and an explanation \citep{shah2025words,gambacorta2024cb,hansen2024can, ash2023text}. We exploit this capability to construct an alternative, semantically richer sentiment measure, which we use primarily to validate the FinBERT-based series across architectures.

We package the procedure into the Central Bank Intelligent Language Agent (CBILA). Each sentence in the corpus is passed through a structured prompt that instructs the model to act as a central-bank communication analyst and to return three outputs: a topic label among Monetary Policy, Inflation, Financial Stability, and similar policy-relevant categories; a temporal-orientation label among forward-looking, backward-looking, or neutral; and a sentiment score in $[-1,1]$ together with a categorical label among positive, negative, neutral, and irrelevant. The prompt also asks for a short rationale (generative models only), which we use only for diagnostic inspection. This reasoning-with-constraints design is intended to improve reproducibility and output stability, two attributes that matter for policy applications \citep{hansen2024can, gambacorta2024cb, gorodnichenko2023voice,ayivodji2023housing, korinek2023generative}.

\begin{figure}[!t]
\centering
\caption{Cross-model correlations of sentence-level sentiment scores\label{fig:llmheatmap}}
\includegraphics[width=0.85\linewidth,height=0.4\textheight]{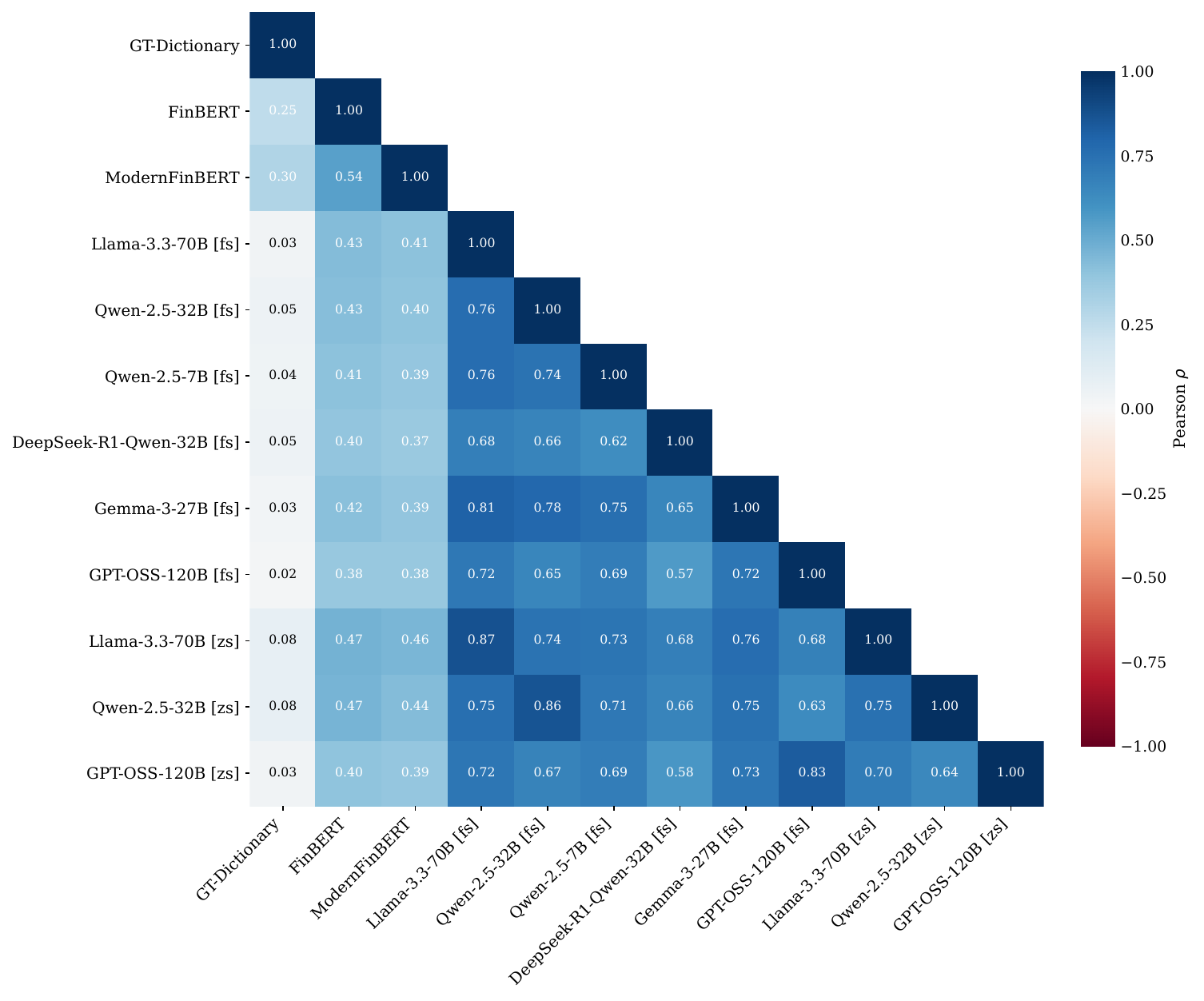}
\par\smallskip
\begin{minipage}{0.95\linewidth}\footnotesize
\textit{Note}: GT-Dictionary, (Modern)FinBERT are included as benchmark sentiment measures. The remaining models are evaluated under both few-shot ([fs]) and zero-shot ([zs]) prompting configurations within the CBILA pipeline.
\end{minipage}
\end{figure}

\begin{figure}[!t]
\centering
\caption{Inter-model agreement (Krippendorff's $\alpha$) for sentiment\label{fig:llminter}}
\includegraphics[width=0.85\linewidth,height=0.4\textheight]{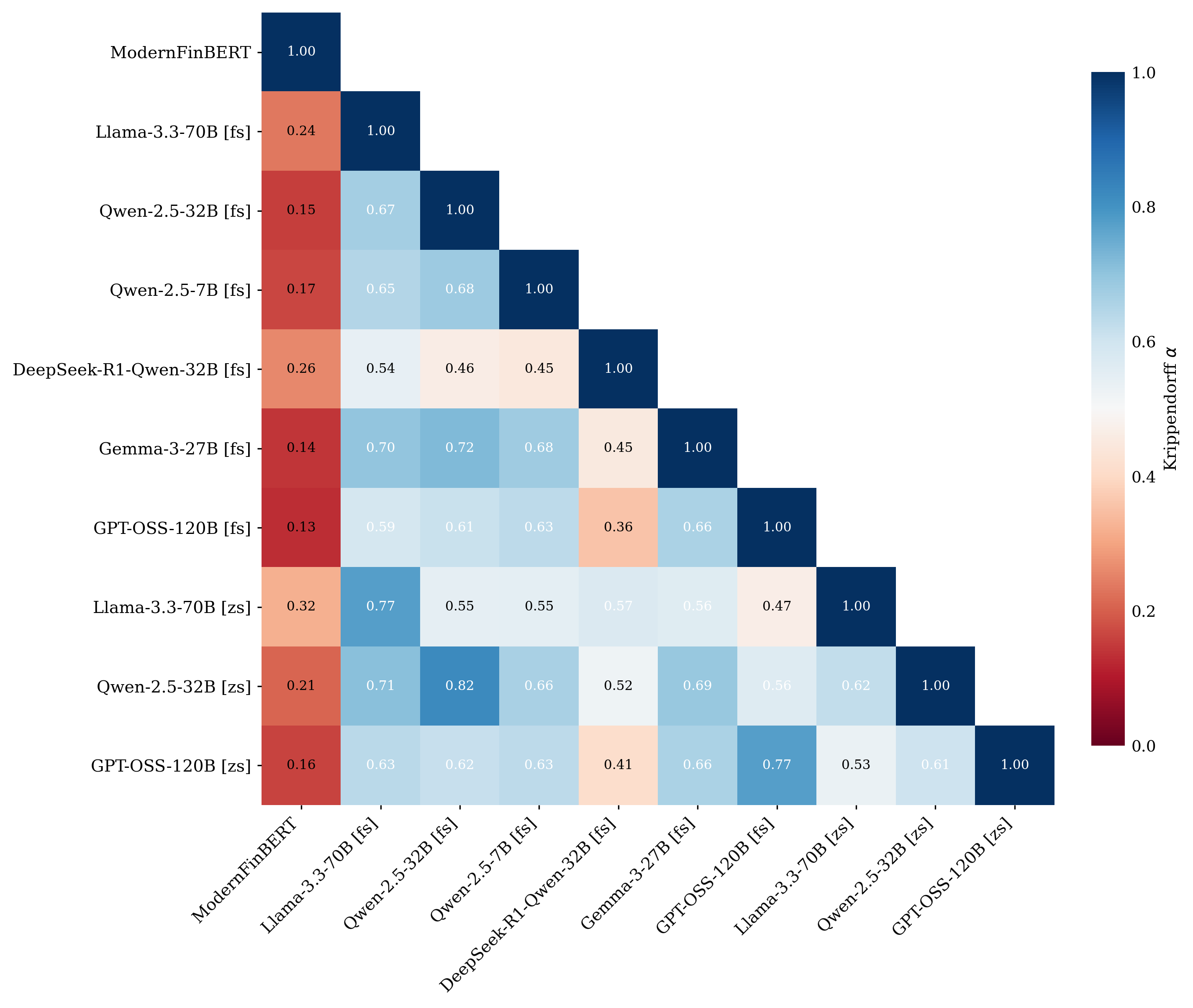}
\par\smallskip
\begin{minipage}{0.95\linewidth}\footnotesize
\textit{Note}: ModernFinBERT is included as a sentiment-only benchmark; the remaining models are evaluated under both few-shot ([fs]) and zero-shot ([zs]) prompting configurations within the CBILA pipeline.
\end{minipage}
\end{figure}

We apply the full CBILA pipeline to six instruction-tuned generative LLMs: DeepSeek-R1 Qwen 32B, Qwen 2.5 (7B and 32B), Gemma 27B, GPT-OSS-120B, and Llama 3.3 70B. For each sentence, these models produce all four CBILA outputs (topic label, temporal orientation, sentiment score and label, and rationale). FinBERT and ModernFinBERT are encoder-only sentence-level sentiment classifiers and cannot be run through the CBILA prompt pipeline; we include them in the cross-model comparison of Figures~\ref{fig:llmheatmap}--\ref{fig:llminter} as sentiment-only benchmarks, to assess the robustness of the sentiment dimension across architectures and training corpora. The rationale output is therefore available only for the six generative models and is used purely for diagnostic inspection. For the generative models, we evaluate both few-shot ([fs]) and zero-shot ([zs]) prompting configurations. The constrained structure of the CBILA prompt is designed to improve classification consistency and reduce stochastic variation across runs. 

Figures~\ref{fig:llmheatmap} and \ref{fig:llminter} report the pairwise correlations of continuous sentiment scores and the Krippendorff $\alpha$ for nominal classifications, respectively. Three substantive findings stand out. First, modern instruction-tuned models agree strongly with each other across both few-shot and zero-shot configurations, with many pairwise correlations exceeding 0.7 and substantial inter-model agreement for the largest models according to Krippendorff $\alpha$. This overall agreement structure remains broadly stable across historical subperiods and alternative treatments of the irrelevant category. Second, agreement with the dictionary-based index is substantially lower, confirming that lexical methods capture only a subset of the semantic content that LLMs read. Third, this agreement holds across prompting configurations and across models trained in markedly different information environments, the U.S.-developed Gemma, GPT-OSS, and Llama families alongside the China-developed DeepSeek-R1 and Qwen families. This cross-corpus convergence is informative in its own right: \citet{cao2026foreign} document substantial divergence between U.S. and Chinese LLMs on firm-level financial predictions, and the absence of comparable divergence in our setting indicates that sentence-level annotation of central-bank communication isolates a semantic signal that is robust to LLM training-corpus heterogeneity. The categorical decision pattern is also stable across architectures, with a dominant role for neutral and negative categories that reflects the informational and risk-focused tone of monetary-policy reporting. Additional robustness checks reported in Appendix Figures~\ref{fig:alpha_irrelevant} and \ref{fig:alpha_subperiods} show that the agreement structure remains qualitatively similar when excluding the irrelevant category from pairwise comparisons and across major monetary-policy subperiods.

These results validate CBILA as a measurement framework. They do, however, raise important concerns about the direct use of CBILA-based measures as regressors. The reason is look-ahead bias, a concern increasingly documented in the recent LLM-forecasting literature \citep{fariae2024ai,crane2025total,alam2026chatmacro,eliseev2026fake}. The LLMs we use are pre-trained on text corpora that extend up to 2023 or 2024, so when we score an article published in, say, 1990, the model's internal representations may implicitly encode information about post-1990 macroeconomic outcomes. This contamination is structural. It is unambiguously problematic for the pseudo-out-of-sample forecasting exercise of Section~\ref{sec:forecasting}, where a CBILA-based predictor would inflate predictive performance through information unavailable to a real-time forecaster. It can also be problematic for causal regressions like the Taylor-rule estimation of Section~\ref{sec:taylor}: the IV strategy with lagged macro instruments does not eliminate the bias if the leakage component of $s_t$ correlates with those instruments through the persistence of the macro process. We document the empirical relevance of this concern for the Taylor rule in Section~\ref{sec:taylor_results}, and provide the formal analysis in Appendix~\ref{app:cbila_taylor}.

By contrast, the dictionary index is essentially free of look-ahead bias because the lexicon is fixed ex ante. The FinBERT measure is also largely safe, since the underlying model was trained on data ending well before the bulk of our sample. Both are therefore appropriate as the reference measures for the three empirical exercises that follow. In this paper, the CBILA series plays the complementary role of validating the qualitative content of the FinBERT signal across LLM architectures. A real-time-safe version of the CBILA pipeline would require temporally aligned language models, such as ChronoBERT/ChronoLLM in \citet{he2025chronologically}, DatedGPT in \citet{yan2026datedgpt}, or related temporal-alignment approaches. These frameworks aim to align the model’s effective information set to a target historical date through explicit knowledge cutoffs, temporally restricted training windows, or time-aware alignment procedures. In such a design, each article would be scored only by a model whose knowledge cutoff predates the article publication date, thereby preserving temporal consistency and substantially mitigating look-ahead contamination. This would make CBILA more suitable as a primary regressor in forecasting, expectations, and Taylor-rule exercises; we leave this extension for future work. An alternative route, used in \cite{stevanovic2026whosaw} for inflation forecasting, is to design prompts that hold the training-leakage bias common across treatments and identify the variation of interest from cross-treatment differences. A parallel effort by the Bank of Canada uses a fine-tuned LLM to classify the tone of BoC communications as dovish, neutral, or hawkish, and documents that the tone of financial-sector commentary shifts in the direction of the tone the central bank uses \citep{wang2026sparks}.

\subsection{Temporal Focus and Forward-Looking Orientation}

A second dimension of monetary-policy communication concerns its temporal orientation, whether narratives in the media emphasize past performance, present conditions, or expectations about the future.
While the literature on central-bank communication has traditionally focused on the tone or sentiment of policy messages, much less attention has been devoted to their temporal horizon.
Yet this aspect is crucial: the degree to which communication looks forward, rather than backward, determines how information about policy intentions is transmitted to markets and the public.
Changes in the temporal balance of media narratives may thus signal shifts in how the public interprets policy credibility and the expected future stance of monetary policy.

To quantify this dimension, we employ a hybrid approach combining grammatical analysis and semantic filtering.
First, we use spaCy’s linguistic model to identify the dominant verb tense of each sentence: past, present, or future.
Second, we refine this classification using a curated dictionary of forward-looking expressions (e.g., will raise, is expected to cut, likely to remain).
Third, we apply FinBERT-FLS, a fine-tuned BERT model trained to detect forward-looking statements (FLS) even when the grammatical form is not explicitly future.
Each sentence is then assigned to one of four mutually exclusive categories (Past, Present, Future, or Other), thereby providing a detailed map of how policy discussions evolve over time (see \cite{ayivodji2023housing} for methodological details).

Two patterns in the temporal decomposition deserve mention. First, present- and past-tense reporting dominate the coverage, but the forward-looking share is non-trivial and rises during regime transitions such as the adoption of inflation targeting in the early 1990s and the introduction of unconventional policies after 2008 (Appendix Figure~\ref{fig:tensepattern}). Second, the forward-looking share tracks the tenure of successive Bank of Canada governors who expanded communication and forward guidance, and falls during periods dominated by short-term operational adjustments (Appendix Figure~\ref{fig:futureovertime}). The temporal horizon of media coverage therefore responds to both macroeconomic regime shifts and institutional communication style.

\begin{figure}[t]
\vspace{0.1cm}
\centering
\caption{Correlation heatmaps of Full Sentiment, Past Focus, Present Focus, and Future Focus.\label{fig:heatmap}}
\includegraphics[width=0.9\linewidth,height=0.45\textheight]{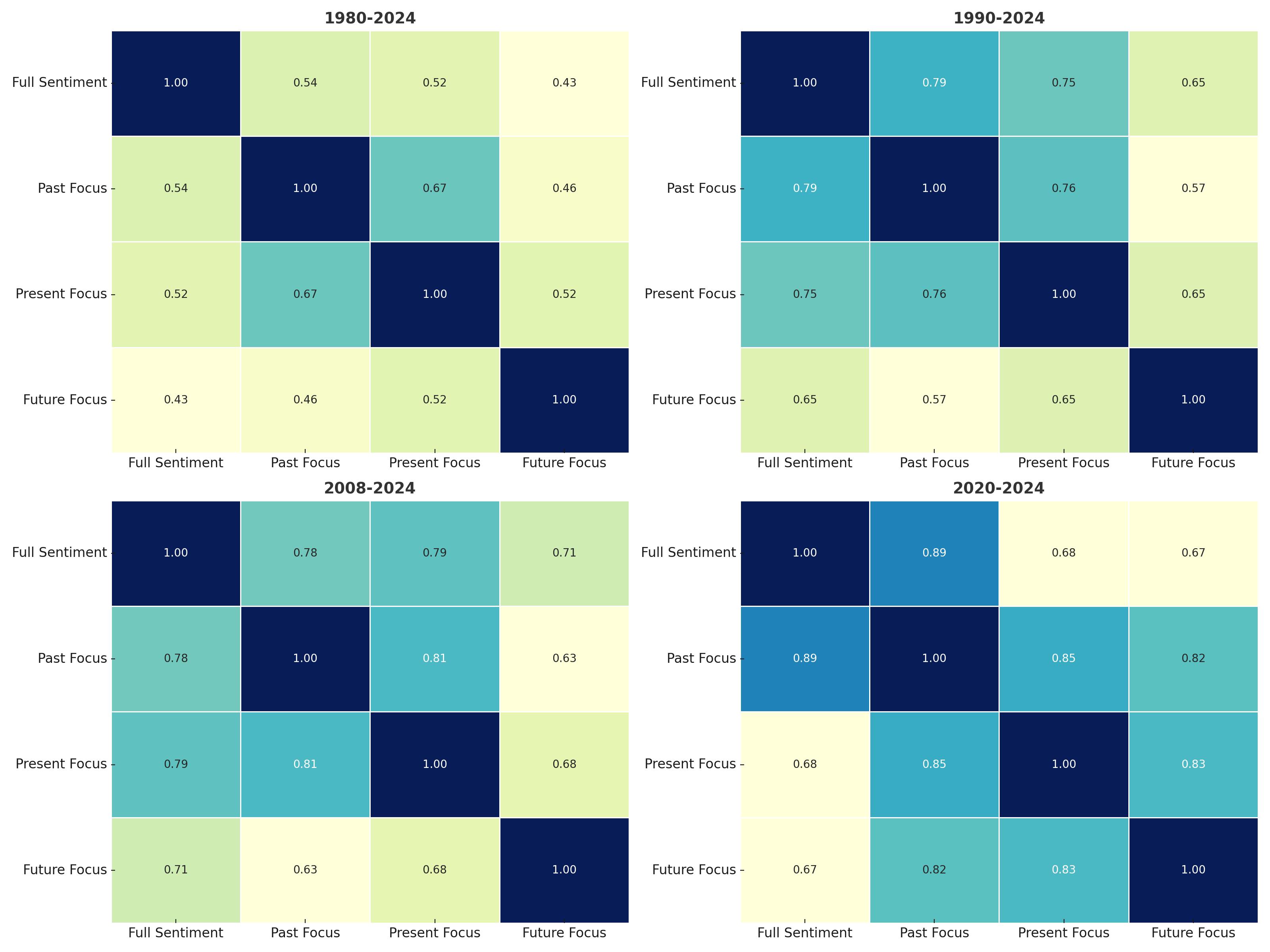}
\end{figure}

Figure~\ref{fig:heatmap} summarizes the correlations between the four temporal dimensions of sentiment (Full Sentiment, Past Focus, Present Focus, and Future Focus) over the full sample (1980–2024) and key subperiods.
Adding this temporal layer reveals that correlations vary across policy regimes.
Notably, the link between Full Sentiment and Future Focus weakens during high-uncertainty phases, suggesting that forward-looking narratives contain additional information not captured by aggregate tone.
This distinction highlights that media optimism about current policy performance and confidence in future actions are conceptually distinct, and only the latter carries predictive power for the evolution of macroeconomic outcomes and policy rates.

Taken together, Figures~\ref{fig:tensepattern}–\ref{fig:heatmap} show that the temporal structure of monetary-policy communication is far from static.
The forward-looking component fluctuates systematically with macroeconomic regimes and institutional practices, serving as a proxy for the public’s attention to the future stance of policy.
By isolating this component, we capture a unique aspect of the monetary-policy transmission mechanism: how expectations are formed and reinforced through the media. The next section explores whether this forward-looking sentiment indeed predicts macroeconomic variables and the estimated parameters of a Taylor-type policy rule, thereby testing its informational and expectational content.

\section{Empirical content of media sentiment}\label{sec:empirical}

The model of Section~\ref{sec:model} delivers a clean prediction about media sentiment and a more open empirical question. The clean prediction follows from the sentiment law of motion in equation~(\ref{eq:sentiment_lom}): movements in $s_t$ should shift the expectations of households and firms beyond what their own persistence already explains. We test this prediction directly in Section~\ref{sec:expectations} using the Bank of Canada household expectations survey. The more open question is whether the same media indicators carry information for realized macroeconomic outcomes. The model itself does not pin down the predictive content of $s_t$ for $\pi$ or $y$ over and above the conventional macro information set, but a coherent reading of the framework suggests that they should: if media coverage is informative enough for the central bank to read it (Section~\ref{sec:taylor}) and for the public to react to it, it should leave a footprint on subsequent realized outcomes. Section~\ref{sec:forecasting} treats this as a separate empirical check, in line with the broader text-as-data forecasting literature.

\subsection{Media sentiment and household expectations}\label{sec:expectations}

The model of Section~\ref{sec:model_narratives} treats media narratives as the channel through which the central bank's communication and the state of fundamentals reach private agents. A direct empirical implication is that the sentiment indicators built in Section~\ref{sec:measure} should help explain short- to medium-horizon expectations of macroeconomic variables, beyond what is already captured by the persistence of these expectations themselves. We test this implication on Canadian household survey data.

We use the Bank of Canada Canadian Survey of Consumer Expectations (CSCE), which provides quarterly measures of household expectations of inflation at multiple horizons, expected wage growth over the next 12 months, and perceptions of past wage growth. The estimation sample runs from 2014Q4 to 2023Q3, that is, 36 quarterly observations. For each expectation series $E_t z_{t,t+h}$ we estimate:
\begin{equation}\label{eq:expectations}
    E_t z_{t,t+h} = \alpha + \gamma\, E_{t-1} z_{t-1,t+h-1} + \beta\, s_t + \epsilon_t,
\end{equation}
where $z \in \{\pi, W\}$ denotes inflation or wage growth, $h$ is the horizon in quarters, and $s_t$ is the FinBERT aggregate sentiment about Canadian monetary policy from Section~\ref{sec:measure}. The lag of the expectation absorbs the persistence and the slow-moving common trend in survey beliefs, so $\beta$ measures the additional information that current media sentiment carries about the path of household expectations. The CSCE backward-looking wage variable does not fit (\ref{eq:expectations}) as a forward expectation; we report the corresponding regression with a one-period-lagged perception on the right-hand side as a perception equation, for comparability with the forward-looking columns.

\begin{table}[htbp]
\centering
\small
\renewcommand{\arraystretch}{1.05}
\caption{Media sentiment and household expectations, 2014Q4--2023Q3}
\label{tab:expectations}
\def\sym#1{\ifmmode^{#1}\else\(^{#1}\)\fi}
\begin{tabular}{l*{6}{c}}
\hline\hline
 & \multicolumn{4}{c}{\textbf{Inflation expectations}} & \multicolumn{2}{c}{\textbf{Wage expectations}} \\
\cline{2-5}\cline{6-7}
 & $E_t[\pi_{t}]$ & $E_t[\pi_{t+4}]$ & $E_t[\pi_{t+8}]$ & $E_t[\pi_{t+20}]$ & $E_t[W_{t,t+4}]$ & $E_t[W_{t-4,t}]$ \\
\hline
$s_t$ & 0.131\sym{**} & 0.204\sym{***} & 0.178\sym{**} & 0.128 & 0.269\sym{*} & 0.200\sym{**} \\
 & (0.000) & (0.011) & (0.025) & (0.421) & (0.010) & (0.082) \\[2pt]
Lagged expectation & 0.940\sym{***} & 0.857\sym{***} & 0.849\sym{***} & 0.680\sym{***} & 0.644\sym{***} & 0.825\sym{***} \\
 & (0.000) & (0.000) & (0.000) & (0.000) & (0.000) & (0.000) \\
\hline
$N$ & 36 & 36 & 36 & 34 & 36 & 35 \\
Adj.\ $R^2$ & 0.897 & 0.947 & 0.870 & 0.346 & 0.718 & 0.520 \\
\hline\hline
\multicolumn{7}{p{14cm}}{\textit{Notes:} OLS estimates of equation~(\ref{eq:expectations}) on the Bank of Canada Canadian Survey of Consumer Expectations (CSCE). $p$-values in parentheses. \sym{*} $p<0.10$, \sym{**} $p<0.05$, \sym{***} $p<0.01$. The dependent variable is the survey-based expectation of inflation at horizon $h$ quarters, expected wage growth over the next 12 months, or perceived wage growth over the past 12 months. The sentiment indicator $s_t$ is the FinBERT aggregate sentiment about Canadian monetary policy described in Section~\ref{sec:measure}. The two wage columns correspond respectively to households' expected wage growth over the next 12 months and perceived wage growth over the past 12 months from the CSCE. More specifically, the two wage columns correspond to the CSCE questions \texttt{CES\_C3\_NEXT\_12} and \texttt{CES\_C3\_PAST\_12}; the forward column $E_t[W_{t,t+4}]$ fits equation~(\ref{eq:expectations}) with $h=4$, while the backward column $E_t[W_{t-4,t}]$ is a perception equation with the lagged perception $E_{t-1}[W_{t-5,t-1}]$ on the right-hand side. \texttt{CES\_C3\_PAST\_12} starts in 2015Q1, one quarter later than \texttt{CES\_C3\_NEXT\_12}, which accounts for the difference between $N=36$ and $N=35$.} \\
\end{tabular}
\end{table}

Table~\ref{tab:expectations} reports the estimates. The sentiment coefficient $\beta$ is positive at every horizon and significant at conventional levels for inflation expectations at the contemporaneous, one-year, and two-year horizons, for the forward 12-month expected wage growth and for the backward 12-month perception of wage growth. The estimate at the five-year horizon for inflation is similar in magnitude to the one-year estimate but loses significance, consistent with the view that media coverage maps to short- and medium-run macroeconomic outlooks rather than to long-run anchored expectations. After controlling for the persistence of beliefs, a one-standard-deviation rise in $s_t$ shifts measured expectations by roughly 0.13 to 0.27 percentage points.

These results corroborate the role assigned to $s_t$ in equation~(\ref{eq:sentiment_lom}). The systematic link between sentiment and survey expectations supports the interpretation of media narratives as an active component of the expectations-formation process. This complements the experimental evidence in \cite{andre2026narratives}, who show that exogenously varying the narrative through which US households interpret inflation causally shifts their expectations of future inflation. The next subsection asks the complementary question of whether sentiment also helps predict realized macroeconomic outcomes, while Section~\ref{sec:taylor} examines whether the central bank's own behavior is consistent with this informational role.

\subsection{Media sentiment as a predictor of realized output and inflation}\label{sec:forecasting}

Beyond the direct test of the model in Section~\ref{sec:expectations}, we ask a complementary question: do the media indicators carry information for realized output and inflation, beyond what conventional macro predictors already deliver? The model of Section~\ref{sec:model} does not formally answer this question, but a positive finding would corroborate the broader informational content of media sentiment, which is a precondition for the central-bank behavior we estimate in Section~\ref{sec:taylor}. We follow the text-as-data forecasting literature \citep{bybee2024business,ash2023text,ayivodji2023housing,ellingsen2022news,andre2026narratives} and run pseudo-out-of-sample MIDAS forecasts of CPI inflation and real GDP growth.

We construct $244$ text-based predictors from the sentiment indicators of Section~\ref{sec:measure}, a topic decomposition of the corpus (twenty LDA topics), a media-based monetary policy uncertainty index, and tense-conditioned crosses of these dimensions. MIDAS regressions at horizons $h \in \{1, 2, 4, 6, 8\}$ quarters are estimated on the 1982Q1--2023Q3 sample (training 1982Q1--2002Q4, test 2003Q1--2023Q3) using the machine-learning competing models of \cite{goulet2022machine}: penalised regressions (LASSO, Ridge, Elastic Net), tree-based methods (Random Forest, XGBoost), neural networks with one and three hidden layers, and three ensembles that aggregate them. Construction of the uncertainty index, the LDA topic decomposition, the full forecasting setup, the model-by-horizon RMSE tables, and the SHAP variable-importance analysis are reported in Appendix~\ref{app:forecasting}.

Figure~\ref{fig:ens_all_sets} reports the relative RMSE of the all-models ensemble for four predictor sets, separately for GDP growth and CPI inflation. A value below one indicates an improvement over the random-walk benchmark.

\begin{figure}[!ht]
\centering
\caption{Text measures outperform the benchmark model in ensemble forecasting\label{fig:ens_all_sets}}
\begin{subfigure}{0.49\textwidth}
  \centering
  \caption{GDP forecasting}
  \includegraphics[width=\linewidth]{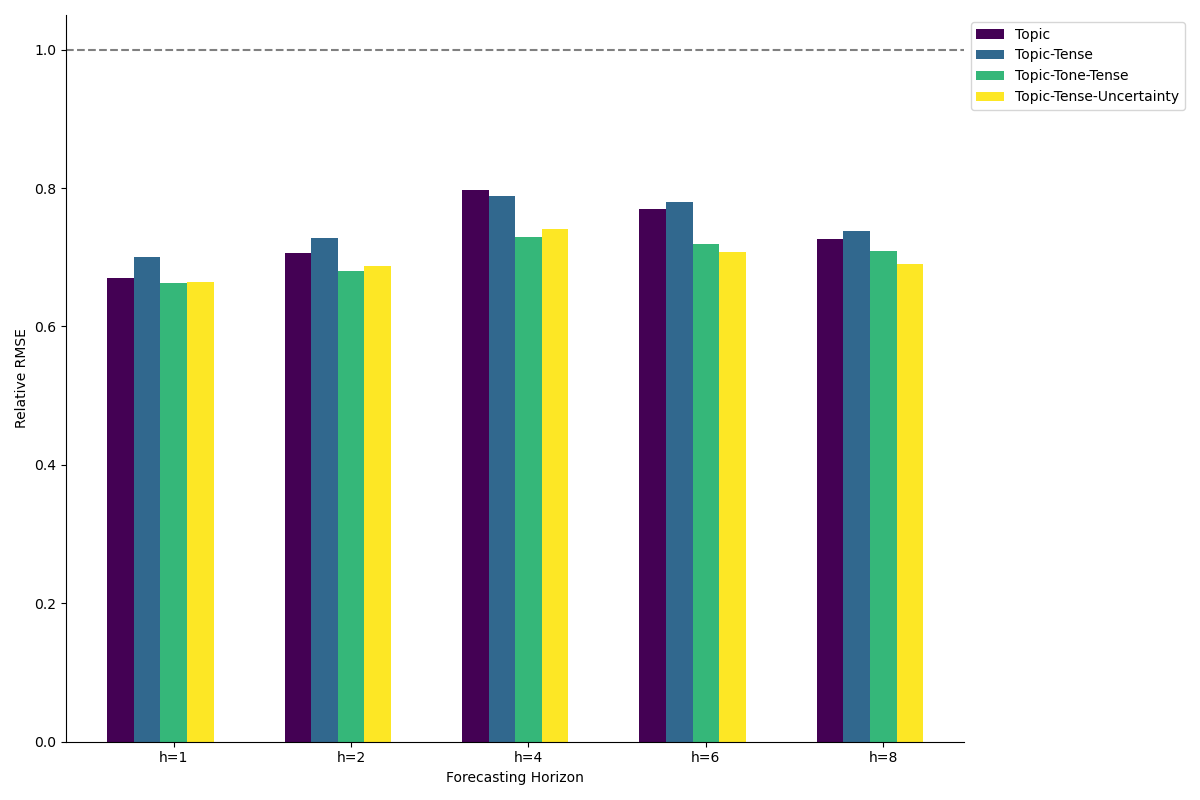}
\end{subfigure}\hfill
\begin{subfigure}{0.49\textwidth}
  \centering
  \caption{CPI forecasting}
  \includegraphics[width=\linewidth]{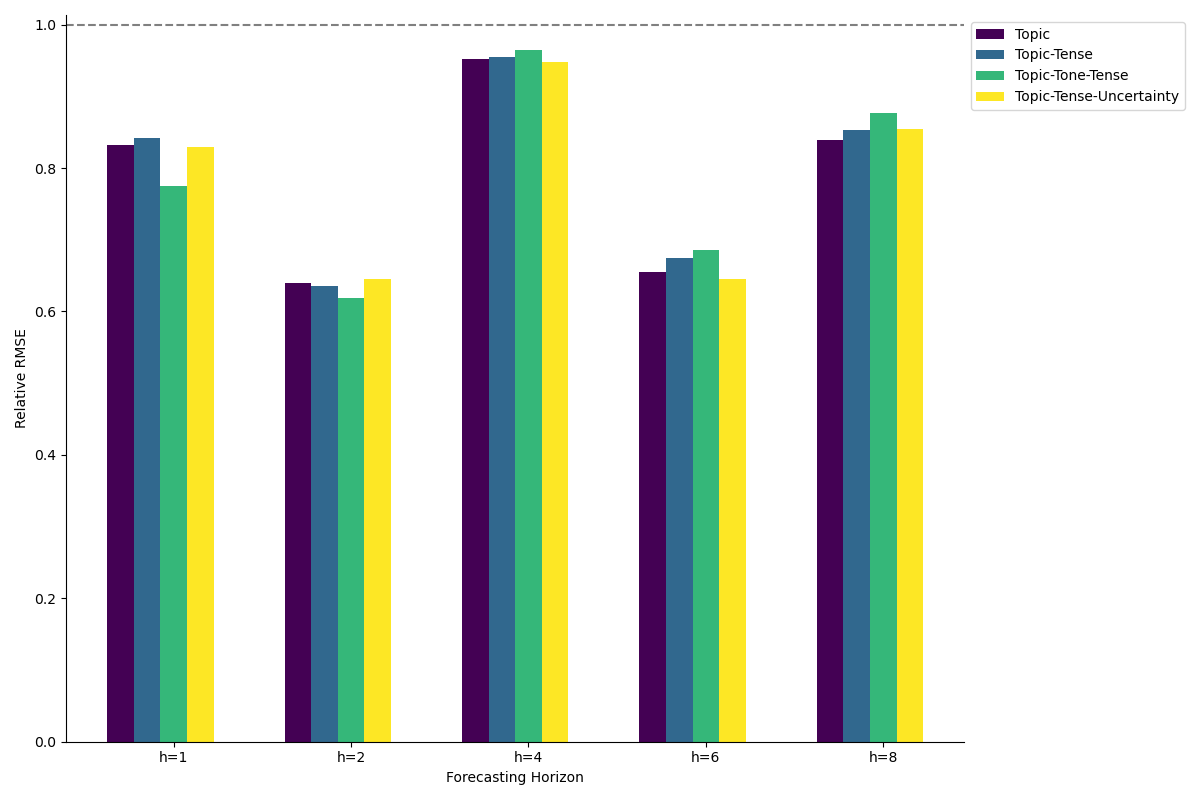}
\end{subfigure}

\vspace{0.3em}
{\scriptsize\textit{Notes}: Each coloured bar corresponds to a distinct predictor set. ``Topic'' denotes the 20 topic shares; ``Topic-Tense'' refers to the topic shares adjusted for tense; ``Topic-Tone-Tense'' (3T) adjusts for both tense and sentiment; ``Topic-Tense-Uncertainty'' (2TU) adjusts for tense and uncertainty.}
\end{figure}

Text-based predictors outperform the random-walk benchmark at most horizons. Across model families and predictor sets, relative RMSE for CPI inflation lies between $0.61$ and $0.96$ at horizons $h \in \{1,2,6,8\}$, with most cells significant under the Diebold-Mariano test, and is essentially flat or slightly above the benchmark at $h = 4$ (relative RMSE between $0.96$ and $1.06$, none significant); for real GDP growth the relative RMSE lies between $0.64$ and $0.77$ across all horizons (Appendix Tables~\ref{tab:main_results_3T_cpi} and \ref{tab:main_results_3T_gdp}). Gains are largest when feature sets combine topics with tense and either sentiment (3T) or uncertainty (2TU); adding tense alone to topics yields little improvement. For CPI inflation the largest gains concentrate at the two- and six-quarter horizons, while for GDP growth the gains are more uniform across horizons. The SHAP variable-importance analysis in Appendix~\ref{app:forecasting_shap} indicates that uncertainty-conditioned topics and their temporal variants (past, present, future) consistently appear among the top predictors, regardless of model family.

\section{The role for monetary policy: text-augmented Taylor rule}\label{sec:taylor}

The behavioral New-Keynesian model of Section~\ref{sec:model} suggests that perceptions of the future stance of monetary policy, embedded in media coverage, can move expectations and economic outcomes through the cognitive-discounting channel. A natural empirical question is whether the central bank itself appears to internalize these perceptions when setting the policy rate. We address this question by estimating forward-looking monetary policy reaction functions for the Bank of Canada, augmented with the media-based sentiment indicators constructed in Section~\ref{sec:measure}.

\subsection{Specification}\label{sec:taylor_spec}

Let $i_t$ denote the BoC overnight policy rate, $E_t \pi_{t+h_\pi}$ the BoC staff inflation forecast at horizon $h_\pi$, from the BoC Staff Economic Projections (SEP) database, $\pi^*_{t+h_\pi}$ the contemporaneous inflation target, and $E_t x_{t+h_x}$ the staff output-gap forecast at horizon $h_x$. Following \cite{pang2024boc}, we set $h_\pi=4$ and $h_x=2$, and we allow the long-run real neutral rate to be time-varying through the proxy $r^*_t$, defined as the 10-year government bond yield net of the inflation nowcast \citep{orphanides2001monetary}. The augmented inertial Taylor rule is
\begin{equation}\label{eq:taylor_aug}
    i_t = c + \rho\, i_{t-1} + \alpha\, r^*_t + \gamma \, E_t\!\left(\pi_{t+4} - \pi^*_{t+4}\right) + \beta\, E_t\!\left(x_{t+2}\right) + \delta\, s_t + \varepsilon_t,
\end{equation}
where $\rho \in (0,1)$ captures interest-rate smoothing, $\alpha$ is the loading on the time-varying neutral real rate, $c$ aggregates the time-invariant component of the neutral rate together with the constant of the rule, and $s_t$ is one of the standardized monetary-policy sentiment indicators built in Section~\ref{sec:measure}. We consider eight sentiment measures: (i) the dictionary-based aggregate sentiment (\textit{Dict}); (ii) the FinBERT aggregate sentiment (\textit{BERT}); (iii) the dictionary sentiment restricted to forward-looking sentences (\textit{Fut-Dict}); the FinBERT sentiment by temporal orientation, broken down into (iv) past (\textit{Past}), (v) present (\textit{Pres}), and (vi) forward-looking (\textit{Fut-BERT}); and the ModernFinBERT counterparts (vii) aggregate (\textit{M-BERT}) and (viii) forward-looking (\textit{Fut M-BERT}). ModernFinBERT is encoder-only with a pre-training cut-off that predates the bulk of the sample, so it carries no look-ahead bias by construction; the CBILA generative-LLM tones do, which is why we keep them separate in Appendix~\ref{app:cbila_taylor}. Specification~(\ref{eq:taylor_aug}) without sentiment ($\delta = 0$) is reported as the \textit{Baseline} column.

As in \cite{pang2024boc}, the estimation sample runs from 1991Q1, when the BoC adopted the inflation-targeting framework, to 2019Q4, the last quarter available at the SEP database, but also unaffected by the COVID-19 break in real-time data and forecasts. This delivers $T=110$ quarterly observations. All sentiment indicators are standardized prior to estimation, so that $\delta$ measures the basis-point response of the policy rate to a one standard-deviation movement in $s_t$.

\subsection{Estimation strategy}\label{sec:taylor_estim}

\cite{pang2024boc}, building on \cite{orphanides2001monetary}, argue that OLS yields consistent estimates of forward-looking Taylor rules when the right-hand-side variables are constructed from real-time staff projections. By construction, those projections are produced under the assumption that the current policy rate stays unchanged, and are therefore orthogonal to the contemporaneous policy shock. The lagged interest rate is also predetermined, and the time-varying neutral rate is built from the 10-year bond yield, which is exogenous to current monetary policy decisions at quarterly frequency.

The same logic does not extend to media sentiment $s_t$. The model of Section~\ref{sec:model_narratives} makes clear that $s_t$ and $r_t$ are jointly determined at date $t$: through equation~(\ref{eq:sentiment_lom}), $s_t$ depends on the contemporaneous output gap and inflation, which in turn react to the contemporaneous policy decision; conversely, through equation~(\ref{eq:taylor_lin}), $r_t$ depends on $s_t$. The simultaneity contaminates $\hat\delta_{\text{OLS}}$ in directions that depend on which monetary policy shock is at work and on which channel of $s_t$ dominates. An unanticipated tightening $\varepsilon^u_t > 0$ raises $r_t$ on impact (directly through the policy rule) and lowers $s_t$ through the fundamentals block, since output and inflation contract endogenously; the resulting negative covariance between $s_t$ and the policy innovation biases $\hat\delta_{\text{OLS}}$ \emph{downward}. An anticipated tightening $\varepsilon^a_t > 0$, by contrast, does not enter the policy rule at $t$ directly; through the lean against the wind derived in Section~\ref{sec:svar_signs}, $r_t$ falls slightly via $\phi_s s_t < 0$, and $s_t$ also falls because the fundamentals block dominates the communication block. The two co-moves generate a positive correlation that biases $\hat\delta_{\text{OLS}}$ \emph{upward}, but by a small amount given the small magnitude of the contemporaneous lean. The net direction of the OLS bias therefore depends on the relative magnitude of the two channels and on the mix of MP shocks in the sample. 

To address this simultaneity, we re-estimate (\ref{eq:taylor_aug}) by efficient two-step GMM in the spirit of \cite{clarida1998monetary,clarida2000monetary}. The lagged interest rate, the staff forecasts and the time-varying neutral rate are taken as predetermined, while $s_t$ is treated as endogenous. Identification of $\delta$ requires instruments that are correlated with $s_t$ but uncorrelated with the contemporaneous policy innovation; we do not require the instruments to project $s_t$ onto its narrative-innovation block, only to be predetermined with respect to the date-$t$ shock. Our base instrument set follows the standard CGG configuration in the Canadian Taylor-rule literature: lags 2 to 5 of the policy rate together with lags 1 to 4 of the staff inflation gap and the staff output gap, giving 12 excluded instruments and 11 overidentifying restrictions. Appendix~\ref{app:cbila_taylor} additionally reports estimates under a recent-lag set (lags 2 to 3 of the policy rate, lags 1 to 2 of the forecasts) and a distant-lag set (lags 4 to 5 of the policy rate, lags 3 to 4 of the forecasts); for the eight sentiment series of Section~\ref{sec:taylor_results} both alternatives deliver $\hat\delta$ quantitatively similar to the base set, but the picture differs for the CBILA generative-LLM tones discussed there. Instrument relevance is assessed with the first-stage $F$ on $s_t$ against the $F > 10$ benchmark of \cite{staiger1997instrumental}. Instrument exogeneity is assessed with Hansen's $J$ test of overidentifying restrictions, computed using a block wild Rademacher bootstrap ($B=500$ resamples, non-overlapping block length $L=4$ matching the HAC bandwidth) rather than the asymptotic $\chi^{2}$ approximation; we make this choice because the bootstrap test of the GMM overidentifying restrictions admits a higher-order asymptotic refinement over the $\chi^{2}$ critical values \citep{hallhorowitz1996}, and the wild scheme is robust to heteroskedasticity of unknown form in IV regressions \citep{davidsonmackinnon2010}.

A more substantive alternative replaces the lagged levels of the policy rate and forecasts with one-step revisions of the staff projections themselves: $E_{t-1}(\pi_{t+4}) - E_{t-2}(\pi_{t+4})$ for inflation, $E_{t-1}(x_{t+2}) - E_{t-2}(x_{t+2})$ for the output gap, and lag 2 of the policy rate. This forecast-revision set isolates the new information that the BoC staff incorporated between $t-2$ and $t-1$. It differs from the lagged-level instruments in two ways. First, revisions are a short-window flow rather than a stock of past information, so they capture the recent updates of the staff's view that are most likely to drive sentiment innovations. Second, because they are differences of two pre-period forecasts, they purge the slow-moving common trends in inflation and activity that link the lagged levels of $i_{t-2},\ldots,i_{t-5}$ to the structural shocks at $t$. This gives a much smaller instrument set (3 instruments, 2 overidentifying restrictions), which sharpens the orthogonality argument at the cost of statistical power.

\subsection{Results}\label{sec:taylor_results}

Table~\ref{tab:gmm_table1} reports the OLS estimates (Panel A) and the GMM estimates with the base instrument set (Panel B) over 1991Q1 to 2019Q4, with Newey-West HAC standard errors ($L=4$). The reference specification without sentiment (Column 1) reproduces a textbook inertial Taylor rule for the BoC: high smoothing ($\rho=0.89$), a positive loading on the time-varying neutral rate ($\alpha=0.12$), a significant response to the inflation gap ($\gamma=0.28$) and to the output gap ($\beta=0.11$). These magnitudes line up with the post-1991 estimates in \cite{pang2024boc}.

\begin{table}[htbp]
\centering
\scriptsize
\setlength{\tabcolsep}{2pt}
\renewcommand{\arraystretch}{0.95}
\caption{Taylor Rule Estimates, 1991Q1--2019Q4}
\label{tab:gmm_table1}
\begin{tabular}{lccccccccc}
\hline\hline
 & (1) & (2) & (3) & (4) & (5) & (6) & (7) & (8) & (9) \\
 & \textbf{Baseline} & \textbf{Dict} & \textbf{BERT} & \textbf{Fut (D)} & \textbf{Past} & \textbf{Pres} & \textbf{Fut (B)} & \textbf{M-BERT} & \textbf{Fut (MB)} \\
\hline
\multicolumn{10}{l}{\textbf{Panel A. OLS (Newey-West HAC standard errors, $L=4$)}} \\
\hline
$\rho$ (Lag rate) & 0.888*** & 0.899*** & 0.893*** & 0.937*** & 0.892*** & 0.905*** & 0.889*** & 0.901*** & 0.904*** \\
 & (0.037) & (0.036) & (0.042) & (0.029) & (0.043) & (0.044) & (0.039) & (0.043) & (0.041) \\
$\alpha$ (Neutral rate) & 0.119*** & 0.112*** & 0.109** & 0.078** & 0.115** & 0.099** & 0.118*** & 0.104** & 0.101** \\
 & (0.039) & (0.036) & (0.047) & (0.032) & (0.047) & (0.047) & (0.046) & (0.046) & (0.043) \\
$\gamma$ (Inflation gap) & 0.275** & 0.199* & 0.263** & 0.247** & 0.275** & 0.277** & 0.276** & 0.276** & 0.285** \\
 & (0.124) & (0.113) & (0.129) & (0.108) & (0.125) & (0.125) & (0.121) & (0.125) & (0.123) \\
$\beta$ (Output gap) & 0.108*** & 0.059* & 0.078 & 0.032 & 0.099** & 0.083** & 0.107*** & 0.073 & 0.073* \\
 & (0.042) & (0.035) & (0.048) & (0.035) & (0.043) & (0.041) & (0.041) & (0.045) & (0.042) \\
$\delta$ Dict &  & 0.238*** &  &  &  &  &  &  &  \\
 &  & (0.059) &  &  &  &  &  &  &  \\
$\delta$ BERT &  &  & 0.085 &  &  &  &  &  &  \\
 &  &  & (0.111) &  &  &  &  &  &  \\
$\delta$ Fut-Dict &  &  &  & 0.303*** &  &  &  &  &  \\
 &  &  &  & (0.058) &  &  &  &  &  \\
$\delta$ Past &  &  &  &  & 0.026 &  &  &  &  \\
 &  &  &  &  & (0.083) &  &  &  &  \\
$\delta$ Pres &  &  &  &  &  & 0.077 &  &  &  \\
 &  &  &  &  &  & (0.062) &  &  &  \\
$\delta$ Fut-BERT &  &  &  &  &  &  & 0.005 &  &  \\
 &  &  &  &  &  &  & (0.060) &  &  \\
$\delta$ M-BERT &  &  &  &  &  &  &  & 0.075 &  \\
 &  &  &  &  &  &  &  & (0.070) &  \\
$\delta$ Fut M-BERT &  &  &  &  &  &  &  &  & 0.081 \\
 &  &  &  &  &  &  &  &  & (0.052) \\
Constant & 0.161*** & 0.023 & 0.108 & -0.032 & 0.154** & 0.138** & 0.161*** & 0.120* & 0.121* \\
 & (0.056) & (0.069) & (0.099) & (0.069) & (0.062) & (0.063) & (0.056) & (0.073) & (0.062) \\
\hline
Obs.  & 110 & 110 & 110 & 110 & 110 & 110 & 110 & 110 & 110 \\
Adj. $R^{2}$  & 0.933 & 0.939 & 0.933 & 0.947 & 0.932 & 0.933 & 0.932 & 0.933 & 0.933 \\
\hline
\multicolumn{10}{l}{\textbf{Panel B. GMM (rich instrument set, 12 instruments, 11 over-identifying restrictions)}} \\
\hline
$\rho$ (Lag rate) & 0.888*** & 0.931*** & 0.941*** & 0.959*** & 0.969*** & 1.014*** & 0.942*** & 0.964*** & 0.984*** \\
 & (0.037) & (0.021) & (0.026) & (0.021) & (0.024) & (0.039) & (0.029) & (0.029) & (0.037) \\
$\alpha$ (Neutral rate) & 0.119*** & 0.076*** & 0.035 & 0.048** & 0.004 & -0.034 & -0.021 & 0.025 & 0.013 \\
 & (0.039) & (0.023) & (0.028) & (0.022) & (0.029) & (0.045) & (0.048) & (0.031) & (0.038) \\
$\gamma$ (Inflation gap) & 0.275** & 0.245*** & 0.240*** & 0.301*** & 0.299*** & 0.289*** & 0.406*** & 0.311*** & 0.400*** \\
 & (0.124) & (0.067) & (0.076) & (0.061) & (0.089) & (0.096) & (0.105) & (0.071) & (0.079) \\
$\beta$ (Output gap) & 0.108*** & 0.030 & -0.041 & 0.016 & -0.077 & -0.056 & -0.036 & -0.039 & -0.046 \\
 & (0.042) & (0.028) & (0.056) & (0.026) & (0.048) & (0.057) & (0.054) & (0.051) & (0.053) \\
$\delta$ Dict &  & 0.298*** &  &  &  &  &  &  &  \\
 &  & (0.068) &  &  &  &  &  &  &  \\
$\delta$ BERT &  &  & 0.369*** &  &  &  &  &  &  \\
 &  &  & (0.120) &  &  &  &  &  &  \\
$\delta$ Fut-Dict &  &  &  & 0.297*** &  &  &  &  &  \\
 &  &  &  & (0.052) &  &  &  &  &  \\
$\delta$ Past &  &  &  &  & 0.427*** &  &  &  &  \\
 &  &  &  &  & (0.097) &  &  &  &  \\
$\delta$ Pres &  &  &  &  &  & 0.417*** &  &  &  \\
 &  &  &  &  &  & (0.119) &  &  &  \\
$\delta$ Fut-BERT &  &  &  &  &  &  & 0.461*** &  &  \\
 &  &  &  &  &  &  & (0.148) &  &  \\
$\delta$ M-BERT &  &  &  &  &  &  &  & 0.265*** &  \\
 &  &  &  &  &  &  &  & (0.074) &  \\
$\delta$ Fut M-BERT &  &  &  &  &  &  &  &  & 0.317*** \\
 &  &  &  &  &  &  &  &  & (0.092) \\
Constant & 0.161*** & -0.013 & -0.049 & -0.005 & 0.034 & 0.023 & 0.249*** & 0.041 & 0.034 \\
 & (0.056) & (0.057) & (0.093) & (0.051) & (0.067) & (0.068) & (0.080) & (0.062) & (0.072) \\
\hline
Obs.  & 110 & 110 & 110 & 110 & 110 & 110 & 110 & 110 & 110 \\
First-stage $F$ & --- & 177.36 & 253.57 & 249.03 & 132.19 & 291.22 & 183.20 & 286.55 & 301.38 \\
Hansen $J$ $p$-value & --- & 0.312 & 0.346 & 0.360 & 0.654 & 0.612 & 0.676 & 0.386 & 0.532 \\
\hline\hline
\multicolumn{10}{p{15.5cm}}{\textit{Notes:} Newey-West HAC standard errors with $L=4$ in parentheses. * $p<0.10$, ** $p<0.05$, *** $p<0.01$. The dependent variable is the BoC overnight policy rate (\%, 110 quarterly observations). Panel B treats the sentiment variable as endogenous in two-step efficient GMM. Excluded instruments in Panel B: lags 2--5 of the policy rate, lags 1--4 of the inflation gap, and lags 1--4 of the two-quarter-ahead output gap. Columns: (1) baseline rule with no sentiment; (2) dictionary tone; (3) FinBERT tone; (4) future dictionary tone; (5)--(7) FinBERT past, present, future tone; (8)--(9) ModernFinBERT overall and future tone. GMM weighting and covariance estimators use the Bartlett kernel with bandwidth $L$. The Hansen $J$ $p$-value is obtained by block wild Rademacher bootstrap with $B=500$ resamples and non-overlapping block length 4 (matching the HAC bandwidth), motivated by the higher-order asymptotic refinement of the bootstrap test in GMM (Hall and Horowitz, 1996) and the robustness of the wild scheme to heteroskedasticity of unknown form in IV settings (Davidson and MacKinnon, 2010).} \\
\end{tabular}
\end{table}

Adding sentiment yields a clear pattern. Under OLS, only the dictionary-based aggregate (\textit{Dict}) and the dictionary forward-looking measure (\textit{Fut-Dict}) enter significantly. The FinBERT measures and their ModernFinBERT counterparts are individually small under OLS and indistinguishable from zero. Once sentiment is treated as endogenous and instrumented (Panel B), all eight sentiment indicators load positively on the policy rate with point estimates between $0.27$ and $0.46$ and significance at the 1\% level. First-stage $F$-statistics are large in every column, well above the Staiger--Stock benchmark, and the block wild bootstrap Hansen $J$ does not reject overidentification in any column.

Once $s_t$ is included and instrumented, the loading on the inflation gap remains positive and significant in most columns, while the output-gap coefficient becomes uniformly insignificant, echoing \cite{pang2024boc}'s point that the BoC's reaction function is dominated by inflation once forward-looking information is properly accounted for. The smoothing coefficient $\rho$ stays close to its baseline value, so sentiment captures a distinct dimension of the rule rather than an alternative source of inertia. The ModernFinBERT columns (\textit{M-BERT} and \textit{Fut M-BERT}) deliver $\hat\delta$ in the same range as FinBERT, with similar significance levels and $J$ $p$-values. This is the expected outcome of moving from FinBERT to a more recent encoder-only architecture that still predates the bulk of the sample.

We then re-estimate the GMM specification with the forecast-revision instruments. Table~\ref{tab:gmm_revisions} delivers the same qualitative message as Panel B of Table~\ref{tab:gmm_table1}: all eight sentiment indicators load positively and significantly. Point estimates are uniformly larger under the forecast-revision instruments and standard errors are wider, reflecting the smaller instrument set. First-stage $F$-statistics remain large, and the block wild bootstrap $J$ does not reject in any column. That two conceptually different instrument sets yield consistent estimates supports interpreting the loading on $s_t$ as a genuine response of the BoC to media sentiment, not an artefact of any particular instrument choice.

\begin{table}[!t]
\centering
\scriptsize
\setlength{\tabcolsep}{2pt}
\renewcommand{\arraystretch}{0.95}
\caption{GMM with Forecast-Revision Instruments, 1991Q1--2019Q4}
\label{tab:gmm_revisions}
\begin{tabular}{lccccccccc}
\hline\hline
 & (1) & (2) & (3) & (4) & (5) & (6) & (7) & (8) & (9) \\
 & \textbf{Baseline} & \textbf{Dict} & \textbf{BERT} & \textbf{Fut (D)} & \textbf{Past} & \textbf{Pres} & \textbf{Fut (B)} & \textbf{M-BERT} & \textbf{Fut (MB)} \\
\hline
$\rho$ (Lag rate) & 0.888*** & 0.931*** & 0.967*** & 0.975*** & 0.965*** & 1.033*** & 1.011*** & 0.986*** & 1.000*** \\
 & (0.037) & (0.035) & (0.056) & (0.040) & (0.060) & (0.068) & (0.096) & (0.052) & (0.062) \\
$\alpha$ (Neutral rate) & 0.119*** & 0.085** & -0.009 & 0.044 & 0.028 & -0.051 & -0.172 & 0.012 & -0.002 \\
 & (0.039) & (0.034) & (0.064) & (0.039) & (0.061) & (0.077) & (0.159) & (0.056) & (0.063) \\
$\gamma$ (Inflation gap) & 0.275** & 0.193* & 0.215 & 0.290*** & 0.286** & 0.319** & 0.631*** & 0.321*** & 0.367*** \\
 & (0.124) & (0.108) & (0.174) & (0.093) & (0.140) & (0.141) & (0.229) & (0.112) & (0.101) \\
$\beta$ (Output gap) & 0.108*** & 0.037 & -0.172 & 0.023 & -0.065 & -0.097 & -0.267 & -0.085 & -0.074 \\
 & (0.042) & (0.047) & (0.107) & (0.044) & (0.081) & (0.081) & (0.182) & (0.090) & (0.089) \\
$\delta$ Dict &  & 0.409*** &  &  &  &  &  &  &  \\
 &  & (0.154) &  &  &  &  &  &  &  \\
$\delta$ BERT &  &  & 0.792*** &  &  &  &  &  &  \\
 &  &  & (0.266) &  &  &  &  &  &  \\
$\delta$ Fut-Dict &  &  &  & 0.376*** &  &  &  &  &  \\
 &  &  &  & (0.132) &  &  &  &  &  \\
$\delta$ Past &  &  &  &  & 0.517*** &  &  &  &  \\
 &  &  &  &  & (0.186) &  &  &  &  \\
$\delta$ Pres &  &  &  &  &  & 0.592*** &  &  &  \\
 &  &  &  &  &  & (0.195) &  &  &  \\
$\delta$ Fut-BERT &  &  &  &  &  &  & 1.270** &  &  \\
 &  &  &  &  &  &  & (0.642) &  &  \\
$\delta$ M-BERT &  &  &  &  &  &  &  & 0.444*** &  \\
 &  &  &  &  &  &  &  & (0.170) &  \\
$\delta$ Fut M-BERT &  &  &  &  &  &  &  &  & 0.440** \\
 &  &  &  &  &  &  &  &  & (0.188) \\
Constant & 0.161*** & -0.076 & -0.337 & -0.077 & 0.018 & -0.026 & 0.184 & -0.071 & -0.047 \\
 & (0.056) & (0.122) & (0.230) & (0.116) & (0.123) & (0.127) & (0.229) & (0.136) & (0.146) \\
\hline
Obs.  & 110 & 110 & 110 & 110 & 110 & 110 & 110 & 110 & 110 \\
First-stage $F$ & --- & 92.34 & 128.95 & 71.42 & 64.07 & 89.15 & 43.12 & 235.60 & 188.79 \\
Hansen $J$ $p$-value & --- & 0.396 & 0.698 & 0.300 & 0.900 & 0.884 & 0.764 & 0.532 & 0.496 \\
\hline\hline
\multicolumn{10}{p{15.5cm}}{\textit{Notes:} Newey-West HAC standard errors with $L=4$ in parentheses. * $p<0.10$, ** $p<0.05$, *** $p<0.01$. The dependent variable is the BoC overnight policy rate. Two-step efficient GMM with sentiment treated as endogenous. Excluded instruments: lag~2 of the policy rate, the one-quarter revision $E_{t-1}(\pi_{t+4})-E_{t-2}(\pi_{t+4})$ in the staff inflation forecast, and the analogous revision $E_{t-1}(x_{t+2})-E_{t-2}(x_{t+2})$ in the staff output-gap forecast (3 instruments, 2 over-identifying restrictions). GMM weighting and covariance estimators use the Bartlett kernel with bandwidth $L$. Column labels as in Table~\ref{tab:gmm_table1}. The Hansen $J$ $p$-value is obtained by block wild Rademacher bootstrap (B=500, block length 4).} \\
\end{tabular}
\end{table}

These results show that media sentiment about Canadian monetary policy enters the BoC's estimated reaction function with a positive and economically meaningful loading, providing a partial-equilibrium counterpart to the behavioral New-Keynesian model of Section~\ref{sec:model}: the central bank's behavior is correlated with the public's perception of the policy stance, beyond what is encoded in standard inflation and activity forecasts, and is consistent with the cognitive-discounting channel through which media sentiment attenuates or amplifies forward guidance. The gap between OLS and GMM estimates, particularly for the FinBERT- and ModernFinBERT-based measures, indicates that simultaneity between sentiment and policy decisions is a first-order concern; treating sentiment as predetermined understates its role in the policy rule.

We also estimated the same specifications with the CBILA generative-LLM tones described in Section~\ref{sec:cbila}. These series carry potential look-ahead bias because the underlying models were pre-trained on text that extends into and beyond our 1991--2019 sample. Appendix~\ref{app:cbila_taylor} documents that the CBILA-based $\hat\delta$ estimates are smaller and less stable across instrument sets than the FinBERT- and ModernFinBERT-based estimates, that the close-vs-distant pattern in the IV estimate diverges sharply for the CBILA series while remaining flat for the no-leakage series, and that the block wild bootstrap Hansen $J$ test rejects overidentification for several CBILA series at conventional levels. We provide a formal account of the underlying IV identification failure in the same appendix. The findings in this section therefore rest on the no-leakage series; the CBILA evidence supports the construction of an alternative central-bank communication index but is not currently suitable as a primary regressor in causal regressions of this type.

\section{The Role of Media Sentiments for Monetary Policy Transmission}\label{sec:svar}

In this section we use the model of Section~\ref{sec:model} to derive the sign restrictions that identify three structural shocks in a Bayesian SVAR: an unanticipated monetary policy shock $\varepsilon^u$, an anticipated monetary policy shock $\varepsilon^a$, and a narrative shock $\varepsilon^s$.

\subsection{Sign restrictions implied by the behavioral NK model}\label{sec:svar_signs}

We trace the impact response of $(r_t, s_t, x_t, \pi_t)$ and the $h$-period-ahead expectations, $E_t[r_{t+h}]$, $E_t[x_{t+h}]$, and $E_t[\pi_{t+h}]$, to each of the three shocks under the expansionary normalisation: unanticipated easing $\varepsilon^u_0 < 0$, anticipated easing $\varepsilon^a_0 < 0$, and positive narrative $\varepsilon^s_0 > 0$. To keep the algebra transparent we work with the simplified version of the system (\ref{eq:IS})--(\ref{eq:taylor_lin}) in which $\rho_s = 0$, $\tau = 1$, and the loadings of the sentiment law of motion (\ref{eq:sentiment_lom}) are normalised to $\lambda_x = \lambda_\pi = \lambda_a = 1$, with all endogenous variables at their steady state at $t = -1$. We retain the Taylor smoothing parameter $\rho_r$, which generates the empirically relevant persistence in the policy rate. Under these simplifications the policy rule and sentiment law of motion at $t = 0$ become
\begin{align}
r_0 &= (1-\rho_r)\bigl(\phi_\pi \pi_0 + \phi_x x_0 + \phi_s s_0\bigr) + \varepsilon^u_0, \label{eq:r0_svar}\\
s_0 &= x_0 + \pi_0 + \varepsilon^a_0 + \varepsilon^s_0, \label{eq:s0_svar}
\end{align}
where the $\rho_r r_{-1}$ and $\varepsilon^a_{-1}$ terms have been eliminated by the steady-state assumption at $t = -1$. The IS and Phillips equations (\ref{eq:IS})--(\ref{eq:NKPC}) are unchanged.

For an unanticipated easing, $\varepsilon^u_0 < 0$ enters the Taylor rule additively and pushes the policy rate down on impact. Forward-looking households anticipate the lower rate path (smoothed by $\rho_r$) and bid up current output through the cognitively discounted IS curve; firms raise current inflation through the Phillips curve. With $x_0, \pi_0 > 0$, the LoM (\ref{eq:s0_svar}) gives $s_0 = x_0 + \pi_0 > 0$. The Taylor-rule bracket $\phi_\pi \pi_0 + \phi_x x_0 + \phi_s s_0$ is therefore positive and partially leans against the direct cut, but $\varepsilon^u_0$ dominates and $r_0 < 0$. Taylor smoothing then transmits the easing forward: the rate follows an AR(1) with persistence $\alpha \in (0, 1)$ (determined by an endogenous fixed point that adjusts $\rho_r$ for the bracket's feedback), and the forward expectations of $r$, $x$ and $\pi$ inherit the impact-period signs and decay geometrically at rate $\alpha$. The full impact pattern is $r_0 < 0$, $x_0 > 0$, $\pi_0 > 0$, $s_0 > 0$, with $E_0[r_h] < 0$, $E_0[x_h] > 0$, $E_0[\pi_h] > 0$ for all $h \ge 0$. The detailed derivation is in Appendix~\ref{app:svar_signs}.

For an anticipated easing, $\varepsilon^a_0 < 0$ enters the LoM (\ref{eq:s0_svar}) at $t = 0$ but the policy rule (\ref{eq:r0_svar}) only at $t = 1$, since it lags by $\tau = 1$. The system has a recursive structure. From $t \ge 2$ on, no further shocks hit and the system reverts to the autonomous stable mode of the unanticipated case, $r_t = \alpha r_{t-1}$ and $z_t = G r_{t-1}$. At $t = 1$, the announcement enters the Taylor rule additively and acts structurally like a one-period rate innovation: $r_1 = \alpha r_0 + \beta_g \varepsilon^a_0$ with the same $\beta_g$ as in the unanticipated case, and $z_1 = G r_0 + g \varepsilon^a_0$ with the same impact loadings $g$. At $t = 0$, the contemporaneous IS, Phillips, Taylor and LoM equations are solved jointly with these one-step-ahead expectations as forward inputs, and with the $\phi_s \varepsilon^a_0$ contribution to the rate that the LoM passes through. Two implications follow. First, the rate leans against the wind on impact, $r_0 > 0$, because forward-looking households and firms bid up current $x_0, \pi_0$ in anticipation of the future cut and the central bank reads the implied rise in sentiment as an incipient overheating signal. Second, the sign of $s_0$ depends on whether the fundamentals block of the LoM dominates the direct loading of the announcement,
\begin{equation}
\lambda_x x_0 + \lambda_\pi \pi_0 \;>\; \lambda_a |\varepsilon^a_0|,
\label{eq:fundamentals_dominance}
\end{equation}
which we adopt as an identifying assumption and verify numerically (Appendix~\ref{app:svar_signs}). Under this assumption, $s_0 > 0$. The inequality is not implied by standard NK calibrations: with $M^h \to 0$ the cognitive-discount damping shrinks $|x_0|, |\pi_0|$ to zero while the direct effect $|\varepsilon^a_0|$ is unchanged, so it fails and $s_0$ flips sign. Under the alternative case $\lambda_a < 0$ of Section~\ref{sec:model_narratives}, $s_0 > 0$ holds unconditionally. The full impact pattern is $r_0 > 0,\; x_0 > 0,\; \pi_0 > 0,\; s_0 > 0$, with $E_0[r_h] < 0,\; E_0[x_h] > 0,\; E_0[\pi_h] > 0$ for $h \ge 1$. The contemporaneous reduced-form policy rule
\begin{equation}
r_0 = (1-\rho_r)\bigl[\widetilde\phi_\pi \pi_0 + \widetilde\phi_x x_0 + \phi_s\, \varepsilon^a_0\bigr],
\qquad \widetilde\phi_\pi \equiv \phi_\pi + \phi_s,\;\; \widetilde\phi_x \equiv \phi_x + \phi_s,
\label{eq:r0_a}
\end{equation}
makes the lean explicit: the first two terms aggregate the fundamentals-driven response (boosted by the sentiment loading $\phi_s$), and the third is the small direct sentiment-pass-through of $\varepsilon^a_0$. In the standard rational-expectations NK model the lean exists only through the small endogenous moves in $\pi_0, x_0$; here it acquires an additional driver, $\phi_s s_0 > 0$, which follows from $\phi_s > 0$ in the policy rule (\ref{eq:taylor_lin}) and is supported by the GMM estimates of Section~\ref{sec:taylor_results}. The detailed derivation is in Appendix~\ref{app:svar_signs}.

For a positive narrative shock, $\varepsilon^s_0 > 0$ enters only the LoM (\ref{eq:s0_svar}) at $t = 0$ and is absent from the IS, Phillips and Taylor equations at every date. Substituting (\ref{eq:s0_svar}) into the Taylor rule (\ref{eq:r0_svar}) with $\varepsilon^u_0 = \varepsilon^a_0 = \varepsilon^a_{-1} = 0$ gives
\begin{equation}
r_0 = (1-\rho_r)\bigl[\widetilde\phi_\pi\, \pi_0 + \widetilde\phi_x\, x_0 + \phi_s\, \varepsilon^s_0\bigr],
\label{eq:r0_eps}
\end{equation}
with $\widetilde\phi_\pi, \widetilde\phi_x$ as in (\ref{eq:r0_a}). The narrative innovation enters the rate \emph{additively} through $\phi_s\, \varepsilon^s_0$, and the lean against the wind operates entirely through this sentiment channel since the standard Taylor inputs $\pi_0, x_0$ are themselves driven endogenously by $r_0$ in the contemporaneous block. Under $\phi_s > 0$ the central bank reads the autonomous rise in sentiment as an incipient overheating signal and tightens: $r_0 > 0$. The IS--NKPC block then contracts current fundamentals, $x_0 < 0$ and $\pi_0 < 0$. Substituting back into (\ref{eq:s0_svar}) gives $s_0 > 0$ because the direct innovation $\varepsilon^s_0$ dominates the equilibrium contraction $|x_0 + \pi_0|$; under the calibration of Table~\ref{tab:calibration} the contraction is on the order of one-sixth of $\varepsilon^s_0$. The forward expectations follow the same geometric decay as the unanticipated case but with the opposite sign on $r_0$. The full impact pattern is $r_0 > 0,\; x_0 < 0,\; \pi_0 < 0,\; s_0 > 0$, with $E_0[r_h] > 0,\; E_0[x_h] < 0,\; E_0[\pi_h] < 0$ for $h \ge 1$. The detailed derivation is in Appendix~\ref{app:svar_signs}.

Table~\ref{tab:svar_signs} shows the sign restrictions used to identify shocks in the SVAR. We retain only the restrictions on $r_t, s_t$ and on the three forward expectations; the contemporaneous responses of $x_t$ and $\pi_t$ are left unrestricted in the SVAR because they are model predictions of small magnitude (cognitive discounting damps them by $(M^h M^f)^\tau$ at the $\tau = 4$ horizon used in the empirics) and imposing their sign would discard rotations that the data would otherwise accept. 

\begin{table}[H]
\centering
\caption{Sign restrictions used in the SVAR.\label{tab:svar_signs}}
\begin{tabular}{lccccc}
\toprule
                                   & $r_t$ & $s_t$ & $E_t[r_{t+h}]$ & $E_t[x_{t+h}]$ & $E_t[\pi_{t+h}]$\\
\midrule
Unanticipated easing $-\varepsilon^u$ & $-$   & $+$   & $-$            & $+$            & $+$ \\
Anticipated easing  $-\varepsilon^a$  & $+$   & $+$   & $-$            & $+$            & $+$ \\
Positive narrative  $\varepsilon^s$   & $+$   & $+$   & $+$            & $-$            & $-$ \\
\bottomrule
\end{tabular}
\par\smallskip
\begin{minipage}{0.95\linewidth}\footnotesize\textit{Note}: Expansionary normalisation. The contemporaneous responses of $x_t$ and $\pi_t$ are unrestricted in the SVAR and not shown here.\end{minipage}
\end{table}

Anticipated and unanticipated MP shocks generate qualitatively similar paths for the realised macro variables, and the sign restrictions of Table~\ref{tab:svar_signs} differ between the two only on the contemporaneous policy rate. Following \citet{damico2023}, we condition the rotation on the joint behaviour of survey expectations and realised macro variables. For each shock $j$ and each candidate rotation $B^d$, the SVAR delivers two objects from the same draw: the impact response of the staff expectation, $B^d_{E,j}$, and the four-period-ahead IRF of the corresponding realised variable, $\widetilde B^d_{4,j}$. Under rational expectations the two coincide, and the shock-by-shock importance weight
\begin{equation}
w_j(B^d) \;\propto\; \exp\!\Bigl(-\tfrac{1}{2\delta}\,\bigl\|B^d_{E,j} - \widetilde B^d_{4,j}\bigr\|^2\Bigr), \qquad j \in \{\varepsilon^a, \varepsilon^u, \varepsilon^s\},
\label{eq:loose_rationality}
\end{equation}
penalises rotations under which they diverge. The hyperparameter $\delta$ controls the tightness of the penalty: $\delta \to 0$ approaches an indicator on exact equality (strict rationality), $\delta \to \infty$ removes it altogether. The weight is most informative for the $\varepsilon^u/\varepsilon^a$ pair: under an unanticipated easing $E[r]$ tracks the immediate fall in the realised rate, while under an anticipated easing the rate rises on impact through the lean but $E[r]$ is already low because the announced cut is priced in; the weight rewards rotations consistent with this divergence. The narrative shock is identifiable on the sign restrictions alone, since its forward-expectation block ($+,-,-$) differs from each MP shock on three sign moments.

Two substantive differences distinguish our identification from theirs. First, the contemporaneous lean against the wind on the policy rate has an additional driver in our model: in their setup the lean operates only through the standard Taylor channel $\phi_\pi \pi_0 + \phi_x x_0$, while we add the sentiment loading $\phi_s s_0$. Second, the friction that damps the forecast response to anticipated shocks differs: a signal-to-noise filter on the announcement in their model (a credibility friction), cognitive discounting on the perceived path in ours (a salience friction). Appendix~\ref{app:svar_dk} reports the technical comparison.

\subsection{Empirical implementation}\label{sec:svar_implementation}

We estimate a quarterly Bayesian VAR(2) on Canadian data over 1986Q4--2019Q4.\footnote{The start date is fixed by the data: 1986Q4 is the first vintage of the BoC staff projections database \citep{champagne2020bocsep} for which the four-quarter-ahead horizon ($h=4$) is available, with target year 1987Q4. The publicly described start of the database is 1987Q1, which corresponds to the first \emph{forward} target ($h=1$) of the 1986Q4 vintage; the vintage itself contains a 1986Q4 nowcast and forecasts for 1987Q1--Q4. With $p=2$ lags in the BVAR, the effective sample for the identified shocks runs from 1987Q2 to 2019Q4. The sample starts five years before the inflation-targeting regime adopted in 1991Q1, which is the start date of the Taylor-rule estimation in Section~\ref{sec:taylor}. The longer SVAR window exploits the additional identifying variation available before 1991Q1: it contains Black Monday, the bond-market crisis and the inflation-targeting transition itself.} The vector of endogenous variables is
\begin{equation}
Y_t = \bigl(E^S_t[r_{t+4}],\;\; E^S_t[\log Y_{t+4}],\;\; E^S_t[\log P_{t+4}],\;\; r_t,\;\; \log Y_t,\;\; \log P_t,\;\; \log H_t,\;\; i^{1\text{-}3y}_t,\;\; s_t\bigr)',
\label{eq:Y_svar}
\end{equation}
where the first three entries are the BoC staff projections at horizon $h=4$ (policy rate, log real GDP, log CPI), constructed from the BoC staff projections; $r_t$ is the BoC overnight rate; $\log Y_t$, $\log P_t$, $\log H_t$ are real GDP, headline CPI (seasonally adjusted via STL, \citealp{cleveland1990stl}) and total hours worked from the large macroeconomic database of \cite{fortin2022large}; $i^{1\text{-}3y}_t$ is the average yield on Government of Canada bonds with 1--3 years to maturity; and $s_t$ is the standardised FinBERT aggregate sentiment of Section~\ref{sec:measure_dict}, identical to the regressor in column 3 of Table~\ref{tab:gmm_table1}.\footnote{All log variables are multiplied by $100$ so that impulse responses read in percent, the policy rate and the bond yield are expressed in percentage points, and the staff projections of real GDP and CPI are in percentage-point deviations from their respective trends ($100 \times$ four-quarter log changes, level form). The sentiment series is standardised to mean zero and unit variance over the estimation sample.}

We consider the BoC staff projections rather than from private-sector survey forecasts (e.g., Consensus Economics Canada or BlueChip Economic Indicators) for three reasons. First, the staff projections are the information set the BoC actually consults when setting the policy rate, and so they are the natural empirical counterpart of the expectation block of the structural Taylor rule (\ref{eq:taylor_lin}). Second, the staff projections cover the full SVAR sample 1986Q4--2019Q4 at quarterly frequency with a consistent methodology, while Canadian private-sector survey forecasts typically start later and report annual averages rather than quarterly horizons, requiring an interpolation that introduces measurement noise into the survey-vs-VAR comparison underlying the loose-rationality weight. Third, staff projections sit upstream of the BoC's communication: they feed into the policy decision and into the Monetary Policy Report rather than reacting to them. Private-sector forecasts, by contrast, are produced after parsing BoC communication, press coverage and competing forecasts; they are downstream of both the communication channel of $s_t$'s law of motion and of the media-sentiment series itself. Using staff projections in the loose-rationality weight therefore avoids conditioning the identification of $\varepsilon^a$ and $\varepsilon^s$ on an external measure that is itself partly endogenous to those shocks.

We estimate the VAR under a Normal-Inverse-Wishart conjugate prior with Zellner-style shrinkage of the autoregressive coefficient matrix toward the OLS estimate, and sample $D_{\text{post}} = 500$ posterior draws of $(\Phi, \Sigma)$. Per posterior draw we generate $N_{\text{rot}} = 8000$ random orthogonal rotations using the QR algorithm of \citet{rubio2010structural}, and retain those for which one rotation column matches each of the three sign patterns of Table~\ref{tab:svar_signs}, weighted by the loose-rationality importance weight (\ref{eq:loose_rationality}) introduced in Section~\ref{sec:svar_signs}. We set the penalty hyperparameter at $\delta = 0.5$, the central calibration of \citet{damico2023}.\footnote{The three weights $(w_{\varepsilon^a}, w_{\varepsilon^u}, w_{\varepsilon^s})$ enter the posterior moments shock-by-shock. The procedure yields $3{,}179$ accepted rotations, with effective sample sizes near $3{,}177$ for each of the three identified shocks. Algorithmic details---prior covariance, exact sign-test, loose-rationality weight formula including the convention used to construct $\widetilde B^d_{4,j}$ for the policy-rate expectation, full diagnostics---are reported in Appendix~\ref{app:svar_algo}.}

\subsection{Impulse responses and the narrative shock}\label{sec:svar_irfs}

\begin{figure}[!t]
\centering
\caption{Impulse responses to the anticipated MP shock $\varepsilon^a$.\label{fig:svar_irf_eta}}
\includegraphics[width=0.8\linewidth,height=0.45\textheight]{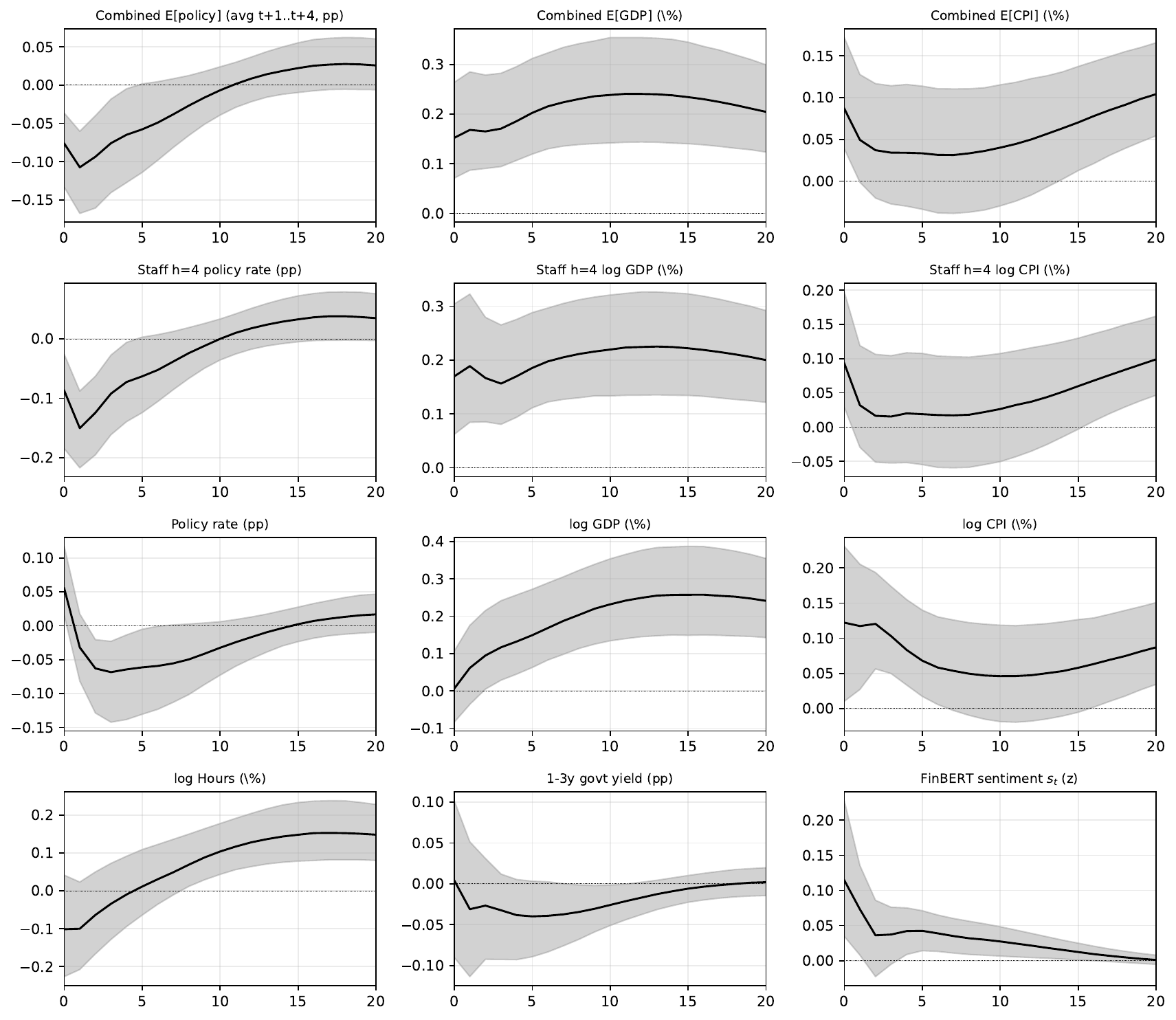}
\par\smallskip
\begin{minipage}{0.95\linewidth}\footnotesize\textit{Note}: Expansionary normalisation. Solid lines are weighted posterior medians; shaded bands are 68\% credible bands across rotations.\end{minipage}
\end{figure}

\begin{figure}[!t]
\centering
\caption{Impulse responses to the unanticipated MP shock $\varepsilon^u$.\label{fig:svar_irf_u}}
\includegraphics[width=0.8\linewidth,height=0.45\textheight]{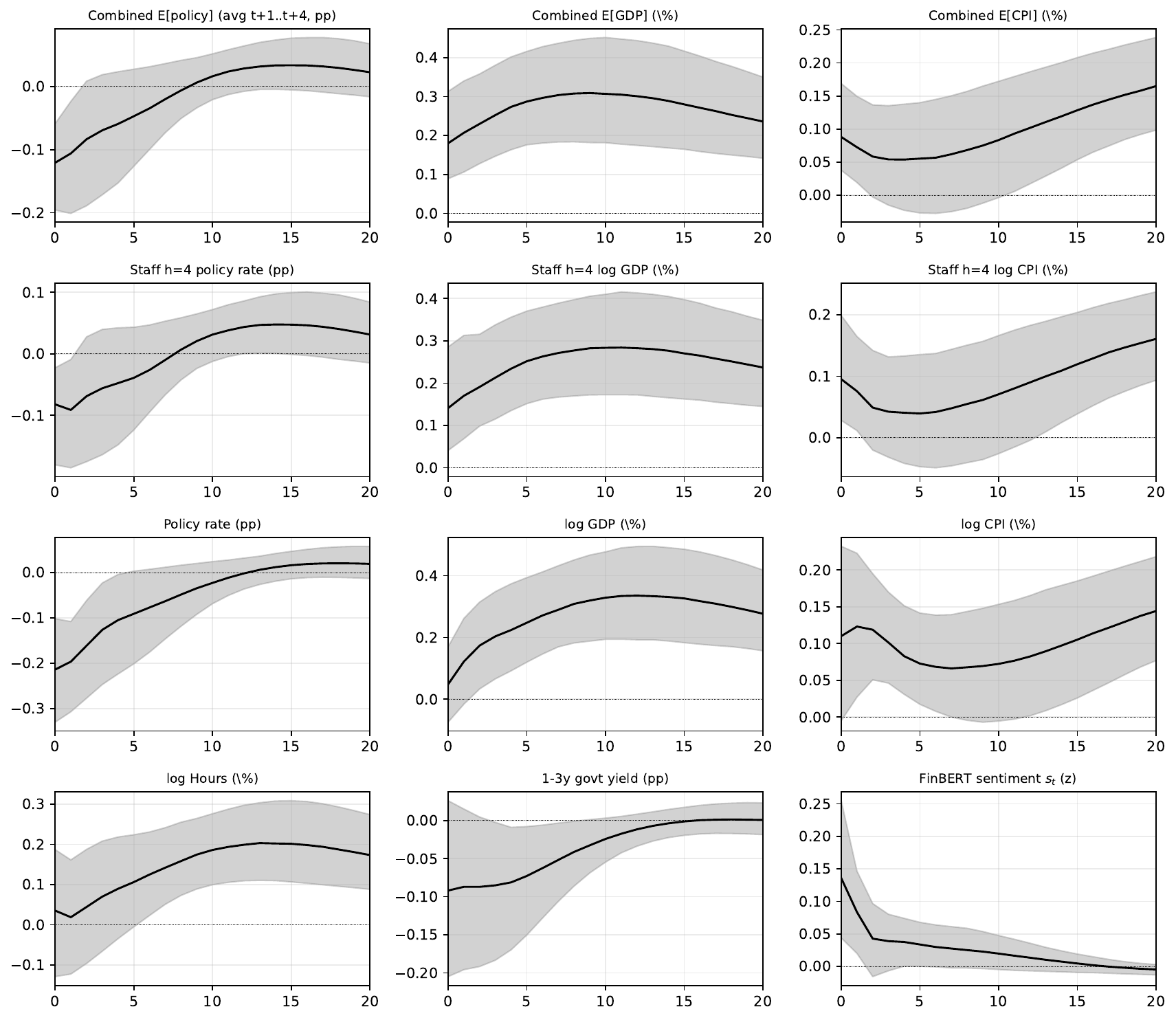}
\par\smallskip
\begin{minipage}{0.95\linewidth}\footnotesize\textit{Note}: Expansionary normalisation. Solid lines are weighted posterior medians; shaded bands are 68\% credible bands across rotations.\end{minipage}
\end{figure}

\begin{figure}[!t]
\centering
\caption{Impulse responses to the narrative shock $\varepsilon^s$.\label{fig:svar_irf_eps}}
\includegraphics[width=0.8\linewidth,height=0.45\textheight]{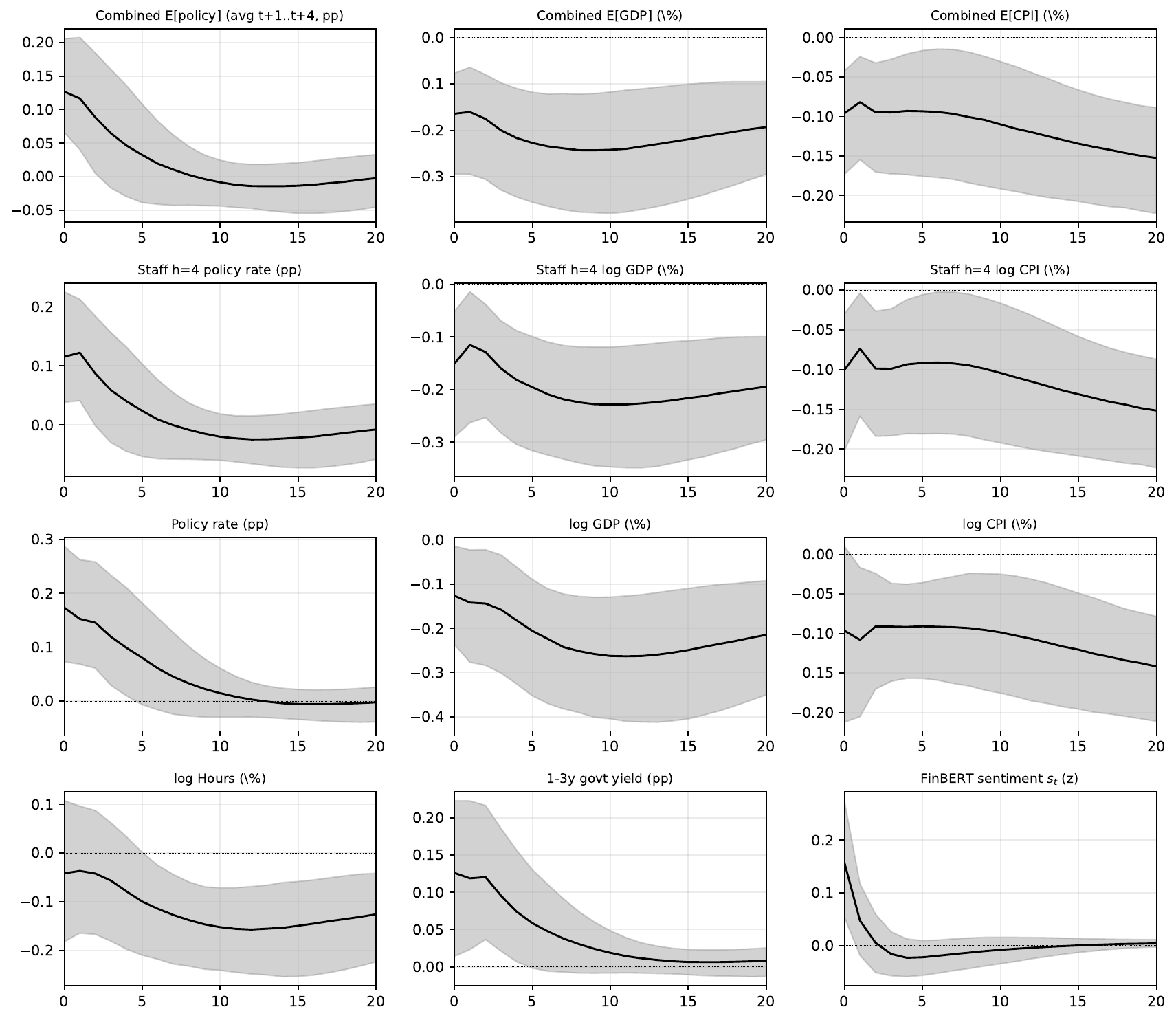}
\par\smallskip
\begin{minipage}{0.95\linewidth}\footnotesize\textit{Note}: Solid lines are weighted posterior medians; shaded bands are 68\% credible bands across rotations.\end{minipage}
\end{figure}

Figures~\ref{fig:svar_irf_eta}--\ref{fig:svar_irf_eps} report the impulse responses of the nine variables to one-standard-deviation realisations of the three identified shocks. In each figure the first row plots the average of $B^d_{E,j}$ and $\widetilde B^d_{4,j}$ defined in (\ref{eq:loose_rationality}); the second row plots $B^d_{E,j}$ alone, that is the impulse responses of the staff projections $E^S_t[r_{t+4}], E^S_t[\log Y_{t+4}], E^S_t[\log P_{t+4}]$. The third row contains the realised macro variables $r_t, \log Y_t, \log P_t$, and the fourth row contains $\log H_t$, the 1--3y bond yield, and the sentiment index $s_t$. Solid lines are weighted posterior medians; shaded bands are 68\% credible bands across rotations.

The anticipated easing shock raises the current rate by approximately 6 basis points on impact, then lowers it to a trough 6 basis points below baseline around $h = 4$ when the announced cut materialises: the lean against the wind derived in (\ref{eq:r0_a}). Log GDP responds with a hump that builds slowly, reaches 0.20\% above baseline by $h \approx 8$ and peaks at 0.26\% around $h \approx 16$, while log CPI rises by 0.12\% on impact and decays slowly. Agents lower their expected rate and raise their expected output and inflation paths. The sentiment variable rises with the announcement and reverts toward zero after the implementation.

The unanticipated easing shock cuts the current rate directly by 21 basis points on impact, gradually reverting to zero, with macro responses roughly 1.5 to 1.7 times larger than under the anticipated shock at the GDP peak and at the rate trough. The sentiment variable rises through the fundamentals channel of (\ref{eq:s0_svar}) once $x_t$ and $\pi_t$ have responded. Together the two MP impulse responses reproduce the canonical pattern of \citet{damico2023}: anticipated and unanticipated shocks are economically distinct, with cognitive discounting dampening the response of the current rate to anticipated announcements while leaving the medium-term macro response broadly comparable.

A positive media-sentiment innovation $\varepsilon^s_t > 0$ raises $s_t$ by $0.15$ standard deviations on impact. The Bank of Canada tightens the current rate by $17$ basis points, log GDP falls by $0.12\%$, log CPI by $0.11\%$, and log hours by $0.04\%$. Expected GDP falls by $0.16$ percentage points, expected CPI by $0.11$ percentage points, and the expected policy rate and the $1$--$3$ year government yield both rise by $0.12$ percentage points. The implied impact ratio $r_0/s_0 \approx 1.1$ percentage points per standard deviation of sentiment is roughly three times larger than the GMM estimate $\hat\delta$ of Section~\ref{sec:taylor_results}, which is expected: the SVAR contemporaneous response includes the indirect feedbacks through the simultaneous response of all other variables in $Y_t$, whereas $\hat\delta$ in (\ref{eq:taylor_aug}) is a partial regression coefficient that holds the other right-hand-side variables fixed.

As assumed in Section~\ref{sec:model}, the narrative shock is interpreted as an autonomous movement in media sentiment unrelated to either macroeconomic fundamentals or monetary-policy announcements. The SVAR results confirm this prior interpretation: a positive narrative innovation induces a contemporaneous tightening through the policy rule, and generates contractionary responses of output and inflation similar to those of a standard monetary-policy tightening, while remaining separately identified from both anticipated and unanticipated monetary-policy shocks through the opposite sign response of $s_t$. The narrative shock is therefore consistent with the idea that autonomous changes in media coverage can affect the economy indirectly through the central bank's reaction function, rather than through changes in the underlying fundamentals themselves. This mechanism is related to the Fed information effect literature \citep{romer2000federal, nakamura2018high}, except that the narrative shock does not reveal information about fundamentals; instead, it shifts the policy response conditional on unchanged fundamentals. \citet{bauer2023alternative} argue empirically that much of the apparent Fed information effect disappears once systematic policy responses are controlled for; our narrative shock can be interpreted as one possible source of the residual co-movement between media narratives and realised policy decisions.

\begin{figure}[!t]
\centering
\caption{Identified shock series.\label{fig:svar_shocks}}
\includegraphics[width=0.9\linewidth,height=0.5\textheight]{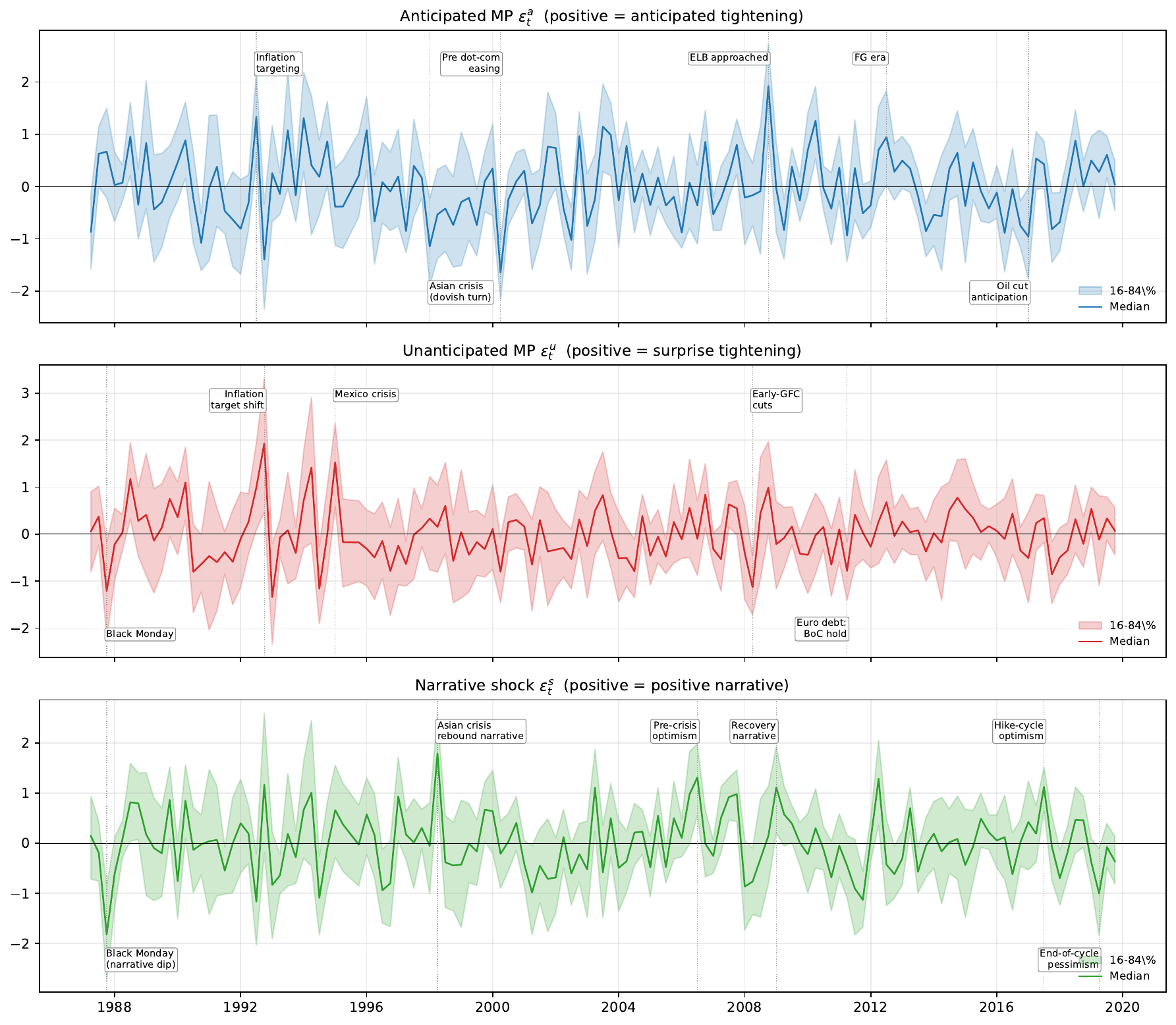}
\par\smallskip
\begin{minipage}{0.95\linewidth}\footnotesize\textit{Note}: Weighted posterior medians with 68\% credible bands. Annotations mark the principal historical episodes in the Bank of Canada's policy environment.\end{minipage}
\end{figure}

Figure~\ref{fig:svar_shocks} reports the median posterior time series of the three identified shocks together with 68\% credible bands. The anticipated MP series shows positive spikes around 1992 and 2002--2003, periods in which staff projections embedded a meaningful tightening path (post-inflation-targeting adoption and post-deflation-scare reflation), and negative spikes in 1998Q1 and 2000Q2 that align with the forward-looking dovish turn around the Asian crisis and the pre-dot-com easing cycle. The 2008Q4 positive spike at the effective lower bound replicates the \citet{damico2023} finding that, in a horizon-four forecast, the announced future policy was less accommodative than agents had been pricing in. The unanticipated MP series displays the volatility cluster of 1987--1995 (Black Monday surprise easing, Mexican peso surprise tightenings, bond-market crisis), the negative spike in 2008Q2 (BoC's pre-emptive cuts in the early phase of the GFC), and the negative episode of 2011Q2 (BoC's hold during the European sovereign debt crisis). The narrative series has its largest negative episode in 1987Q4 (Black Monday narrative pessimism), positive spikes in 1998Q2 (post-Asian-crisis recovery narrative), 2006Q3 (pre-crisis optimism), 2009Q1 (post-trough recovery narrative) and 2017Q3 (narrative around the BoC's resumption of hikes), and a negative spike in 2019Q2 that captures end-of-cycle pessimism. The largest narrative episodes do not coincide with the largest MP episodes, which is the visual counterpart of the orthogonality result reported in Section~\ref{sec:svar_implementation}.

\paragraph{External validation.} Two concerns require attention. The first is the relation of $\varepsilon^a$ and $\varepsilon^u$ to the standard Canadian monetary surprise of \citet{champagne2018identifying}, hereafter CS, which we expect to preserve the anticipated--unanticipated distinction. The second is the orthogonality of the narrative shock $\varepsilon^s$ to standard non-monetary sources of macroeconomic fluctuation: aggregate uncertainty, news-based policy uncertainty, financial-market volatility, and oil-price movements. 

\begin{figure}[!t]
\centering
\caption{Correlations of the identified shocks with external benchmarks.\label{fig:shock_corr}}
\includegraphics[width=0.95\linewidth]{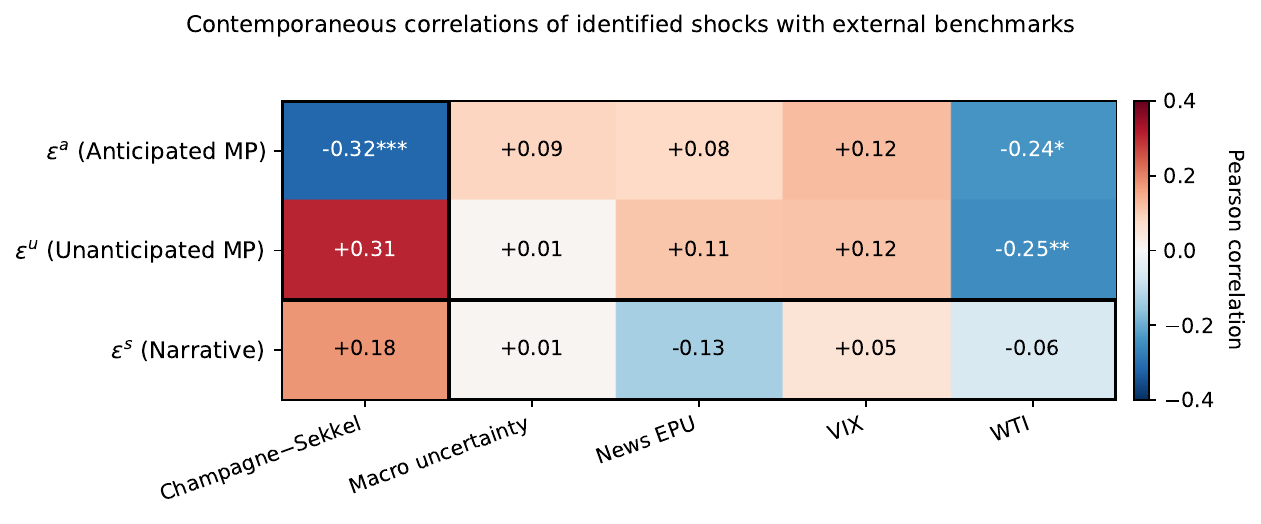}
\par\smallskip
\begin{minipage}{0.97\linewidth}\footnotesize\textit{Note}: Pearson correlations on the 1991Q1--2019Q4 sample ($T=115$). Stars from Newey--West HAC standard errors with $L=4$: $^{*}, ^{**}, ^{***}$ for $p<0.10$, $0.05$, $0.01$. The leftmost column reports the relation between our SVAR MP shocks and the Champagne--Sekkel surprise; the bottom row reports the relation between the narrative shock $\varepsilon^s$ and four external benchmarks: the Canadian macroeconomic uncertainty index of \citet{moran2022macroeconomic}, the Canadian news-based economic policy uncertainty index of \citet{baker2016measuring}, the log first difference of the VIX, and the quarterly change of the WTI crude oil price.
\end{minipage}
\end{figure}

The leftmost column Figure~\ref{fig:shock_corr} reports the relation between our two monetary-policy shocks and the CS surprise shock. The two correlations have similar magnitude but opposite sign, which is consistent with the SVAR separating two empirically distinct components of monetary policy that the univariate CS series aggregates into a single measure. The bottom row of Figure~\ref{fig:shock_corr} reports the relation between the narrative shock and four standard non-monetary alternatives. None of these correlations is statistically significant, suggesting that the narrative shock is not capturing aggregate uncertainty, news-based policy uncertainty, financial-market volatility, or oil-price movements.\footnote{The correlations of $\varepsilon^a$ and $\varepsilon^u$ with the WTI change are $-0.24$ and $-0.25$, marginally significant. The SVAR includes the staff projections $E^S_t[r_{t+4}], E^S_t[\log Y_{t+4}], E^S_t[\log P_{t+4}]$, which embed the staff's expected output and inflation paths and therefore the role assigned to anticipated oil-price movements in those paths. The structural Taylor rule of Section~\ref{sec:taylor} maps these expectations into the systematic component of the policy rate; the part of that mapping driven by anticipated oil movements is absorbed by the projection block of the SVAR. The residual correlation of $-0.25$ may capture the part of contemporaneous oil-price movements that lies outside the staff information set at the projection date.}

\begin{figure}[!t]
\centering
\caption{Local-projection responses to a one-SD monetary-policy shock.\label{fig:lp_compare}}
\includegraphics[width=0.95\linewidth]{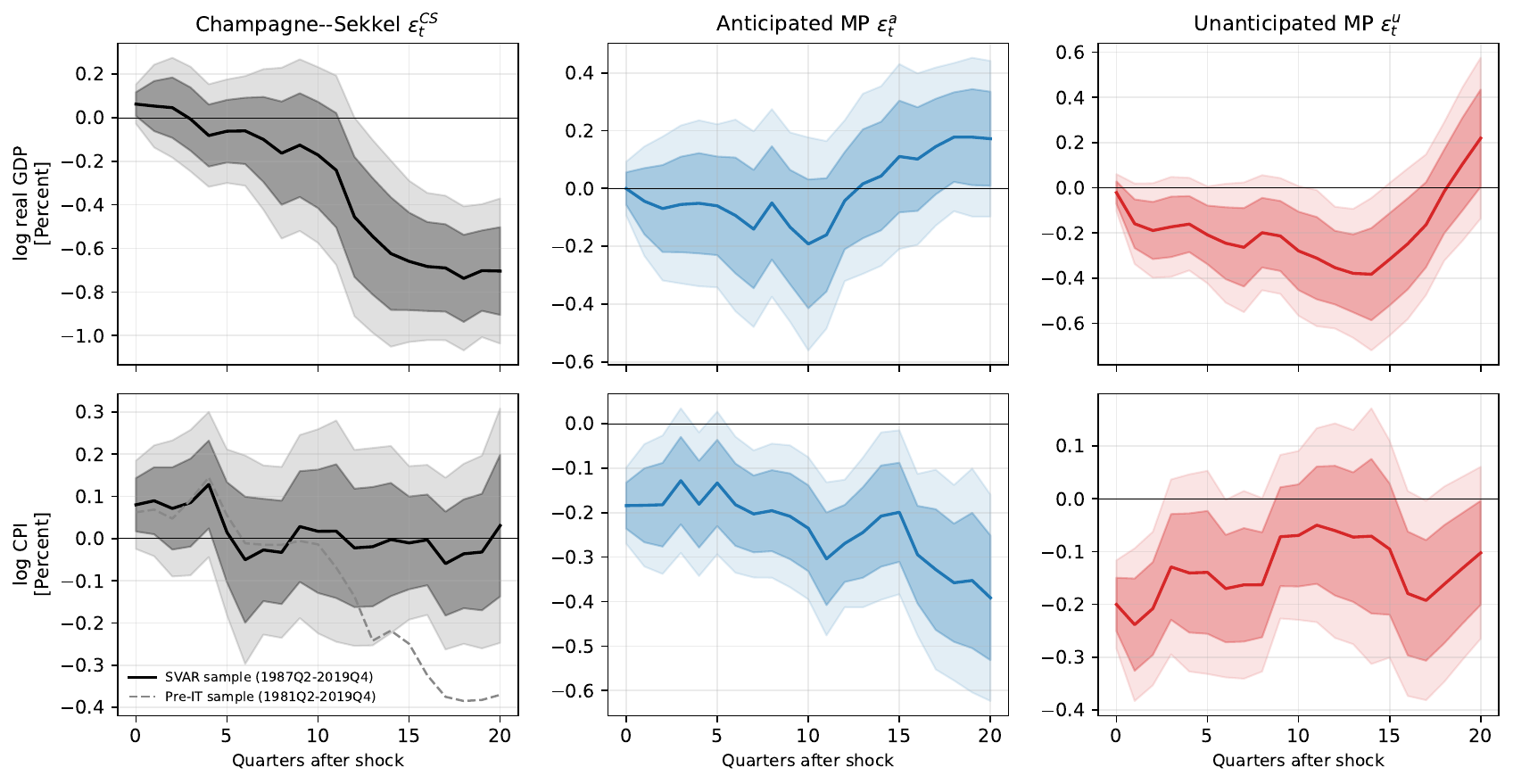}
\par\smallskip
\begin{minipage}{0.97\linewidth}\footnotesize\textit{Note}: $\beta_h$ from equation~(\ref{eq:lp_jorda}) at horizons $h=0,\ldots,20$ quarters with $68\%$ (dark) and $90\%$ (light) HAC bands. Sign convention: positive shock equals monetary tightening. Estimation sample: 1987Q2--2019Q4. The dashed grey line in the lower-left panel reports the same regression for the CS shock on the longer 1981Q2--2019Q4 sample.
\end{minipage}
\end{figure}

In addition, we compare the macroeconomic transmission of $\varepsilon^{CS}$, $\varepsilon^a$ and $\varepsilon^u$ using \citet{jorda2005estimation} local projections under a single specification:
\begin{equation}
y_{t+h} - y_{t-1} = \alpha + \beta_h\, \varepsilon_t + \sum_{k=1}^{12} \gamma_k\, \varepsilon_{t-k} + \delta'\, x_{t-1} + u_{t+h},
\label{eq:lp_jorda}
\end{equation}
where $\varepsilon_t = \{\varepsilon^{CS}, \varepsilon^a,\varepsilon^u \}$, with $x_{t-1} = (\log\mathrm{GDP}^{CA}, \log\mathrm{GDP}^{US}, \log\mathrm{WTI})_{t-1}$, following the implementation of \citet{moran2025chocs}. Figure~\ref{fig:lp_compare} reports the responses of log real GDP and log CPI for the three shocks; for the Champagne--Sekkel CPI we additionally report the response on the longer 1981Q2--2019Q4 sample, which extends back into the pre-IT disinflation. All three shocks generate contractionary responses of output and prices, but the timing and persistence differ substantially across identification schemes. The Champagne--Sekkel shock produces the largest and most persistent decline in GDP, while the unanticipated SVAR shock $\varepsilon^u$ generates a more immediate but comparatively stable contraction. By contrast, the anticipated shock $\varepsilon^a$ has more muted effects on real activity. For prices, the responses to $\varepsilon^a$ and $\varepsilon^u$ are negative throughout, whereas the Champagne--Sekkel shock displays a short-horizon price puzzle on the post-IT sample that disappears once the pre-IT disinflation period is included.

These differences are informative. The Champagne--Sekkel series is identified from high-frequency market surprises and therefore combines multiple components of monetary-policy news into a single measure. By contrast, the SVAR explicitly separates anticipated and unanticipated monetary-policy shocks while controlling for the narrative channel through the inclusion of media sentiment. The comparison therefore suggests that part of the heterogeneity in the local-projection responses may reflect the interaction between policy surprises, forward-guidance effects, and media-driven interpretations of policy decisions. In particular, the decomposition helps isolate the component of policy communication that operates through anticipated future policy paths from the residual contemporaneous surprise component, while conditioning on the central bank's reaction to media narratives. The resulting responses are consistent with the idea that some of the co-movement typically attributed to monetary-policy shocks may instead reflect information or narrative effects embedded in broad market-based surprise measures.

\subsection{Media sentiment as a transmission channel for monetary policy}\label{sec:svar_transmission}

Beyond contributing as a separate structural shock, $s_t$ may also act as a transmission channel for the two MP shocks. Table~\ref{tab:svar_fevd} reports the FEVD shares of the three identified shocks for each variable across horizons. Since only three of the nine structural shocks are identified, the residual share attributed to the remaining rotations is mechanically large and should not be interpreted structurally; the relevant comparison is therefore the relative contribution of the identified shocks. Two patterns emerge. First, the realised policy rate loads more heavily on the unanticipated monetary-policy shock than on the anticipated component, while the narrative shock also contributes meaningfully to policy-rate fluctuations. Second, the macro block attributes a persistent share of output, prices, and hours fluctuations to the narrative shock, with the contribution increasing over longer horizons. Narrative shocks therefore appear to be a non-negligible source of macroeconomic variation in the system.

\begin{table}[!t]
\centering
\caption{Forecast-error variance decomposition.\label{tab:svar_fevd}}
\footnotesize
\begin{tabular}{l|ccc|ccc|ccc|ccc}
\toprule
Variable & \multicolumn{3}{c}{$h=1$} & \multicolumn{3}{c}{$h=4$} & \multicolumn{3}{c}{$h=8$} & \multicolumn{3}{c}{$h=20$} \\
 & $\varepsilon^a$ & $\varepsilon^u$ & $\varepsilon^s$ & $\varepsilon^a$ & $\varepsilon^u$ & $\varepsilon^s$ & $\varepsilon^a$ & $\varepsilon^u$ & $\varepsilon^s$ & $\varepsilon^a$ & $\varepsilon^u$ & $\varepsilon^s$ \\
\midrule
$E^S_t[r_{t+4}]$ & 3.5 & 1.9 & 3.3 & 4.6 & 2.8 & 3.8 & 5.0 & 3.4 & 4.1 & 4.3 & 3.8 & 3.7 \\
$E^S_t[\log Y_{t+4}]$ & 6.0 & 4.5 & 3.2 & 6.2 & 8.0 & 4.8 & 7.5 & 11.8 & 7.4 & 10.2 & 15.6 & 10.2 \\
$E^S_t[\log P_{t+4}]$ & 1.4 & 2.0 & 2.1 & 1.4 & 1.7 & 2.9 & 1.4 & 1.9 & 3.0 & 2.2 & 4.6 & 5.8 \\
$r_t$ & 1.0 & 11.7 & 7.7 & 1.4 & 8.1 & 6.1 & 1.7 & 7.5 & 5.6 & 2.0 & 7.0 & 5.3 \\
$\log Y_t$ & 1.5 & 3.2 & 4.8 & 2.1 & 6.0 & 4.8 & 4.0 & 10.0 & 7.3 & 8.1 & 14.7 & 9.7 \\
$\log P_t$ & 7.2 & 7.1 & 5.7 & 6.6 & 6.6 & 5.2 & 4.4 & 4.9 & 4.8 & 3.4 & 5.8 & 6.5 \\
$\log H_t$ & 5.0 & 4.2 & 3.9 & 3.6 & 4.3 & 3.6 & 3.9 & 6.8 & 5.5 & 6.2 & 12.2 & 8.2 \\
$i^{1\text{-}3y}_t$ & 1.9 & 4.2 & 6.5 & 1.9 & 4.8 & 6.6 & 2.3 & 5.6 & 6.6 & 2.5 & 5.5 & 6.3 \\
$s_t$ & 4.8 & 6.4 & 6.9 & 5.8 & 7.3 & 7.0 & 6.4 & 7.4 & 7.1 & 6.7 & 7.5 & 7.1 \\
\bottomrule
\end{tabular}
\par\smallskip
\begin{minipage}{0.95\linewidth}\footnotesize\textit{Note}:
Median posterior shares (in percent) of the forecast-error variance attributable to the anticipated MP shock $\varepsilon^a$,
the unanticipated MP shock $\varepsilon^u$ and the narrative shock $\varepsilon^s$. The remaining share, attributable to the six unidentified rotations, is omitted.\end{minipage}
\end{table}

\begin{figure}[!t]
\centering
\caption{Historical decomposition.\label{fig:svar_hd}}
\includegraphics[width=0.9\linewidth,height=0.4\textheight]{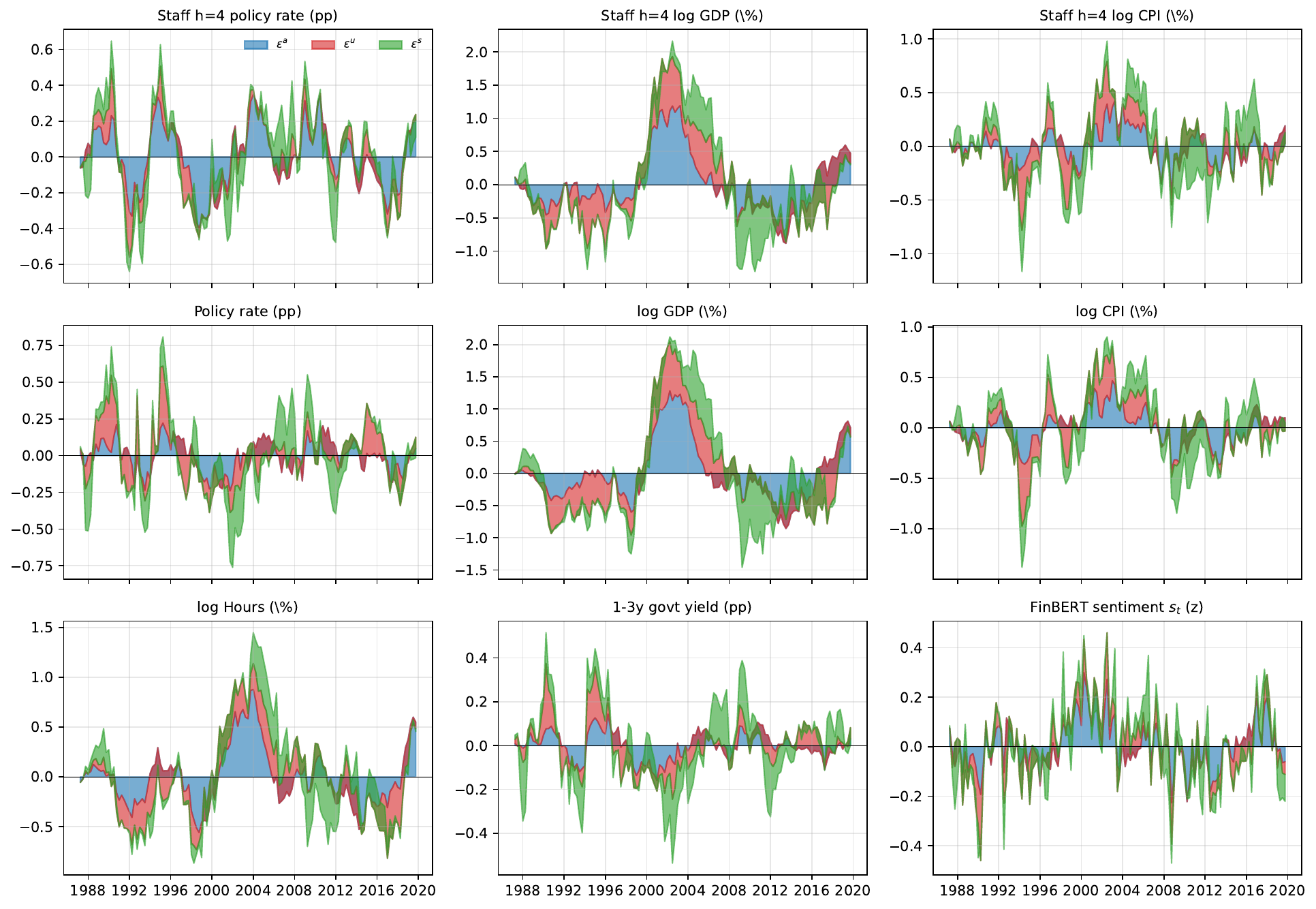}
\par\smallskip
\begin{minipage}{0.95\linewidth}\footnotesize\textit{Note}: Stacked contributions of the three identified shocks to each variable at each date. The sum of the three coloured bands equals the total identified contribution.\end{minipage}
\end{figure}

The historical decomposition (Figure~\ref{fig:svar_hd}) is consistent with the IRF pattern implied by the model, in which negative narrative innovations contribute positively to macroeconomic activity while positive narrative innovations are contractionary through the policy rule. The 1987Q4 episode is associated with positive contributions of narrative shocks to GDP and hours during the subsequent expansion, whereas the cumulative effect of the positive narrative shocks of the mid-2000s contributes negatively to activity during the 2008--2010 period, with a noticeable effect on GDP at the trough of the recession. The 2017--2019 tightening cycle around the Bank of Canada's rate-hike resumption also appears clearly in both media sentiment and the realised policy rate.

To evaluate the impact of media sentiments for the monetary policy transmission we apply the channel-shutdown counterfactual of \citet{bernanke1997systematic}, adapted by \citet{barsky2011news} to news shocks. The BGW exercise evaluates how the transmission of monetary-policy shocks changes once the lagged reduced-form feedback from $s_t$ to the remaining variables is removed, while preserving the identified contemporaneous impact matrix and the internal dynamics of $s_t$ itself. The counterfactual therefore does not alter the structural identification of the shocks or the contemporaneous policy reaction; it isolates the dynamic propagation channel operating through media sentiment. Comparing the baseline and counterfactual impulse responses then measures the extent to which media narratives amplify or attenuate the transmission of monetary-policy shocks over time.

\begin{figure}[!t]
\centering
\caption{Counterfactual IRFs to the anticipated MP shock $\varepsilon^a$.\label{fig:svar_cf_eta}}
\includegraphics[width=0.95\linewidth]{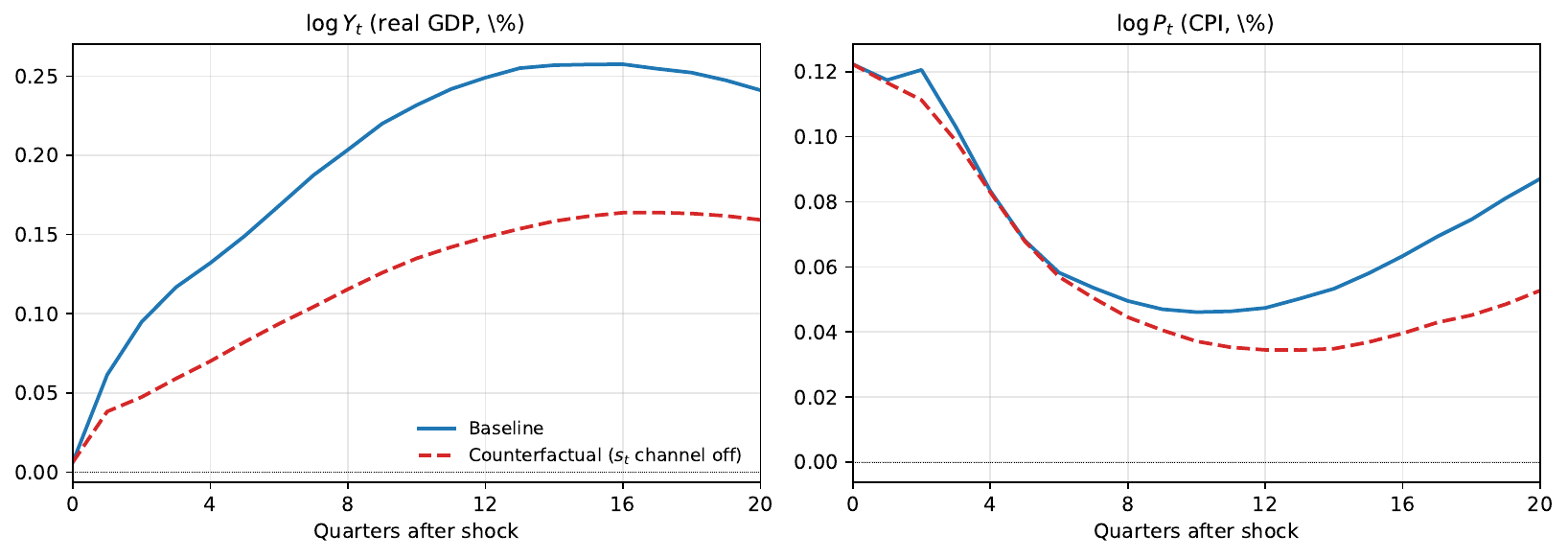}
\par\smallskip
\begin{minipage}{0.95\linewidth}\footnotesize\textit{Note}: Expansionary normalisation. Solid blue: baseline median IRF. Dashed red: counterfactual in which the lagged feedback of $s_t$ on the other variables is set to zero.\end{minipage}
\end{figure}

Figure~\ref{fig:svar_cf_eta} reports the GDP and CPI responses to an anticipated expansionary monetary-policy shock in the baseline SVAR (solid blue) and in the counterfactual that shuts down the lagged feedback of $s_t$ on the other variables (dashed red). These are the empirical counterparts of the two macro variables on which Figure~\ref{fig:irf} reports the predictions of the structural model. In the structural IRFs of Section~\ref{sec:svar_signs}, an anticipated easing produces a slow-building expansion of output and inflation that peaks near the announced implementation date and operates partly through media-sentiment feedback under $\phi_s > 0$. The SVAR responses match this pattern: log $Y_t$ rises gradually, peaks around $h = 14$ at $0.26\%$ in the baseline, and the log $P_t$ response is positive on impact and stays positive throughout the response horizon. Shutting down the sentiment channel attenuates both responses --- the GDP peak falls to $0.16\%$ and the CPI response shifts down in the medium term --- without eliminating the expansion. The empirical pattern therefore confirms the prediction of Figure~\ref{fig:irf}: media sentiment is a quantitatively non-trivial propagation channel for monetary-policy transmission to real activity and prices.

\section{Conclusion}\label{sec:conclusion}

In this paper we embed the media into a behavioral New-Keynesian model with cognitive discounting. Media sentiment $s_t$ enters the central bank's reaction function and follows a law of motion that decomposes coverage into persistence, fundamentals, central-bank communication, and a residual narrative component. The first three blocks make $s_t$ an endogenous outcome of the system; only the last is plausibly orthogonal to monetary policy. The structure pins down the identification problem in empirical estimation.

We construct media indicators of monetary-policy sentiment from a corpus of more than 50,000 articles published in major Canadian newspapers between 1977 and 2024. The indicators span four dimensions --- topic, tone, temporal orientation, and uncertainty --- and rely on three classes of tools: dictionary methods, transformer-based encoder models (FinBERT and ModernFinBERT), and a structured generative-LLM framework (CBILA). Encoder-only models are the workhorse for the historical sample because their pre-training cutoff predates the bulk of the data; the CBILA series carry potential look-ahead bias and are kept separate.

Media sentiment shifts household inflation and wage expectations beyond their own persistence and improves out-of-sample forecasts of CPI inflation and real GDP growth. Once $s_t$ is treated as endogenous and estimated by GMM with appropriate instruments, the encoder-based sentiment measures enter the Bank of Canada's estimated Taylor rule with a positive and significant loading; the OLS--GMM gap confirms that simultaneity between sentiment and policy is a first-order concern. The CBILA generative-LLM tones fail standard IV validity checks because of their look-ahead bias.

A Bayesian SVAR identified by sign restrictions derived from the model isolates an anticipated MP shock, an unanticipated MP shock, and a narrative shock. The unanticipated easing cuts the policy rate on impact and is followed by a hump-shaped expansion of GDP and prices, while the anticipated easing produces a smaller impact response of the rate, a slow build-up of GDP that peaks near the announced implementation date, and a positive CPI response, consistent with the cognitive-discounting mechanism. The narrative shock contributes a non-trivial share of medium-horizon macroeconomic variance. A counterfactual also confirms that media sentiment is a quantitatively relevant propagation channel for monetary-policy transmission.

\clearpage
\begingroup
\setstretch{1.0}
\bibliographystyle{chicago}
\bibliography{references}
\endgroup

\clearpage
\appendix
\renewcommand{\thesection}{\Alph{section}}
\renewcommand{\thesubsection}{\Alph{section}.\arabic{subsection}}
\renewcommand{\thefigure}{\thesection\arabic{figure}}
\renewcommand{\thetable}{\thesection\arabic{table}}
\setcounter{section}{0}
\setcounter{page}{0}
\setcounter{table}{0}
\setcounter{figure}{0}

\begin{center}
\textbf{\Large \textsc{Online Appendix to ``Monetary Policy in the Media Spotlight: Sentiments, Signals, and Economic Impact''}}\\[1em]

\begin{minipage}[t]{0.24\linewidth}
\centering
Firmin Ayivodji\\
\textit{IMF}
\end{minipage}
\hfill
\begin{minipage}[t]{0.24\linewidth}
\centering
Etienne Briand\\
\textit{UQAM}
\end{minipage}
\hfill
\begin{minipage}[t]{0.24\linewidth}
\centering
Kevin Moran\\
\textit{Université Laval}\\
\textit{CIRANO}
\end{minipage}
\hfill
\begin{minipage}[t]{0.24\linewidth}
\centering
Dalibor Stevanovic\\
\textit{UQAM}\\
\textit{CIRANO}
\end{minipage}\\[1em]

\small{\textcolor{blue}{\today}}
\end{center}

\vspace{2em}

\section*{\large Contents}
\vspace{0.5em}

\makeatletter
\newcommand{\localtoc}{%
  \begingroup
  \makeatletter
  \@starttoc{apptoc}%
  \makeatother
  \endgroup
}
\let\oldsection\section
\renewcommand{\section}[1]{%
  \oldsection{#1}%
  \addcontentsline{apptoc}{section}{\protect\numberline{\thesection}#1}%
}
\let\oldsubsection\subsection
\renewcommand{\subsection}[1]{%
  \oldsubsection{#1}%
  \addcontentsline{apptoc}{subsection}{\protect\numberline{\thesubsection}#1}%
}
\makeatother

\begingroup
\hypersetup{linkcolor=mediumtealblue}  
\localtoc
\endgroup

\newpage
\section{Additional figures on the news corpus and topic decomposition}
\label{app:A}
\setcounter{page}{0}
\global\long\def\thetable{A.\arabic{table}}
\setcounter{table}{0}
\global\long\def\thefigure{A.\arabic{figure}}
\setcounter{figure}{0} 

\renewcommand{\thepage}{A.\arabic{page}}
\pagenumbering{arabic} 

\pagestyle{fancy}
\fancyhf{}
\fancyfoot[C]{A.\thepage}
\renewcommand{\headrulewidth}{0pt} 
\renewcommand{\footrulewidth}{0pt} 

\begin{figure}[ht]
\vspace{0.1cm}
    \centering
      \caption{Articles distribution across news sources \label{fig:dist_articles}}
    \includegraphics[width=0.99\linewidth]{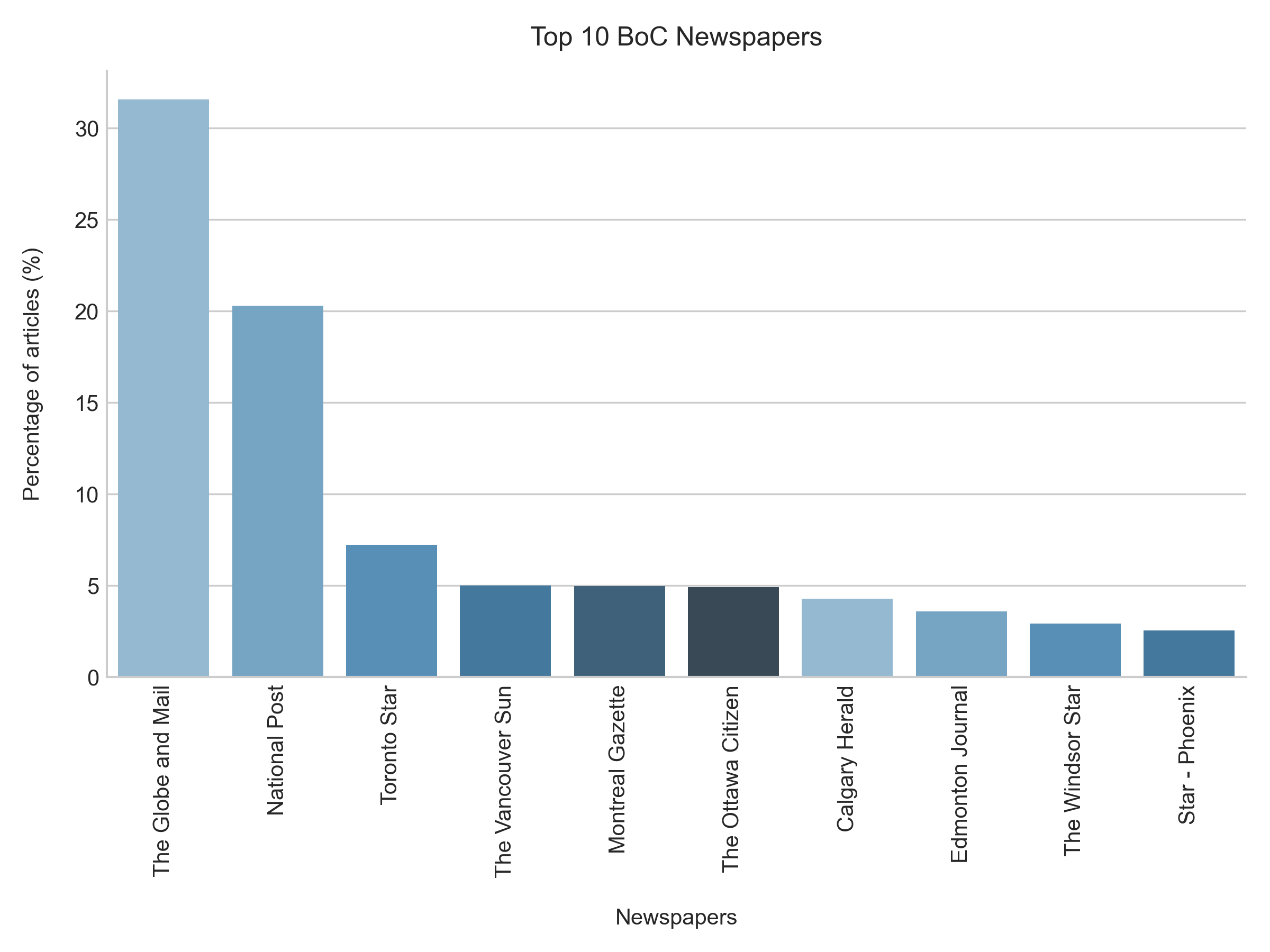}  
\end{figure}

\begin{figure}[ht]

  \centering
  \caption{LDA output: Terms within topics ranked by probability \label{fig:topic_heatmap}}
    \includegraphics[width=0.99\linewidth]{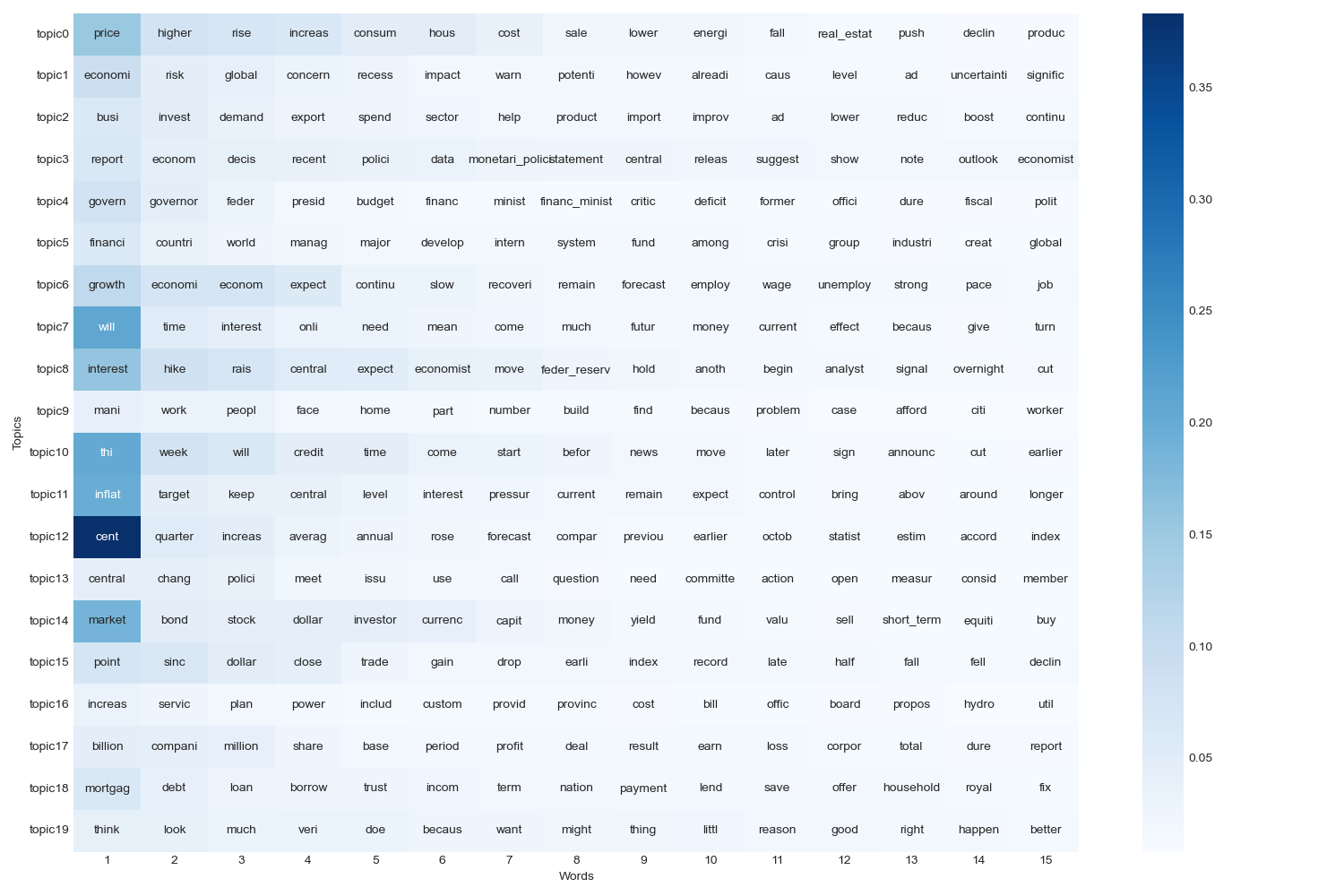} 
\end{figure}

\clearpage
\begin{figure}[ht]
\centering
\caption{Word distribution of selected narratives and monthly narratives distributions \label{wc2}}
\begin{subfigure}{.32\textwidth}
    \centering
    \caption{Monetary policy}
    \includegraphics[width=0.99\linewidth]{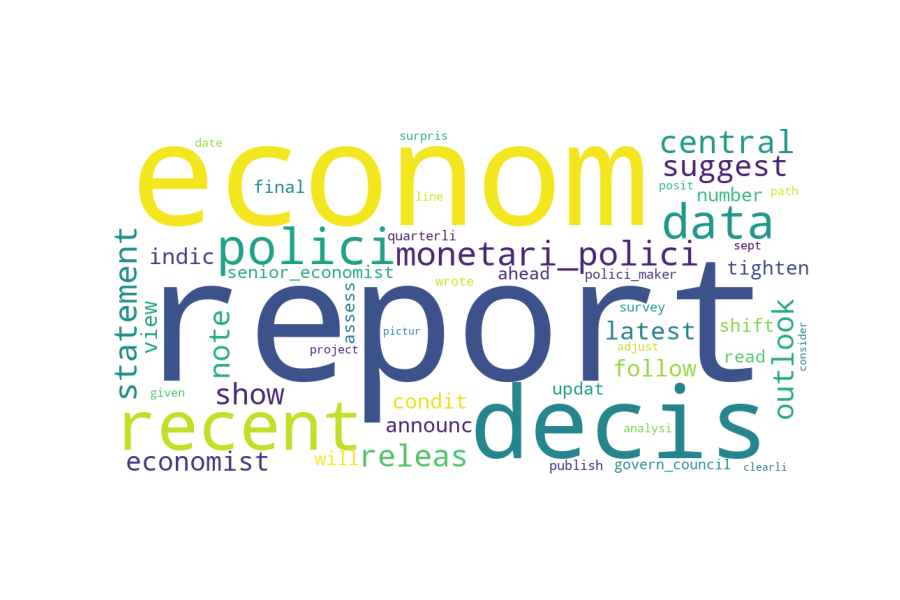}
\end{subfigure}\hfill
\begin{subfigure}{.32\textwidth}
    \centering
    \caption{Output/Demand}
    \includegraphics[width=0.99\linewidth]{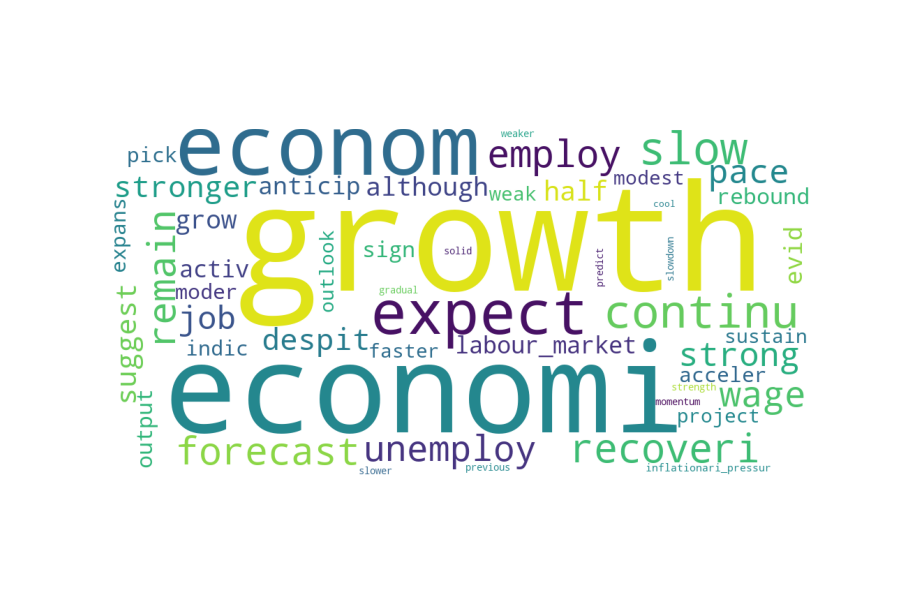}
\end{subfigure}\hfill
\begin{subfigure}{.32\textwidth}
    \centering
    \caption{Housing}
    \includegraphics[width=0.99\linewidth]{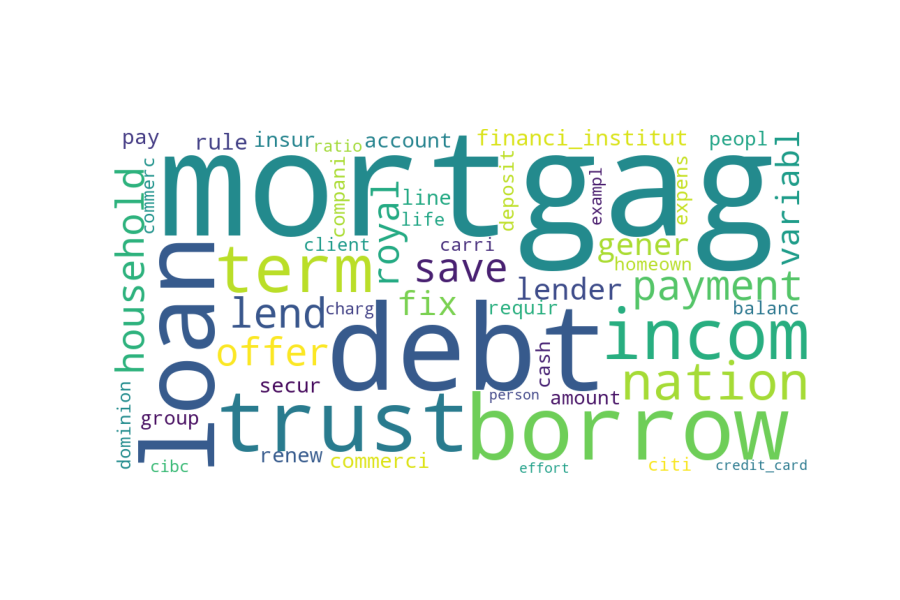}
\end{subfigure}

\bigskip

\begin{subfigure}{.32\textwidth}
    \centering
    \includegraphics[width=0.99\linewidth]{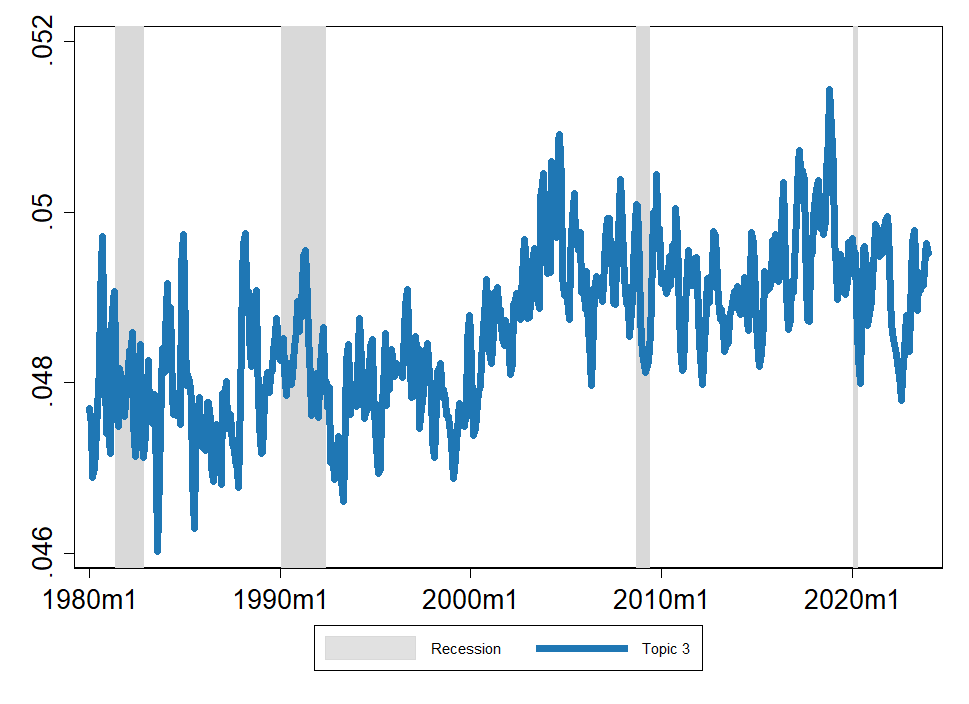}
\end{subfigure}\hfill
\begin{subfigure}{.32\textwidth}
    \centering
    \includegraphics[width=0.99\linewidth]{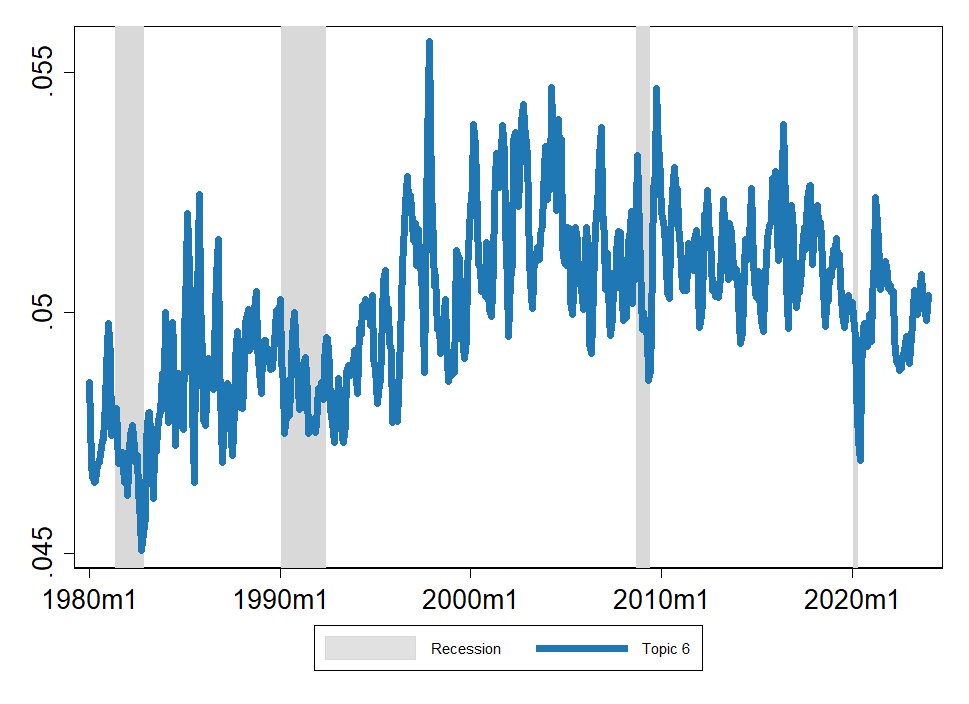}
\end{subfigure}\hfill
\begin{subfigure}{.32\textwidth}
    \centering
    \includegraphics[width=0.99\linewidth]{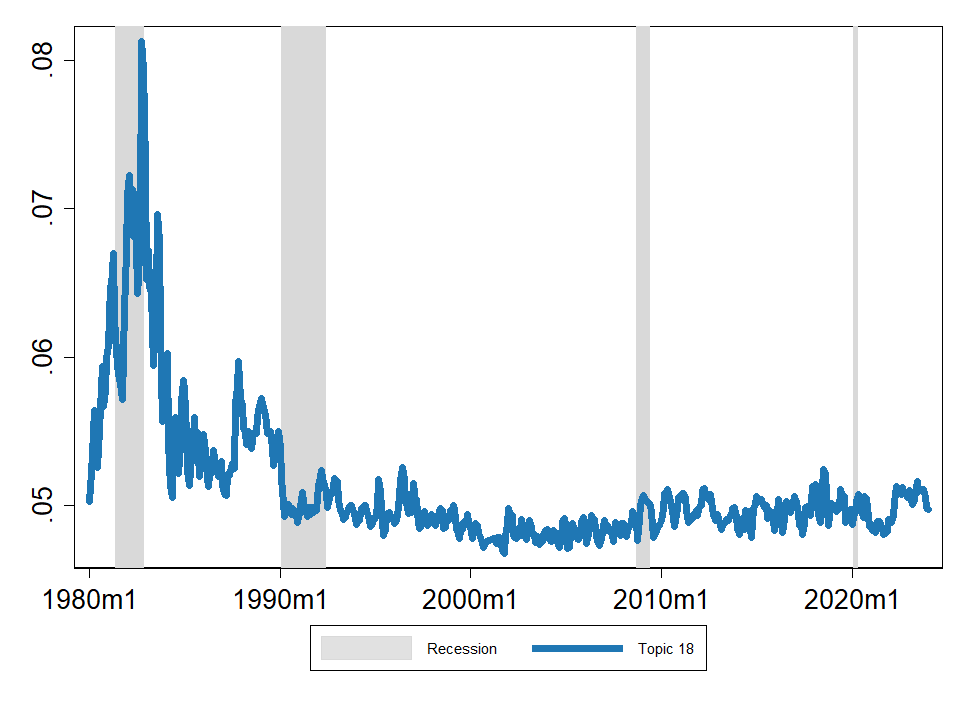}
\end{subfigure}

\vspace{0.3em}
{\scriptsize\textit{Notes}: This figure depicts three out of 20 topics given by the estimated LDA model. The upper row is the word distribution of each topic; the lower row is the topic distribution at monthly frequency. The size of a term represents its probability within a given topic. Position and color convey no information.}
\end{figure}

\begin{figure}[H]
\centering
\caption{Share of sentences by temporal orientation in monetary-policy news, 1980--2024.\label{fig:tensepattern}}
\includegraphics[width=0.95\linewidth]{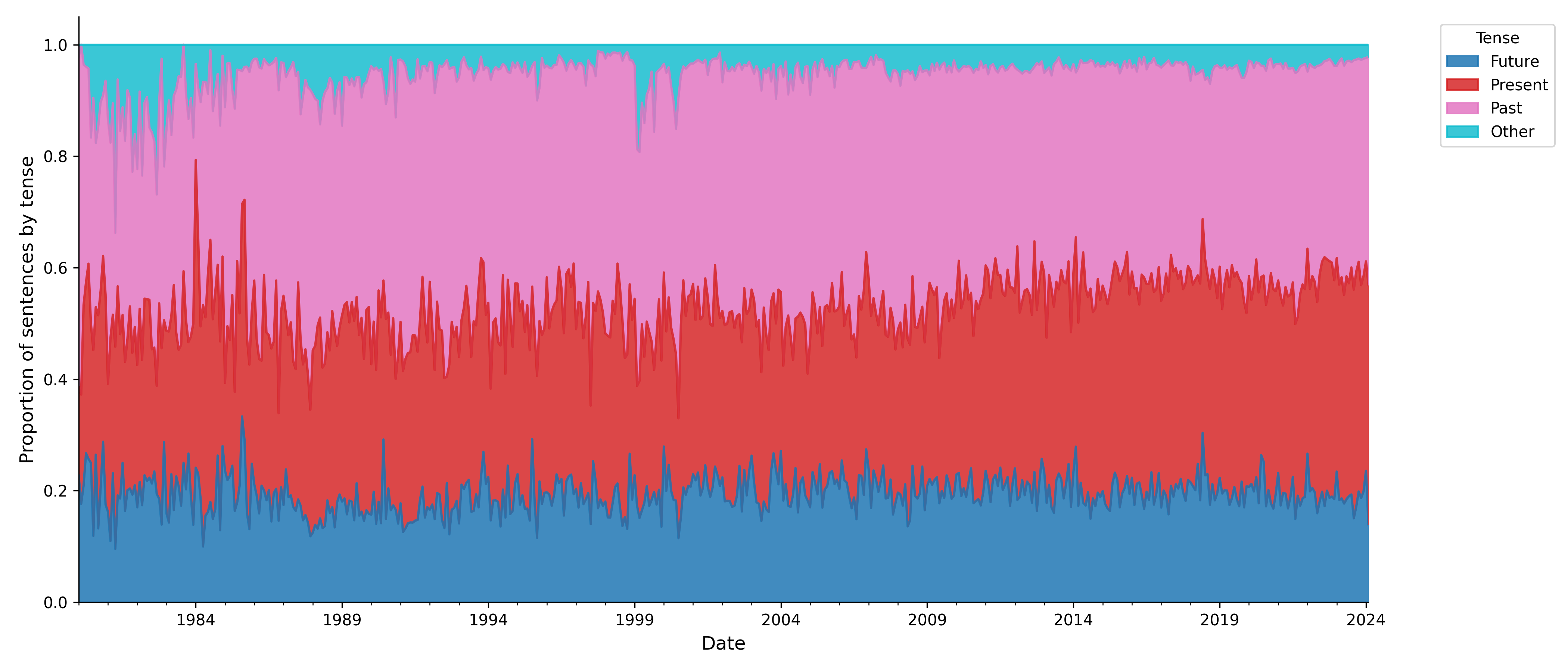}
\par\smallskip
\begin{minipage}{0.95\linewidth}\footnotesize\textit{Note}: Share of sentences classified as past-, present-, or future-tense in the Canadian monetary-policy news corpus, monthly frequency, 1980--2024. The classification combines spaCy's grammatical tense detection with a curated forward-looking dictionary and FinBERT-FLS as described in Section~\ref{sec:cbila}.\end{minipage}
\end{figure}

\begin{figure}[H]
\centering
\caption{Forward-looking media orientation and BoC governors' tenure.\label{fig:futureovertime}}
\includegraphics[width=0.99\linewidth]{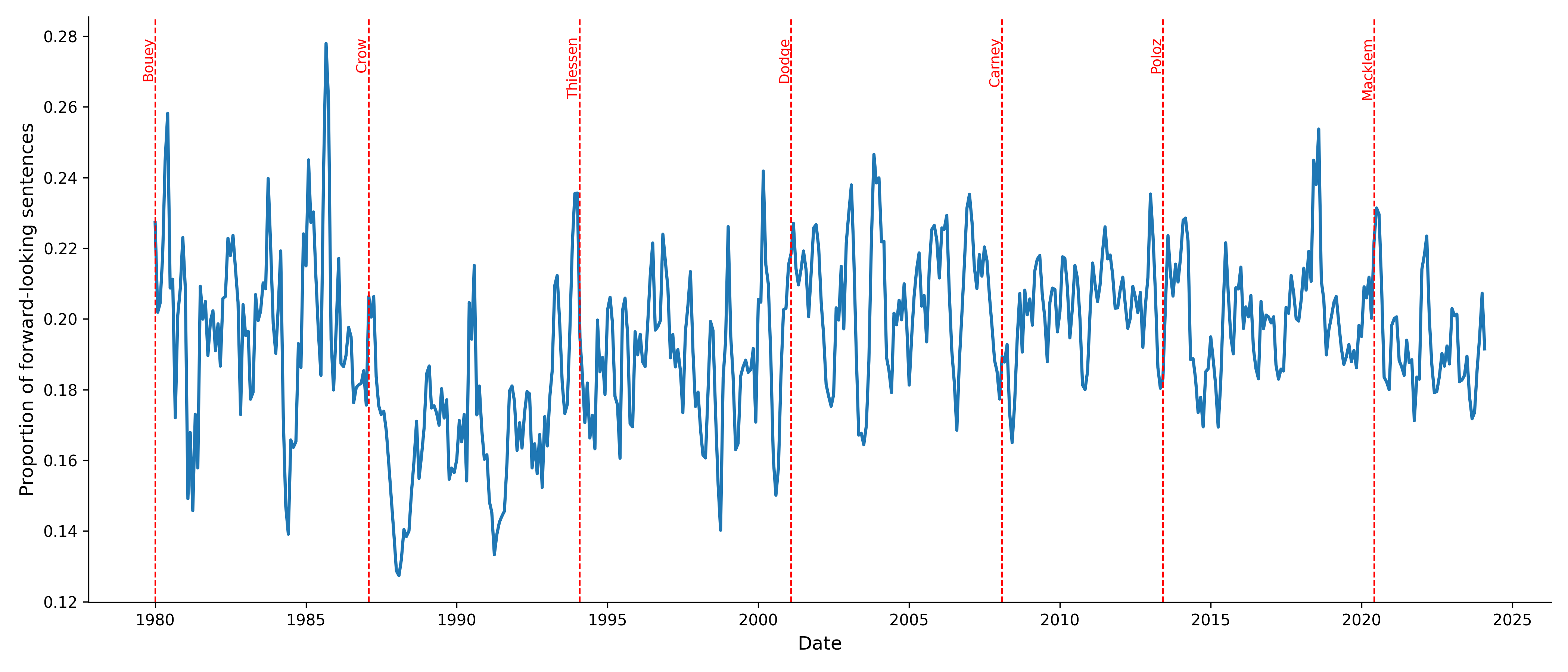}
\par\smallskip
\begin{minipage}{0.95\linewidth}\footnotesize\textit{Note}: Three-month moving average of the share of forward-looking sentences in the Canadian monetary-policy news corpus, overlaid with the tenure of successive Bank of Canada governors.\end{minipage}
\end{figure}

\clearpage
\section{Forecasting details}\label{app:forecasting}
\label{sec:shap}

This appendix collects the construction of the additional text-based features used in the forecasting exercise of Section~\ref{sec:forecasting} (a media-based monetary policy uncertainty index and a topic decomposition of the corpus), the forecasting setup and the competing model families, the full set of forecast-accuracy results for CPI inflation and real GDP growth, and the variable-importance analysis based on SHAP values.

\subsection{Construction of additional text-based features}\label{app:forecasting_features}

In addition to the sentiment indicators of Section~\ref{sec:measure}, we construct two further dimensions of the corpus that enter the predictor sets defined below: a media-based monetary policy uncertainty (MPU) index and a topic decomposition of the news articles. Both are built from the same corpus as the sentiment indicators.

Our measurement of the uncertainty index is based on the local co-occurrence of terms denoting uncertainty. To obtain the uncertainty terms, we begin with the four seed terms `uncertain', `uncertainty', `risk', and `risks' as proposed by \cite{cieslak2023policymakers}. We then use a word embedding model, specifically the Continuous Bag-of-Words model \citep{mikolov2013distributed}, applied to media monetary policy sentences to generate an expanded set of terms. A word embedding model represents each unique term in a corpus as a relatively low-dimensional vector in a vector space. Words whose vectors lie close together in the vector space share similar meanings. In general, the neighbors are synonyms of the seeds, such as `unclear' and `unsure', or terms reflecting worries and concerns, such as `threat', `fear', and `wary'. The nearest neighbors can also contain generic terms not related to uncertainty. We therefore further organize the lists using our domain expertise, and after removing irrelevant terms,
\[
\operatorname{Uncertainty \quad index}_t=\frac{n_t^{\text {Uncertainty }}}{n_t^{\text {Words }}}\times 100
\]
where $n_t^{\text{Uncertainty}}$ is the count of uncertainty or risk-related words and $n_t^{\text{Words}}$ is the total number of words in the corpus at time $t$. The index is aggregated across all articles published in quarter $t$ before normalisation, and it is standardized to have a mean of 100 for the period between 1995 and 2011 as proposed by \cite{baker2016measuring}.

\begin{figure}[!ht]
\centering
    \caption{Media Uncertainty Index (3-Month Moving Average)\label{gr_unc}}
\includegraphics[width=0.6\textwidth]{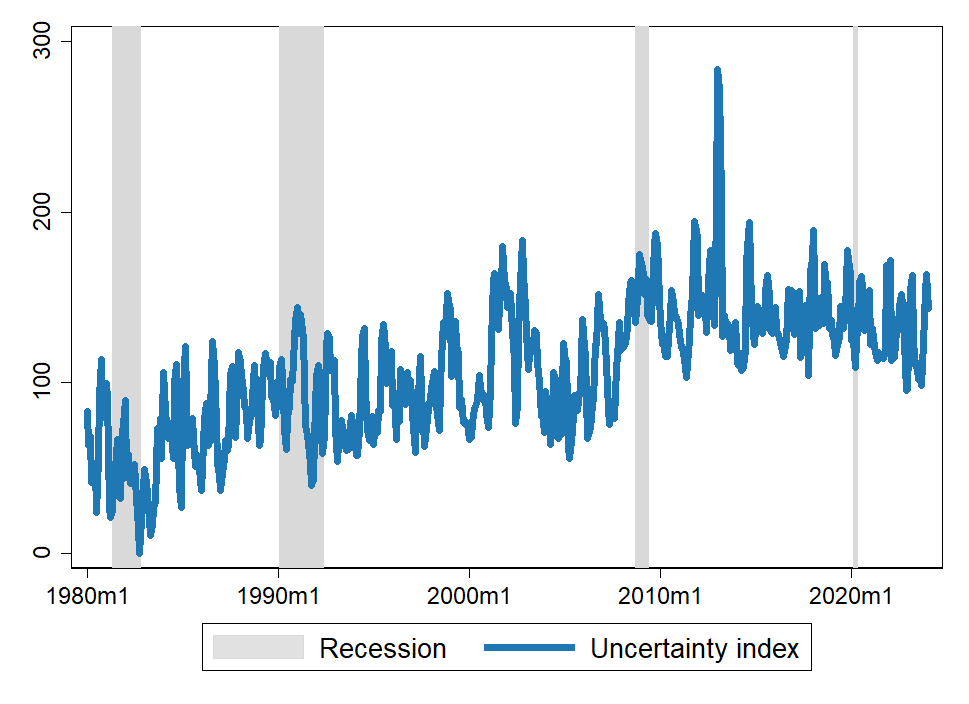}
\end{figure}

Figure~\ref{gr_unc} demonstrates an increasing trend in the uncertainty index since 1980, peaking significantly in January 2013. This notable peak corresponds to heightened concerns over key economic risks prevalent during this period. Identified risks included a potential sharp correction in housing prices and the instability within China's banking sector, both of which posed substantial threats to the economic landscape. These concerns were exacerbated by global economic conditions and the anticipated impacts of adjustments in U.S. monetary policy on long-term interest rates. The surge in media coverage during this time reflects an increase in uncertainty, capturing the growing apprehension regarding these vulnerabilities and their potential repercussions on financial stability.

We summarize the corpus into 20 latent topics with Latent Dirichlet Allocation \citep{blei2003latent}, applied at the article level after standard preprocessing (sentence tokenization, stopword removal, stemming, bigram and trigram extraction, and TF-IDF weighting). Hyperparameters take their conventional values.\footnote{The topic-document prior is set at 2.5 and the term-topic prior at $1/166$.} Aggregating the article-level topic distributions over time gives a monthly series of topic shares $\theta_{k,t}$ for $k=1,\ldots,20$. We assign descriptive names to each of the 20 topics by passing the top keywords and a sample of representative documents to a large language model and using its output as a label; the full list of topics and their top keywords is in Figure~\ref{fig:topic_heatmap}.

\begin{figure}[ht]
\centering
\caption{Word distribution of narratives and monthly narratives distributions\label{wc}}
\begin{subfigure}{.32\textwidth}
    \centering
    \caption{Financial conditions}
    \includegraphics[width=0.9\linewidth]{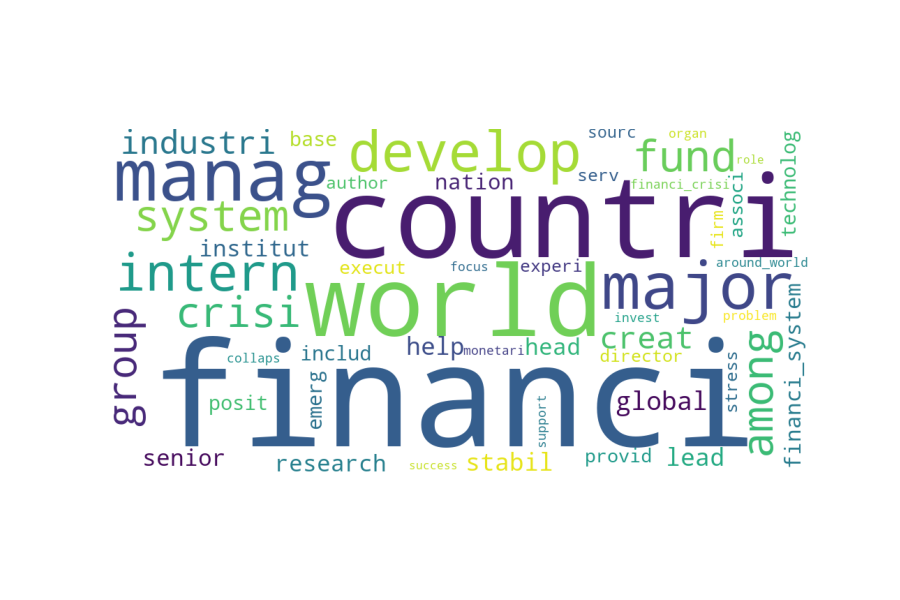}
\end{subfigure}\hfill
\begin{subfigure}{.32\textwidth}
    \centering
    \caption{Inflation target}
    \includegraphics[width=0.9\linewidth]{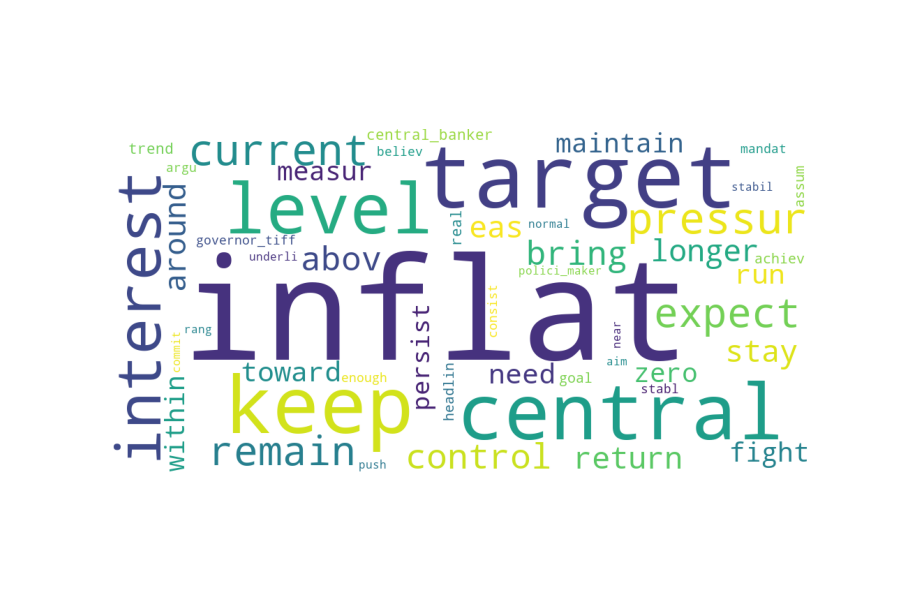}
\end{subfigure}\hfill
\begin{subfigure}{.32\textwidth}
    \centering
    \caption{Interest rates}
    \includegraphics[width=0.9\linewidth]{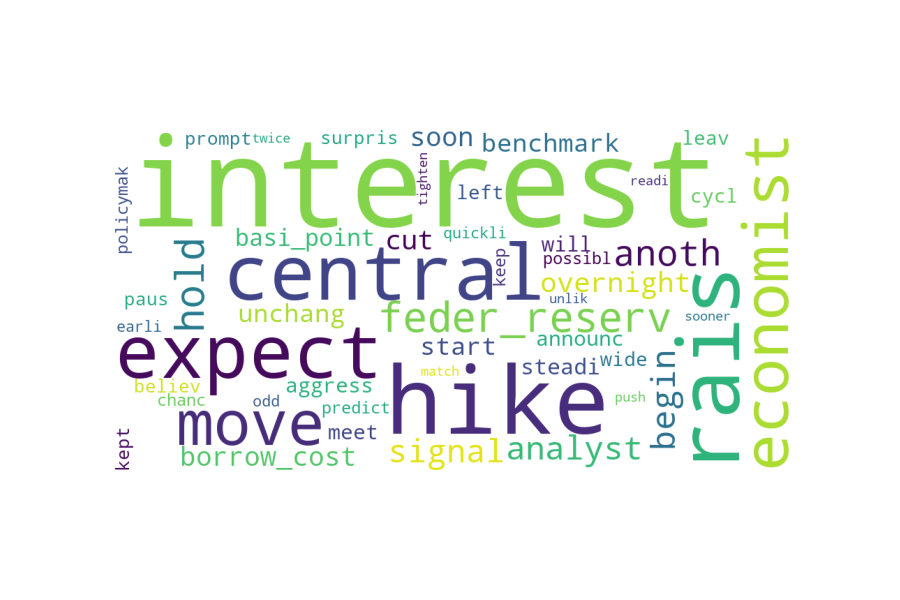}
\end{subfigure}

\bigskip

\begin{subfigure}{.32\textwidth}
    \centering
    \includegraphics[width=0.9\linewidth]{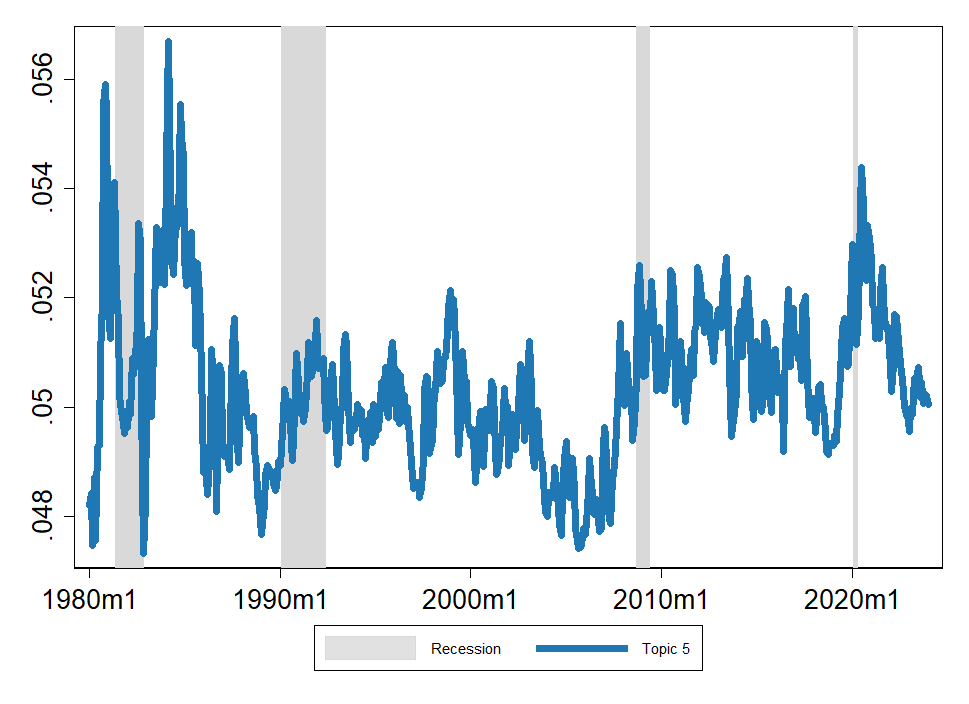}
\end{subfigure}\hfill
\begin{subfigure}{.32\textwidth}
    \centering
    \includegraphics[width=0.9\linewidth]{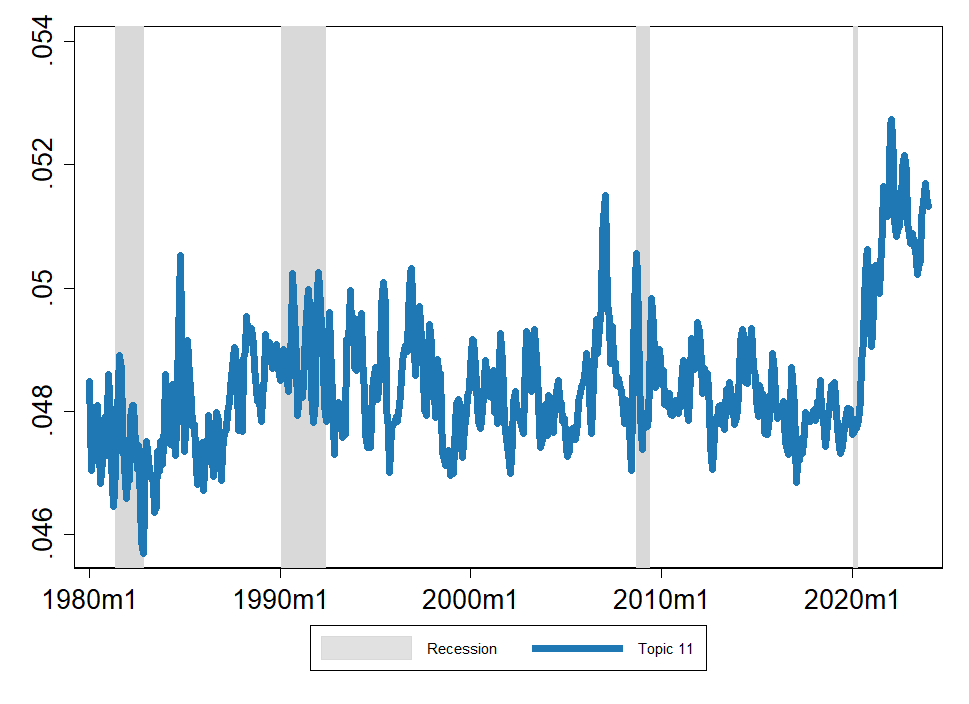}
\end{subfigure}\hfill
\begin{subfigure}{.32\textwidth}
    \centering
    \includegraphics[width=0.9\linewidth]{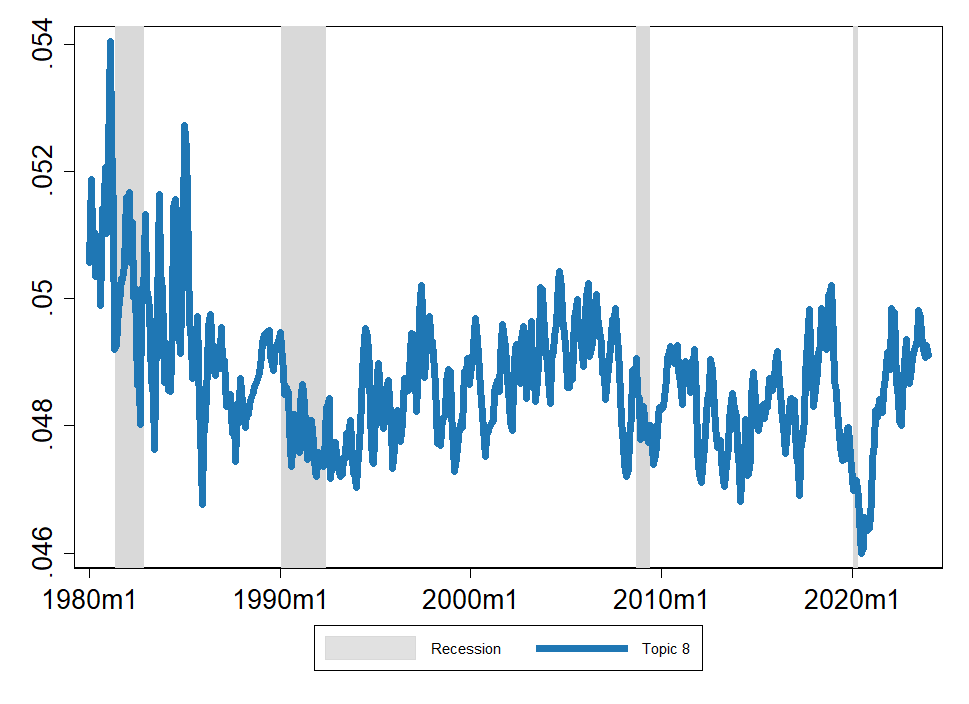}
\end{subfigure}

\vspace{0.3em}
{\scriptsize\textit{Notes}: This figure depicts three out of 20 topics given by the estimated LDA model. The upper row is the word distribution of each topic; the lower row is the topic distribution at monthly frequency. The size of a term represents its probability within a given topic. Position and color convey no information.}
\end{figure}

Figure~\ref{wc} illustrates three of the 20 estimated topics: \textit{Financial Conditions}, \textit{Inflation Target}, and \textit{Interest Rates}. For each topic, the upper row shows the word distribution and the lower row the monthly share. \textit{Financial Conditions} spikes during the 2008 financial crisis and at the start of the COVID-19 pandemic, with leading terms such as ``Financi'', ``Manag'', ``World'', ``Financi\_System'', and ``Fund''. \textit{Inflation Target} loads on ``Inflat'', ``Target'', ``Level'', and ``Interest''; it remains stable during routine periods and surges during episodes of high inflation or rapid policy-rate moves. \textit{Interest Rates} loads on ``Interest'', ``Hike'', ``Rais'', ``Expect'', and ``Borrow\_Cost'', and peaked sharply after the high inflation crisis in 1980 before declining during the financial crisis and at the beginning of the COVID-19 pandemic.

\begin{figure}[!ht]
\centering
\caption{Inter-model agreement with and without the irrelevant category
\label{fig:alpha_irrelevant}}
\includegraphics[width=0.99\linewidth]
{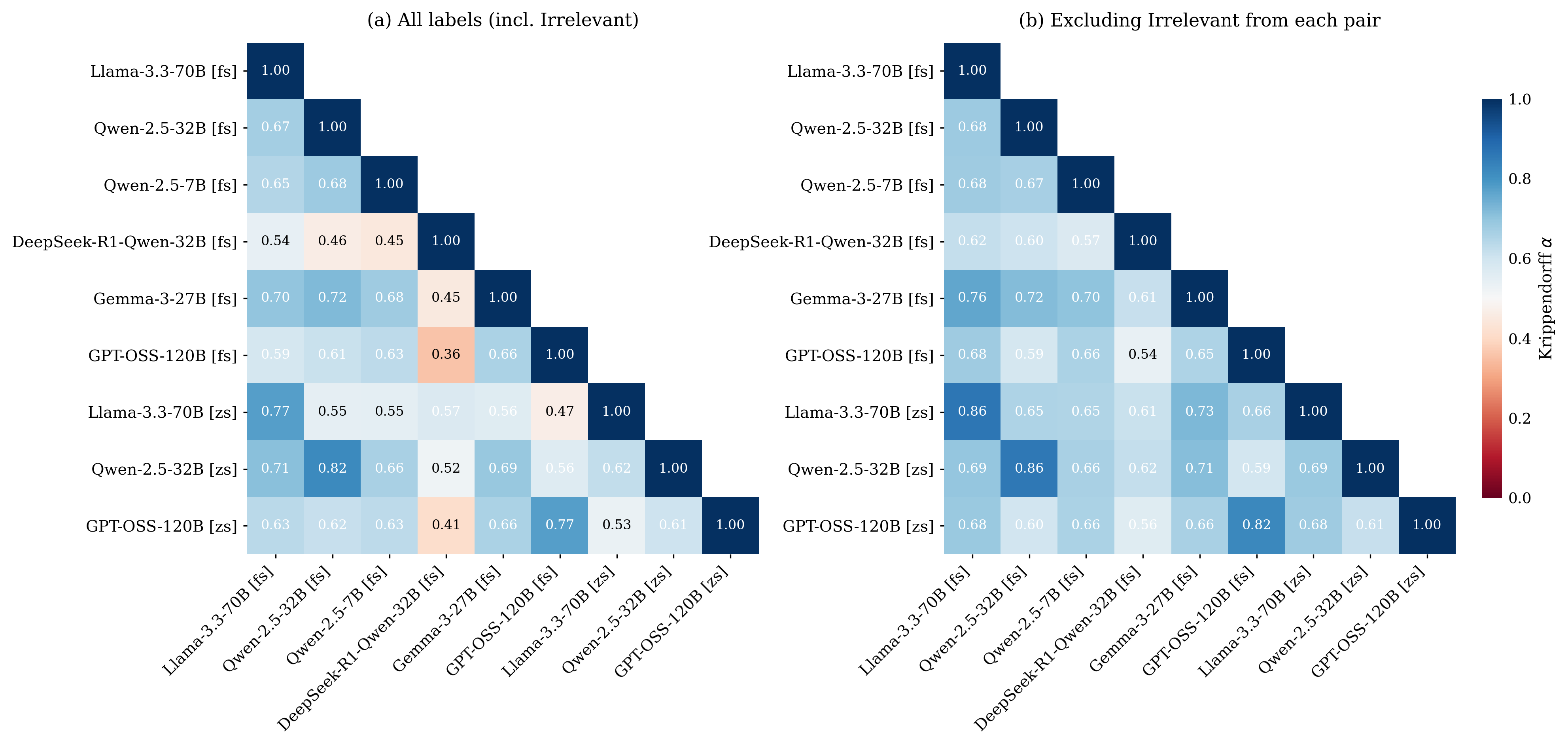}

\par\smallskip
\begin{minipage}{0.95\linewidth}\footnotesize
\textit{Note}: Panel (a) reports pairwise Krippendorff $\alpha$ including all sentiment labels, including the irrelevant category. Panel (b) recomputes pairwise agreement after excluding observations classified as irrelevant by either model in the pair. The overall agreement structure remains broadly stable across specifications.
\end{minipage}
\end{figure}

\begin{figure}[!ht]
\centering
\caption{Inter-model agreement across monetary-policy subperiods
\label{fig:alpha_subperiods}}
\includegraphics[width=0.99\linewidth]
{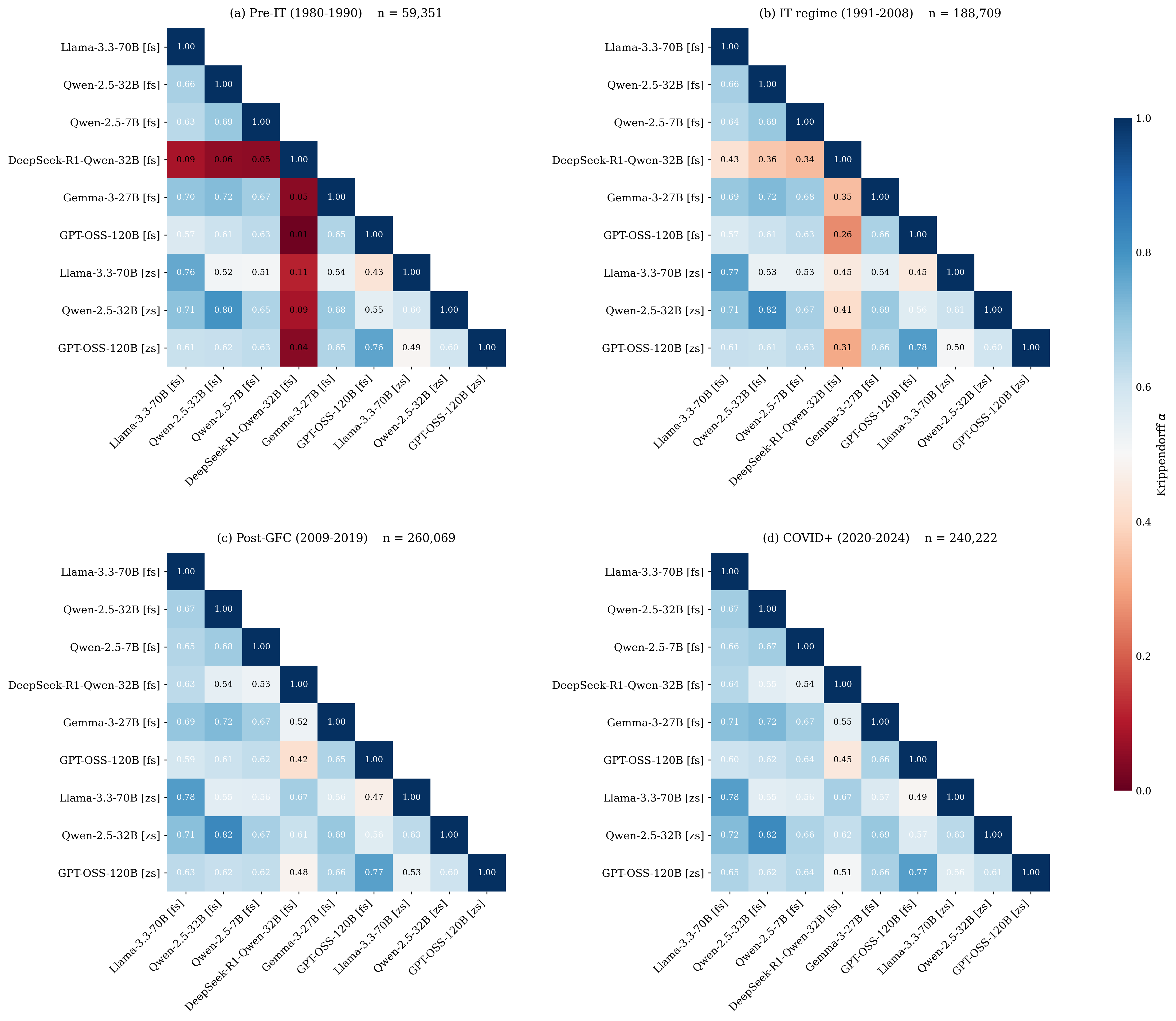}

\par\smallskip
\begin{minipage}{0.95\linewidth}\footnotesize
\textit{Note}: The figure reports pairwise Krippendorff $\alpha$ across four monetary-policy subperiods: pre-inflation-targeting (1980--1990), the inflation-targeting regime (1991--2008), post-global financial crisis (2009--2019), and the COVID and post-COVID period (2020--2024). Agreement remains comparatively strong across prompting configurations and model families in most subperiods, although DeepSeek-R1 exhibits weaker concordance in earlier decades.
\end{minipage}
\end{figure}

\subsection{Forecasting setup}\label{app:forecasting_setup}

We evaluate the predictive content of the text-based indicators by performing a pseudo-out-of-sample forecasting exercise. We consider the following unrestricted MIDAS model, which also includes the lags of the dependent variable, written as
\begin{equation}
    y_{t+h}=\beta_0+\sum_{j=0}^{p-1} \lambda_j y_{t-j}+\sum_{j=1}^N \sum_{k=0}^{K-1} \boldsymbol{b}_{j, k} L^{k / m} x_{j, t}^{(m)}+\varepsilon_{t+h}.
\end{equation}
To deal with the high-dimensional setting (244 predictors), we rely on machine-learning predictive modeling:
\begin{align}
y_{t+h}=G_{ h, t}\left(y_{t-p: t}^{AR},  \boldsymbol{x}_{t} \right)+ \varepsilon_{ t+h},
\end{align}
where the monthly predictors are
\[
\boldsymbol{x}_{t} = \left[x_{1, t}, x_{1, t-1/3}, \ldots, x_{1, t-(K-1)/3}, \ldots, x_{N, t}, \ldots, x_{N, t-(K-1)/3}\right]^{\prime},
\]
$\boldsymbol{G}_{h, t}$ is the mapping between the predictors and the target variable, and $\varepsilon_{i, t+h}$ is a zero-mean random error.

The target variable $y_{t}$ is the log difference of the CPI or Real GDP at time $t$. $h=1, \ldots, H$ represents the forecast horizon. The predictors consist of different sets of text measures:
\begin{itemize}
\item Media narratives: $20$ topics
\item Tense narratives: $80$ topic-tense (2T) categories ($20 \times 4$ tense categories)
\item Topic-tense-uncertainty: $80$ 2TU categories
\item Topic-tense-tone: $80$ 3T categories
\end{itemize}

The forecasting horizons are \textbf{$h=1, 2, 4, 6, 8$} quarters. The sample period is 1982Q1--2023Q3 (167 observations) where the training sample is 1982Q1--2002Q4, and the test period is 2003Q1--2023Q3. The hyperparameters are tuned with K-fold cross-validation. Benchmark models are the random walk (RW) and the autoregressive (AR). All forecasts are computed using a rolling window scheme with $R=84 - p + 1$ observations.

We consider three groups of competing models. Penalized regressions address the high dimension of the predictor set through L1 (LASSO), L2 (Ridge), and convex-combination (Elastic Net) penalties on a linear specification. Tree-based methods include Random Forest \citep{breiman2001random}, which averages predictions across an ensemble of decision trees grown on bootstrapped samples and random feature subsets, and Extreme Gradient Boosting (XGB), which builds trees sequentially to minimize the residuals of previous ones. Neural networks include a shallow feedforward architecture with one hidden layer (NN1) and a deeper version with three hidden layers (NN3). Three ensemble forecasts aggregate the previous models with a meta-learner: Ens-Linear pools the penalized regressions, Ens-NonLinear pools XGB, RF, NN1, and NN3, and Ens-All pools all of the above. We refer the reader to \cite{goulet2022machine} for a textbook treatment.

\subsection{Detailed forecast accuracy results}\label{app:forecasting_results}

Forecast accuracy is measured by the root mean squared error (RMSE) over the test sample, and the relative RMSE in Tables~\ref{tab:main_results_3T_cpi} and \ref{tab:main_results_3T_gdp} is reported with respect to the random walk benchmark.

\begin{table}[H]
\centering
    \begin{threeparttable}
\caption{Relative RMSE Forecasting Comparison CPI Inflation using 3T\label{tab:main_results_3T_cpi}}
\begin{tabular}{lccccc}
\hline
\hline
 & \multicolumn{5}{c}{\textbf{Relative RMSE}} \\
\cline{2-6}
\textbf{Models} & \textbf{h=1} & \textbf{h=2} & \textbf{h=4} & \textbf{h=6} & \textbf{h=8} \\
\hline
MFML LASSO & 0.7517\(^{***}\) & 0.6374\(^{***}\) & 0.9624 & 0.6669\(^{***}\) & 0.8585 \\
MFML Ridge & 0.7449\(^{***}\) & 0.6101\(^{***}\) & 1.0068 & 0.7253\(^{***}\) & 0.8880 \\
MFML Enet & 0.7513\(^{***}\) & 0.6365\(^{***}\) & 0.9624 & 0.6667\(^{***}\) & 0.8535 \\
\hline
MFML XGB & 0.8602\(^{**}\) & 0.6527\(^{***}\) & 1.0569 & 0.7452\(^{***}\) & 0.9553 \\
MFML RF & 0.8239\(^{***}\) & 0.6461\(^{***}\) & 0.9799 & 0.6896\(^{***}\) & 0.9076 \\
MFML NN1 & 0.9297 & 0.6265\(^{***}\) & 1.0401 & 0.6831\(^{***}\) & 0.9034 \\
MFML NN3 & 0.8236\(^{***}\) & 0.6462\(^{***}\) & 0.9579 & 0.7011\(^{***}\) & 0.9572 \\
\hline
Ens-Linear & 0.7386\(^{***}\) & 0.6217\(^{***}\) & 0.9703 & 0.6805\(^{***}\) & 0.8545 \\
Ens-NonLinear & 0.8291\(^{***}\) & 0.6311\(^{***}\) & 0.9696 & 0.6922\(^{***}\) & 0.9037 \\
Ens-All & 0.7754\(^{***}\) & 0.6193\(^{***}\) & 0.9652 & 0.6853\(^{***}\) & 0.8771 \\
\hline
\end{tabular}
 \begin{tablenotes}[flushleft]  {\scriptsize
     \textit{Notes}: Relative RMSE (i.e.\ RW) for each model and forecasting horizon. A number below 1 means that the competing model outperforms the benchmark. \(^{***}\), \(^{**}\), \(^{*}\) stand for 1\%, 5\% and 10\% significance of Diebold-Mariano (DM) test.}
     \end{tablenotes}
   \end{threeparttable}
\end{table}

For CPI inflation (Table~\ref{tab:main_results_3T_cpi}), text-based predictors deliver relative RMSE between 0.61 and 0.96 at horizons $h \in \{1, 2, 6, 8\}$, with most cells significant at the 1\% level. Gains are largest at the two- and six-quarter horizons (around 0.6 to 0.7) and weaken at four quarters, where the relative RMSE clusters between 0.96 and 1.06 across model families and none of the cells reaches statistical significance: at this horizon several specifications fall slightly below the random-walk benchmark. At eight quarters the relative RMSE returns into the 0.85--0.96 range. The four model families (penalized regressions, tree-based methods, neural networks, and ensembles) deliver broadly similar performance.

\begin{table}[H]
\centering
    \begin{threeparttable}
\caption{Relative RMSE Forecasting Comparison for GDP using 3T\label{tab:main_results_3T_gdp}}
\begin{tabular}{lccccc}
\hline
\hline
 & \multicolumn{5}{c}{\textbf{Relative RMSE}} \\
\cline{2-6}
\textbf{Models} & \textbf{h=1} & \textbf{h=2} & \textbf{h=4} & \textbf{h=6} & \textbf{h=8} \\
\hline
MFML LASSO & 0.6822 & 0.6934 & 0.7715 & 0.7597 & 0.7582\(^{*}\) \\
MFML Ridge & 0.6873 & 0.7114\(^{*}\) & 0.7552\(^{*}\) & 0.7092 & 0.7475\(^{*}\) \\
MFML Enet & 0.6833 & 0.6933 & 0.7716 & 0.7580 & 0.7582\(^{*}\) \\
\hline
MFML XGB & 0.6421 & 0.6722\(^{*}\) & 0.7204\(^{*}\) & 0.7213 & 0.7058\(^{*}\) \\
MFML RF & 0.6780 & 0.6752\(^{*}\) & 0.7277\(^{*}\) & 0.7055 & 0.6960\(^{*}\) \\
MFML NN1 & 0.6610 & 0.6804\(^{*}\) & 0.7499\(^{*}\) & 0.7240 & 0.7190\(^{*}\) \\
MFML NN3 & 0.6630 & 0.6791\(^{*}\) & 0.7257\(^{*}\) & 0.7020 & 0.7127\(^{*}\) \\
\hline
Ens-Linear & 0.6826 & 0.6959\(^{*}\) & 0.7616\(^{*}\) & 0.7403 & 0.7517\(^{*}\) \\
Ens-NonLinear & 0.6547 & 0.6735\(^{*}\) & 0.7217\(^{*}\) & 0.7064 & 0.7028\(^{*}\) \\
Ens-All & 0.6627 & 0.6807\(^{*}\) & 0.7293\(^{*}\) & 0.7190 & 0.7089\(^{*}\) \\
\hline
\end{tabular}
 \begin{tablenotes}[flushleft]  {\scriptsize
     \textit{Notes}: Relative RMSE (i.e.\ RW) for each model and forecasting horizon. A number below 1 means that the competing model outperforms the benchmark. \(^{***}\), \(^{**}\), \(^{*}\) stand for 1\%, 5\% and 10\% significance of Diebold-Mariano (DM) test.}
     \end{tablenotes}
   \end{threeparttable}
\end{table}

For GDP growth (Table~\ref{tab:main_results_3T_gdp}), every model beats the random walk at every horizon, with relative RMSE between 0.64 and 0.77 and a tighter spread than for CPI. Gains are more uniform across horizons. Statistical significance is concentrated at the 10\% level rather than the 1\% level, consistent with smaller residual differences when all competitors do well.

\subsection{Variable importance}\label{app:forecasting_shap}

\begin{figure}[ht]
\caption{Variable importance measures for selected models in CPI forecasting at horizon $h=2$\label{fig:SHAP}}
  \begin{subfigure}[b]{0.5\linewidth}
    \centering
        \caption*{Panel A: Elastic Net}
    \includegraphics[width=0.99\linewidth]{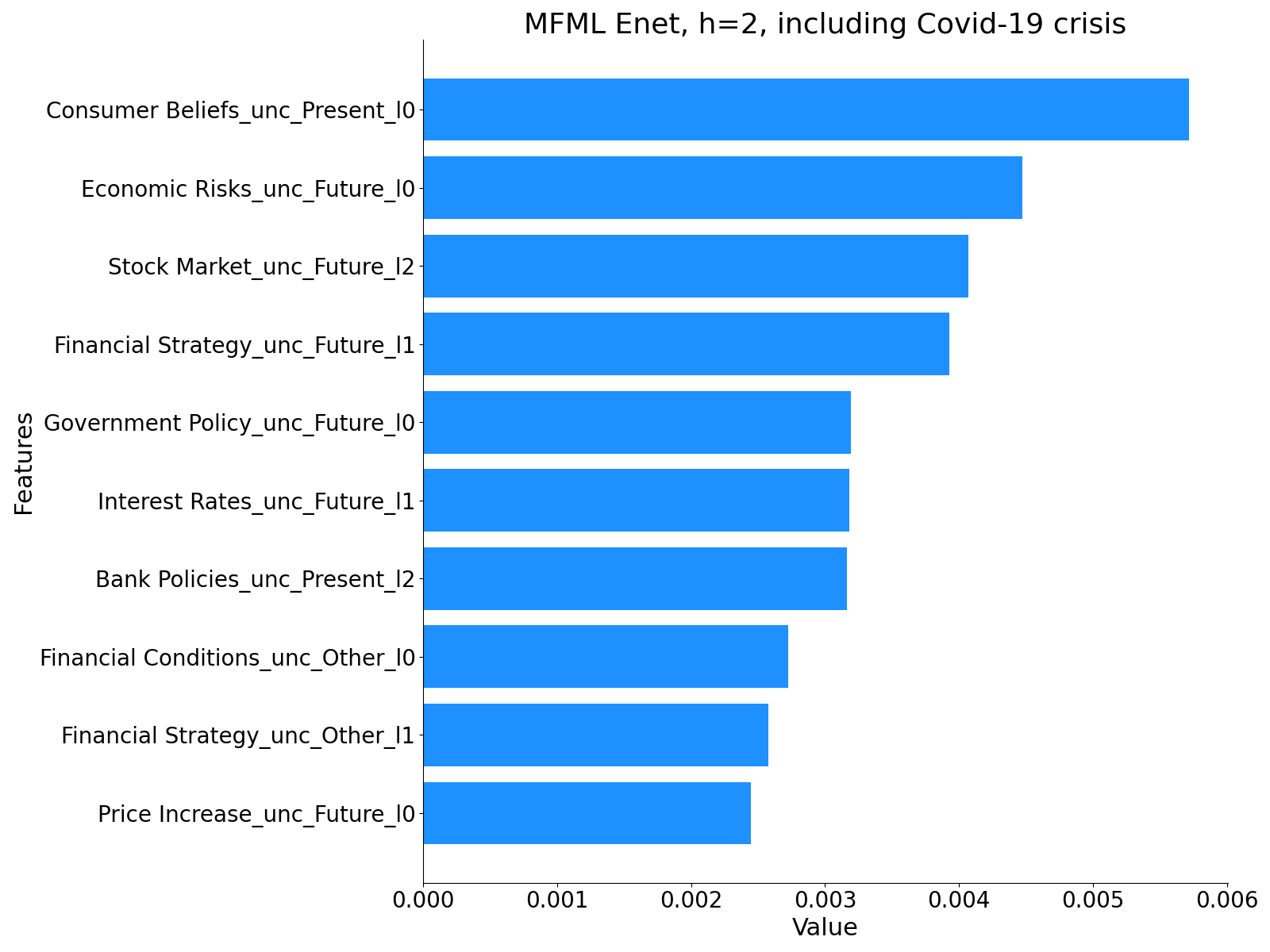}
  \end{subfigure}
  \begin{subfigure}[b]{0.5\linewidth}
    \centering
        \caption*{Panel B: Random Forest}
    \includegraphics[width=0.99\linewidth]{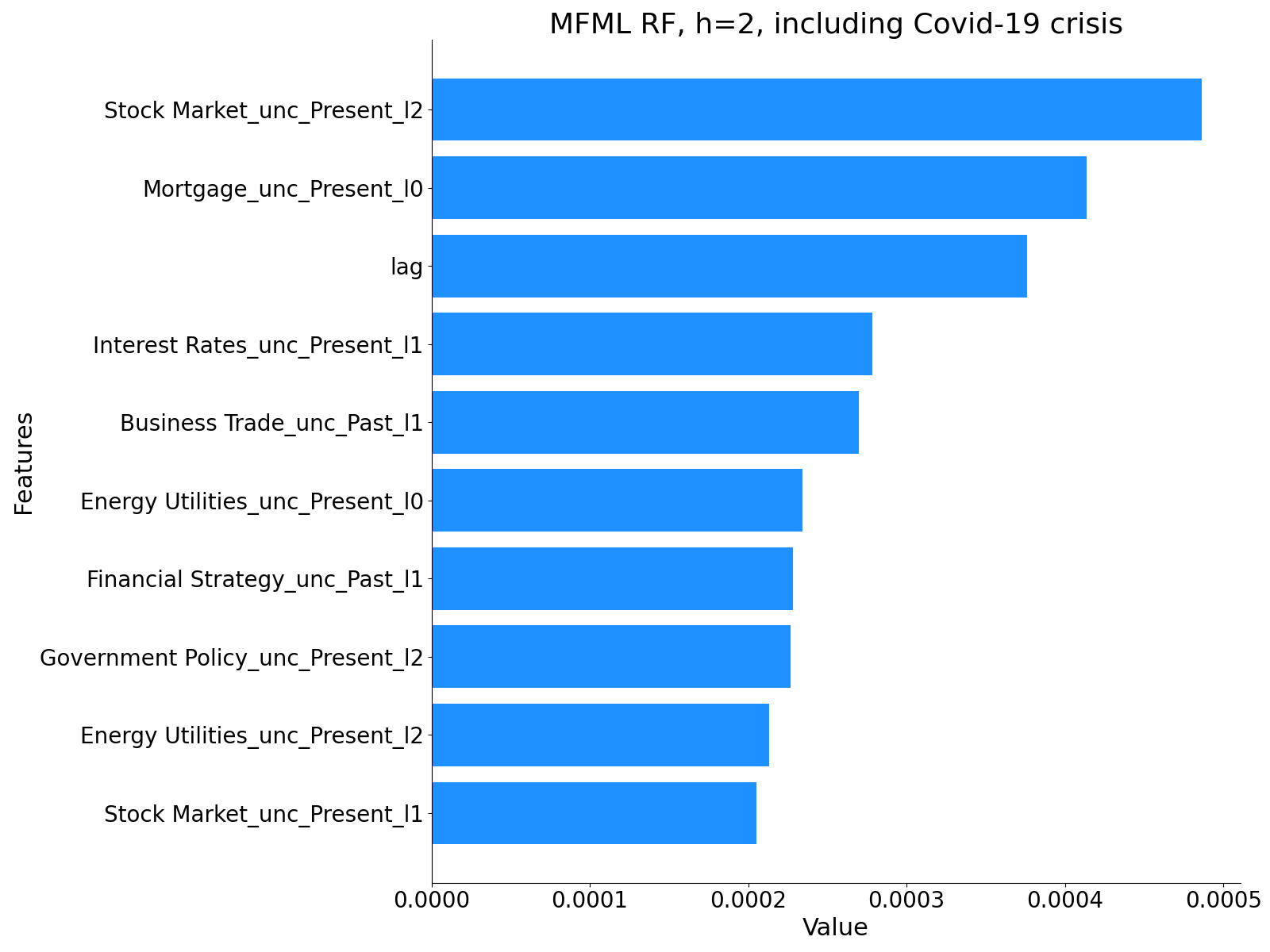}
  \end{subfigure}
  \begin{subfigure}[b]{0.5\linewidth}
    \centering
    \caption*{Panel C: Extreme Gradient Boosting}
    \includegraphics[width=0.99\linewidth]{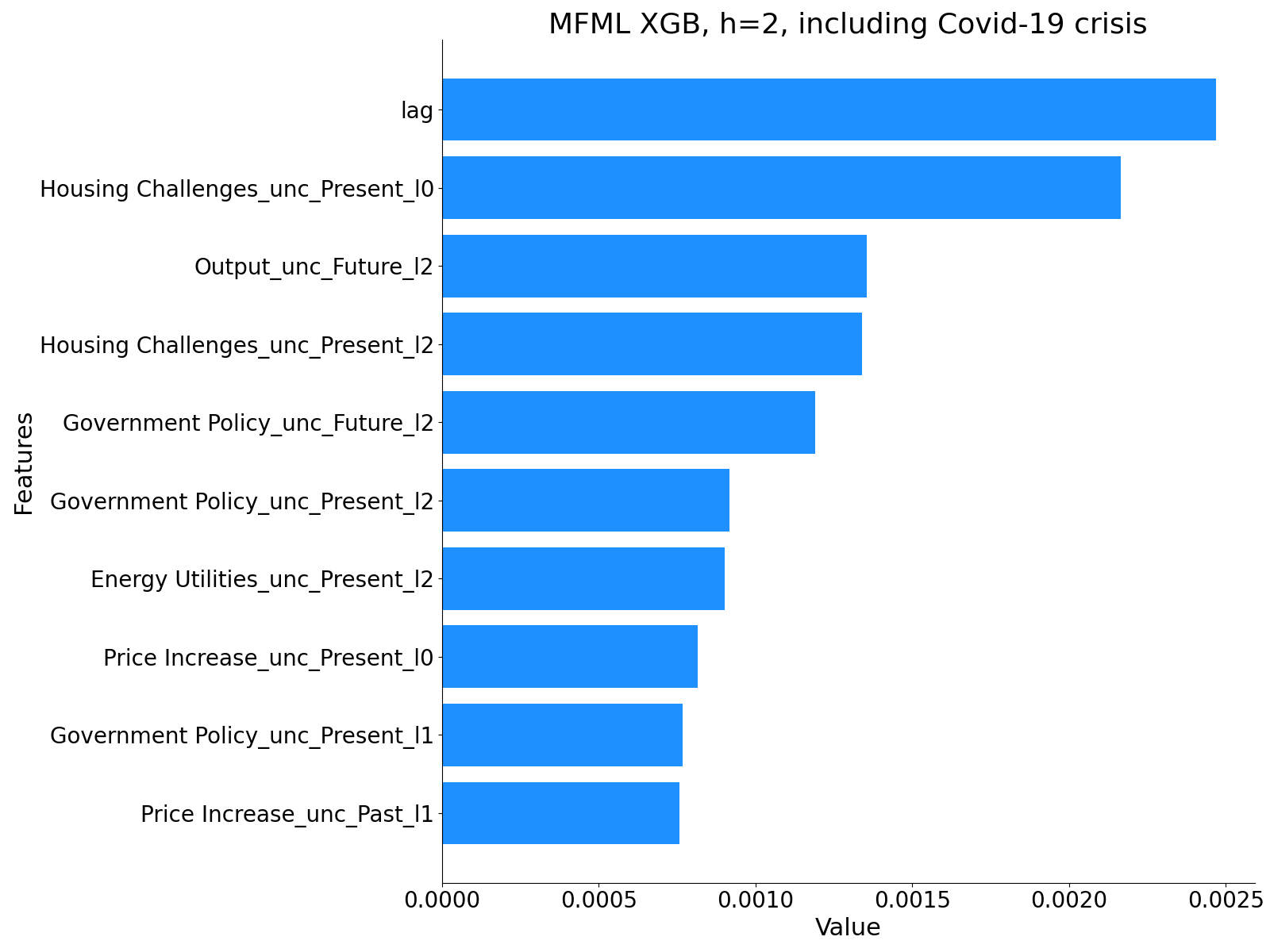}
  \end{subfigure}
  \begin{subfigure}[b]{0.5\linewidth}
    \centering
        \caption*{Panel D: Neural Network}
    \includegraphics[width=0.99\linewidth]{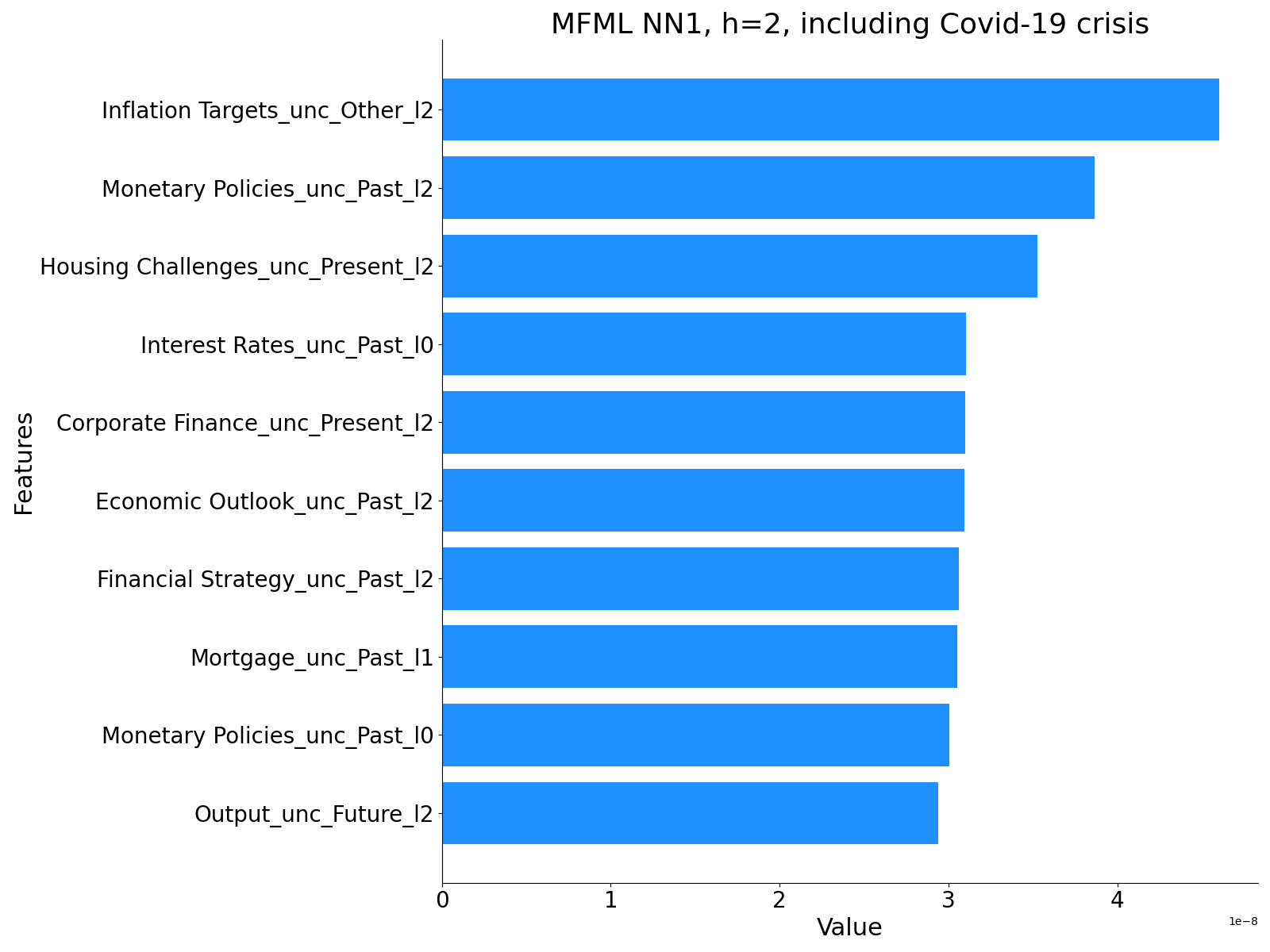}
  \end{subfigure}
\end{figure}

\begin{figure}[ht]
\caption{Variable importance measures for selected models in GDP forecasting at horizon $h=2$\label{fig:SHAP_gdp}}
  \begin{subfigure}[b]{0.5\linewidth}
    \centering
        \caption*{Panel A: Elastic Net}
    \includegraphics[width=0.99\linewidth]{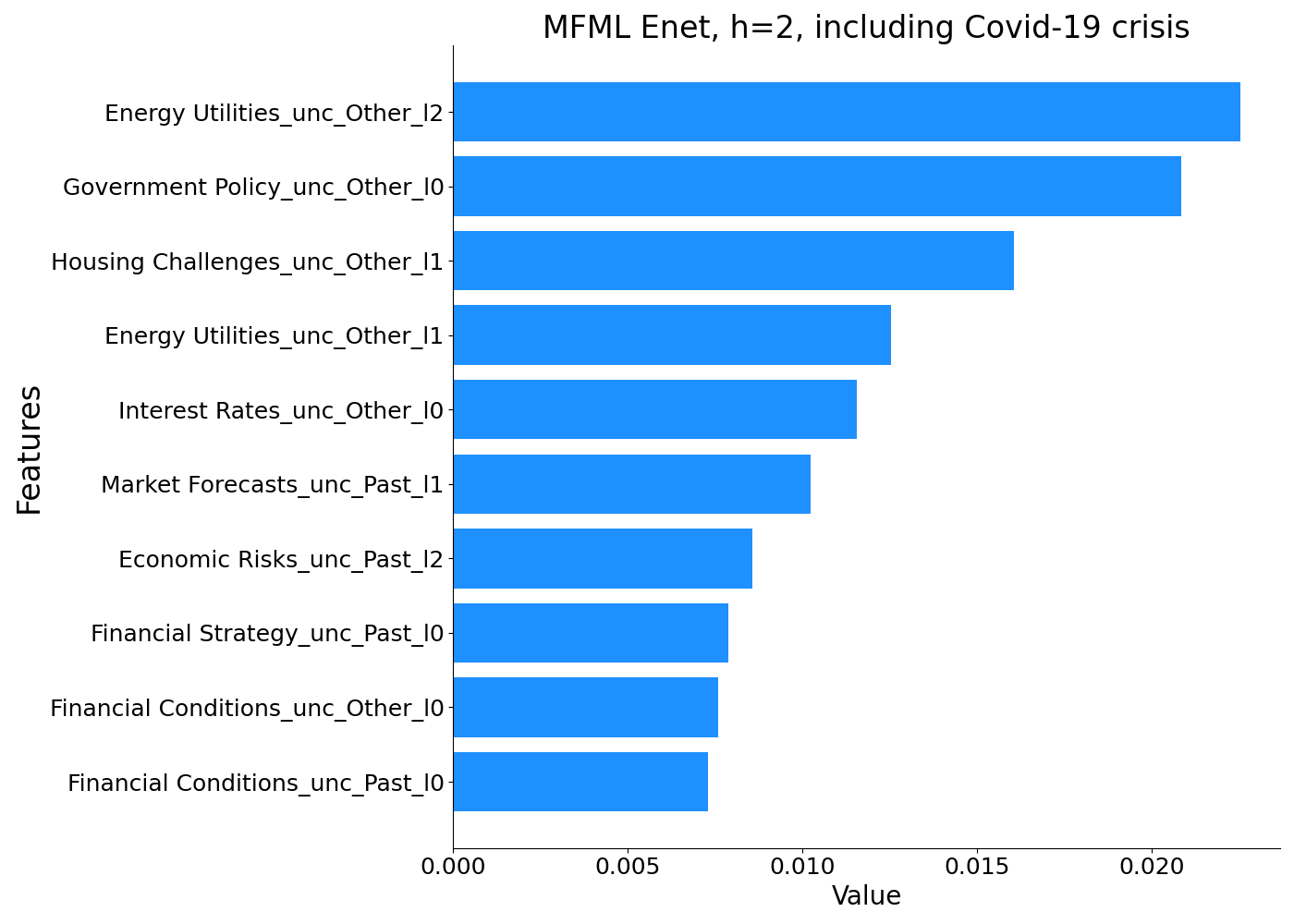}
  \end{subfigure}
  \begin{subfigure}[b]{0.5\linewidth}
    \centering
        \caption*{Panel B: Random Forest}
    \includegraphics[width=0.99\linewidth]{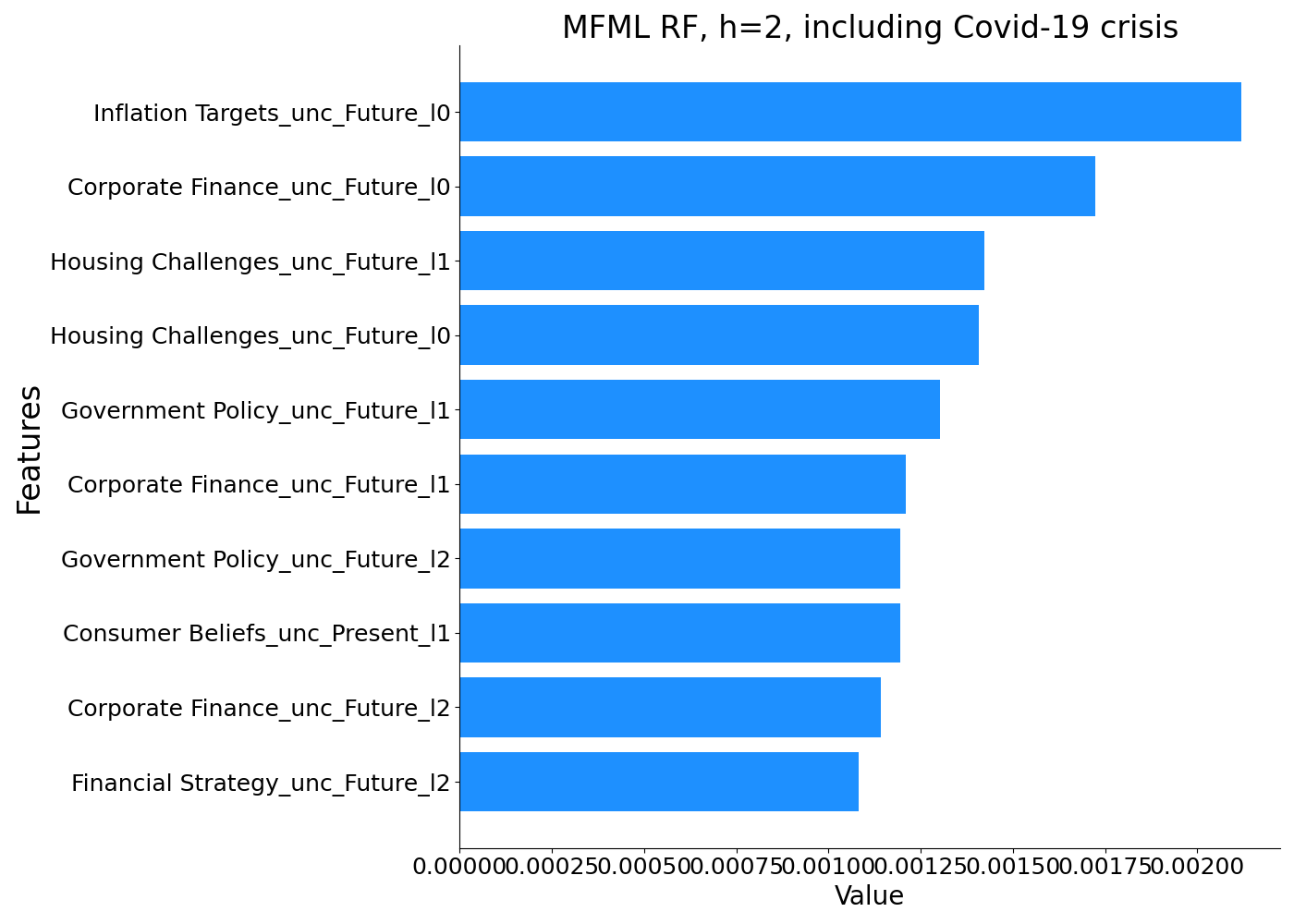}
  \end{subfigure}
  \begin{subfigure}[b]{0.5\linewidth}
    \centering
    \caption*{Panel C: Extreme Gradient Boosting}
    \includegraphics[width=0.99\linewidth]{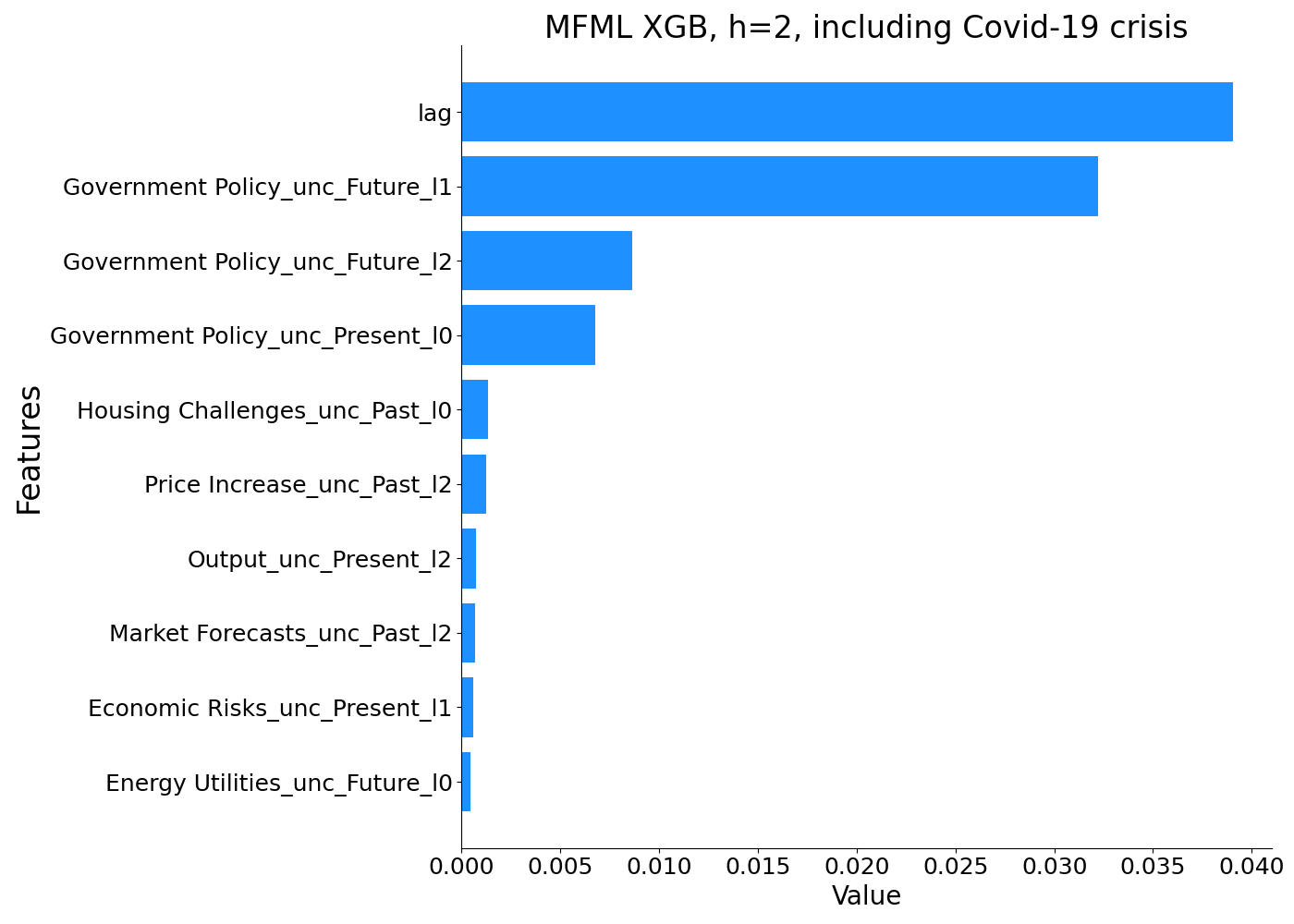}
  \end{subfigure}
  \begin{subfigure}[b]{0.5\linewidth}
    \centering
        \caption*{Panel D: Neural Network}
    \includegraphics[width=0.99\linewidth]{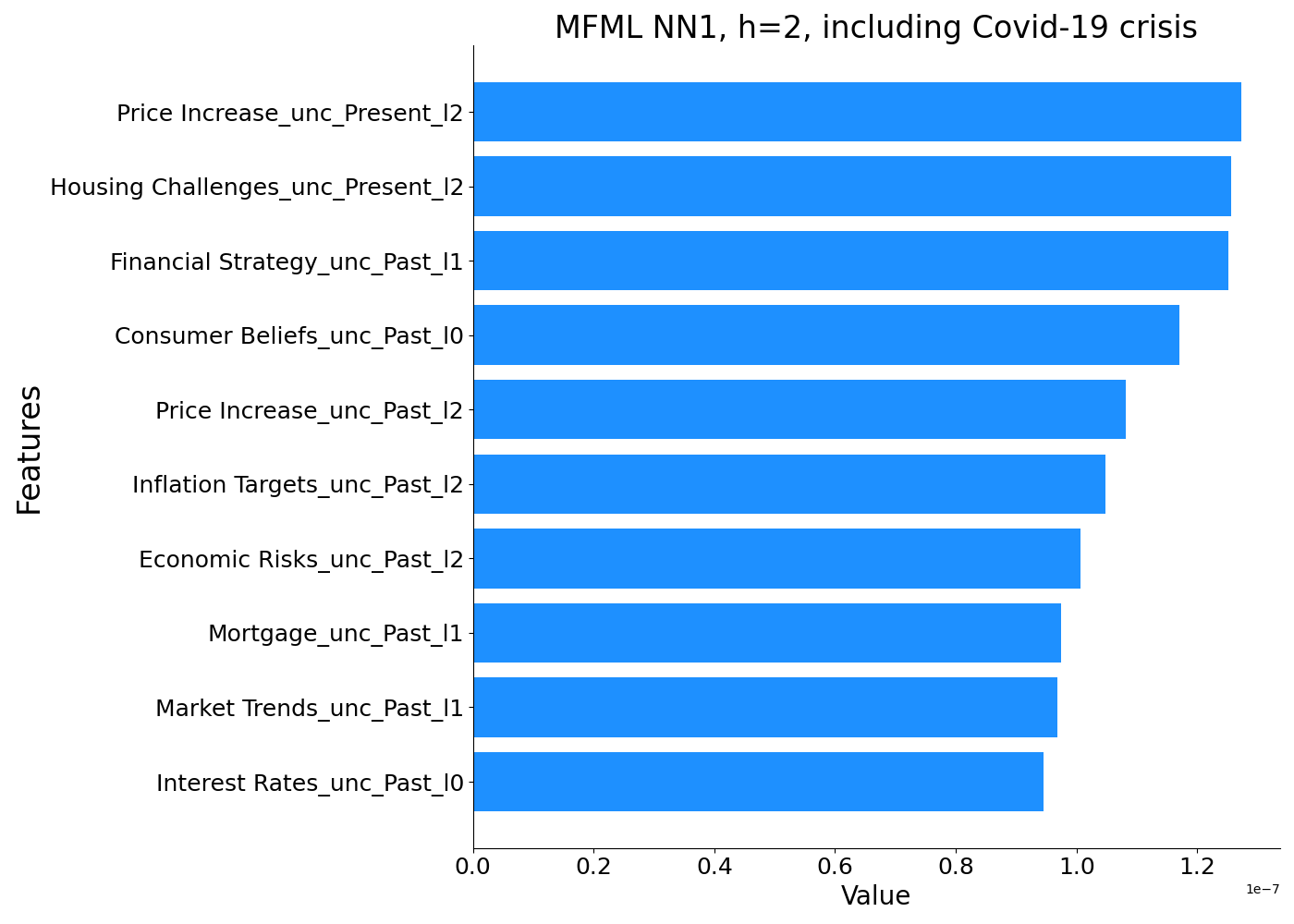}
  \end{subfigure}
\end{figure}

It is of economic interest to know the importance of the individual predictors in the models underlying the forecasts. To this end, we compute variable importance measures for the fitted models via the approach of SHAP (SHapley Additive exPlanations) values methodology, as detailed in \cite{lundberg2017unified}. In our time-series context, we denote the SHAP value of model $m$ corresponding to the $k$-th predictor for the $t$-th observation in the training sample by $\phi_{k,t}^{(m)}$. This SHAP value indicates the contribution of the $k$-th predictor to the prediction of the target variable for the $t$-th observation, measured as the deviation from the mean of the target over the training sample. Figures~\ref{fig:SHAP} and \ref{fig:SHAP_gdp} depict the top 10 predictors in terms of variable importance for selected models from shrinkage, neural networks and tree-based methods.

The analysis reveals several key predictors across different models. Notably, variables related to uncertainty and various temporal dimensions (past, present, and future) consistently appear among the top predictors. This suggests that incorporating topics, tense, and uncertainty is crucial for accurate CPI forecasting. For instance, in the Random Forest model, variables such as stock-market-unc-present, interest-rates-unc-present, and government-policies-unc-present are highly important. In the Elastic Net model, predictors like consumer-beliefs-unc-present, economic-risks-unc-future, and government-policy-unc-future are significant. The Extreme Gradient Boosting model highlights housing-challenges-unc-present and price-increase-unc-past among its top predictors. Finally, the Neural Network model emphasizes inflation-target-unc-other and monetary-policies-unc-past. These findings underscore the importance of considering both the type of topic and the temporal context in which it is discussed. The consistent appearance of uncertainty-related predictors across all models indicates that market perceptions of uncertainty play a vital role in shaping CPI forecasts. The combination of these elements -- topics, tense, and uncertainty -- provides a comprehensive framework for understanding and predicting CPI movements based on narrative data.

\section{CBILA Taylor-rule estimates and the look-ahead bias diagnostic}\label{app:cbila_taylor}

This appendix complements Section~\ref{sec:taylor}. We re-estimate the Taylor rule using the CBILA generative-LLM tones described in Section~\ref{sec:cbila}, document the additional fragility relative to the FinBERT- and ModernFinBERT-based estimates of Section~\ref{sec:taylor_results}, and report a formal analysis of why IV with lagged macro instruments does not eliminate the bias from look-ahead leakage. The sample, exogenous regressors, instrument sets, two-step efficient GMM with Bartlett kernel of bandwidth $L=4$, and HAC inference are identical to Section~\ref{sec:taylor}; the only change is the sentiment regressor.

The CBILA spec includes ten series. ModernFinBERT (\textit{M-FinBERT}) is kept for direct comparison with Section~\ref{sec:taylor_results}. The remaining nine series are the CBILA tones produced by six generative LLMs under two prompting configurations: six few-shot tones ([fs]) for Llama 70B, Qwen 32B, Qwen 7B, DeepSeek-R1 32B, Gemma 27B and GPT-OSS 120B; and three zero-shot tones ([zs]) for Llama 70B, Qwen 32B and GPT-OSS 120B. All series are standardized over the estimation sample.

\subsection{Taylor-rule estimates with the CBILA tones, overall and future-tense}\label{app:cbila_results}

Tables~\ref{tab:appC_cbila_overall_rich} and \ref{tab:appC_cbila_overall_rev} report the OLS and GMM estimates with the overall CBILA tones, under the rich instrument set and the forecast-revision set respectively. Tables~\ref{tab:appC_cbila_fut_rich} and \ref{tab:appC_cbila_fut_rev} report the future-tense counterparts.

\begin{table}[htbp]
\centering
\scriptsize
\setlength{\tabcolsep}{2pt}
\renewcommand{\arraystretch}{0.95}
\caption{Taylor Rule Estimates --- CBILA spec, overall tone, 1991Q1--2019Q4}
\label{tab:appC_cbila_overall_rich}
\begin{threeparttable}
\begin{tabular}{lccccccccccc}
\hline\hline
 & (1) & (2) & (3) & (4) & (5) & (6) & (7) & (8) & (9) & (10) & (11) \\
 & \textbf{Baseline} & \textbf{M-BERT} & \textbf{Llama} & \textbf{Qwen32} & \textbf{Qwen7} & \textbf{DSeek} & \textbf{Gemma} & \textbf{GPT} & \textbf{Llama$^z$} & \textbf{Qwen32$^z$} & \textbf{GPT$^z$} \\
\hline
\multicolumn{12}{l}{\textbf{Panel A. OLS (Newey--West HAC standard errors, $L=4$)}} \\
\hline
$\rho$ (Lag rate)        & 0.888*** & 0.901*** & 0.835*** & 0.870*** & 0.863*** & 0.892*** & 0.848*** & 0.850*** & 0.859*** & 0.866*** & 0.854*** \\
                          & (0.037)  & (0.043)  & (0.054)  & (0.049)  & (0.048)  & (0.041)  & (0.049)  & (0.049)  & (0.051)  & (0.048)  & (0.049)  \\
$\alpha$ (Neutral rate)  & 0.119*** & 0.104**  & 0.173*** & 0.142*** & 0.144*** & 0.118*** & 0.171*** & 0.168*** & 0.150*** & 0.144*** & 0.163*** \\
                          & (0.039)  & (0.046)  & (0.055)  & (0.053)  & (0.049)  & (0.039)  & (0.054)  & (0.054)  & (0.053)  & (0.052)  & (0.054)  \\
$\gamma$ (Inflation gap) & 0.275**  & 0.276**  & 0.204*   & 0.251**  & 0.251**  & 0.272**  & 0.205*   & 0.216*   & 0.239**  & 0.252**  & 0.221*   \\
                          & (0.124)  & (0.125)  & (0.122)  & (0.122)  & (0.115)  & (0.120)  & (0.120)  & (0.119)  & (0.121)  & (0.119)  & (0.117)  \\
$\beta$ (Output gap)     & 0.108*** & 0.073    & 0.190*** & 0.145*** & 0.153*** & 0.103**  & 0.175*** & 0.177*** & 0.161*** & 0.151*** & 0.170*** \\
                          & (0.042)  & (0.045)  & (0.048)  & (0.042)  & (0.041)  & (0.043)  & (0.038)  & (0.042)  & (0.045)  & (0.043)  & (0.043)  \\
$\delta$ (Sentiment)     & ---      & 0.075    & -0.171*  & -0.081   & -0.105   & 0.022    & -0.147*  & -0.146*  & -0.108   & -0.094   & -0.128   \\
                          &          & (0.070)  & (0.099)  & (0.097)  & (0.085)  & (0.054)  & (0.086)  & (0.080)  & (0.096)  & (0.091)  & (0.081)  \\
Constant                  & 0.161*** & 0.120*   & 0.248*** & 0.188*** & 0.215*** & 0.141**  & 0.195*** & 0.203*** & 0.213*** & 0.201*** & 0.200*** \\
                          & (0.056)  & (0.073)  & (0.078)  & (0.070)  & (0.076)  & (0.066)  & (0.066)  & (0.066)  & (0.074)  & (0.071)  & (0.066)  \\
\hline
Obs.                      & 110   & 110   & 110   & 110   & 110   & 110   & 110   & 110   & 110   & 110   & 110   \\
Adj. $R^{2}$              & 0.933 & 0.933 & 0.936 & 0.933 & 0.934 & 0.932 & 0.936 & 0.936 & 0.934 & 0.933 & 0.935 \\
\hline
\multicolumn{12}{l}{\textbf{Panel B. GMM (rich instrument set, 12 instruments, 11 over-identifying restrictions)}} \\
\hline
$\rho$ (Lag rate)        & 0.888*** & 0.964*** & 0.929*** & 0.960*** & 0.961*** & 0.930*** & 0.927*** & 0.942*** & 0.950*** & 0.954*** & 0.938*** \\
                          & (0.037)  & (0.029)  & (0.041)  & (0.033)  & (0.036)  & (0.027)  & (0.043)  & (0.039)  & (0.035)  & (0.035)  & (0.038)  \\
$\alpha$ (Neutral rate)  & 0.119*** & 0.025    & 0.080*   & 0.037    & 0.045    & 0.091*** & 0.076    & 0.060    & 0.053    & 0.048    & 0.065    \\
                          & (0.039)  & (0.031)  & (0.041)  & (0.037)  & (0.038)  & (0.025)  & (0.048)  & (0.044)  & (0.036)  & (0.036)  & (0.042)  \\
$\gamma$ (Inflation gap) & 0.275**  & 0.311*** & 0.274*** & 0.283*** & 0.297*** & 0.298*** & 0.276*** & 0.293*** & 0.292*** & 0.276*** & 0.292*** \\
                          & (0.124)  & (0.071)  & (0.071)  & (0.082)  & (0.080)  & (0.063)  & (0.072)  & (0.074)  & (0.073)  & (0.078)  & (0.074)  \\
$\beta$ (Output gap)     & 0.108*** & -0.039   & 0.087    & 0.032    & 0.036    & 0.079**  & 0.091    & 0.063    & 0.045    & 0.045    & 0.069    \\
                          & (0.042)  & (0.051)  & (0.055)  & (0.050)  & (0.054)  & (0.034)  & (0.056)  & (0.061)  & (0.055)  & (0.049)  & (0.058)  \\
$\delta$ (Sentiment)     & ---      & 0.265*** & 0.051    & 0.163**  & 0.153*   & 0.099    & 0.054    & 0.095    & 0.122    & 0.137*   & 0.082    \\
                          &          & (0.074)  & (0.081)  & (0.082)  & (0.081)  & (0.061)  & (0.091)  & (0.088)  & (0.075)  & (0.078)  & (0.083)  \\
Constant                  & 0.161*** & 0.041    & 0.150**  & 0.112**  & 0.094    & 0.109*   & 0.170*** & 0.149**  & 0.120*   & 0.114*   & 0.149**  \\
                          & (0.056)  & (0.062)  & (0.070)  & (0.056)  & (0.069)  & (0.064)  & (0.057)  & (0.059)  & (0.063)  & (0.062)  & (0.059)  \\
\hline
Obs.                      & 110 & 110    & 110    & 110    & 110    & 110    & 110    & 110    & 110    & 110    & 110    \\
First-stage $F$           & --- & 286.55 & 412.76 & 417.12 & 493.99 & 599.50 & 259.28 & 243.01 & 517.28 & 368.23 & 370.71 \\
Hansen $J$ $p$-value      & --- & 0.386  & 0.024  & 0.280  & 0.240  & 0.088  & 0.032  & 0.060  & 0.178  & 0.218  & 0.042  \\
\hline\hline
\end{tabular}
\begin{tablenotes}[para,flushleft]
\scriptsize
\item \textit{Notes:} Two-step efficient GMM with sentiment treated as endogenous. Newey--West HAC standard errors with $L=4$ in parentheses. * $p<0.10$, ** $p<0.05$, *** $p<0.01$. Dependent variable: BoC overnight policy rate; sample 1991Q1--2019Q4, $T=110$ obs. GMM weighting and covariance use the Bartlett kernel with bandwidth $L$. The Hansen $J$ $p$-value is obtained by block wild Rademacher bootstrap with $B=500$ resamples and non-overlapping block length 4. Column headers identify the LLM family; $^z$ denotes zero-shot prompting, while few-shot prompting is the default. M-BERT abbreviates ModernFinBERT and DSeek abbreviates DeepSeek-R1.
\end{tablenotes}
\end{threeparttable}
\end{table}

\begin{table}[htbp]
\centering
\scriptsize
\setlength{\tabcolsep}{2pt}
\renewcommand{\arraystretch}{0.95}
\caption{GMM with Forecast-Revision Instruments --- CBILA spec, overall tone, 1991Q1--2019Q4}
\label{tab:appC_cbila_overall_rev}
\begin{threeparttable}
\begin{tabular}{lccccccccccc}
\hline\hline
 & (1) & (2) & (3) & (4) & (5) & (6) & (7) & (8) & (9) & (10) & (11) \\
 & \textbf{Baseline} & \textbf{M-BERT} & \textbf{Llama} & \textbf{Qwen32} & \textbf{Qwen7} & \textbf{DSeek} & \textbf{Gemma} & \textbf{GPT} & \textbf{Llama$^z$} & \textbf{Qwen32$^z$} & \textbf{GPT$^z$} \\
\hline
$\rho$ (Lag rate)        & 0.888*** & 0.986*** & 1.185*** & 1.110*** & 1.161*** & 1.060*** & 1.192*** & 1.183*** & 1.186*** & 1.162*** & 1.189*** \\
                          & (0.037)  & (0.052)  & (0.252)  & (0.091)  & (0.126)  & (0.136)  & (0.198)  & (0.176)  & (0.145)  & (0.123)  & (0.170)  \\
$\alpha$ (Neutral rate)  & 0.119*** & 0.012    & -0.158   & -0.147   & -0.148   & 0.060    & -0.228   & -0.238   & -0.199   & -0.179   & -0.248   \\
                          & (0.039)  & (0.056)  & (0.262)  & (0.124)  & (0.136)  & (0.070)  & (0.264)  & (0.248)  & (0.176)  & (0.152)  & (0.239)  \\
$\gamma$ (Inflation gap) & 0.275**  & 0.321*** & 0.701**  & 0.546**  & 0.529**  & 0.205    & 0.787**  & 0.746**  & 0.648**  & 0.560**  & 0.769**  \\
                          & (0.124)  & (0.112)  & (0.349)  & (0.228)  & (0.259)  & (0.174)  & (0.352)  & (0.340)  & (0.289)  & (0.260)  & (0.341)  \\
$\beta$ (Output gap)     & 0.108*** & -0.085   & -0.288   & -0.313   & -0.356   & -0.091   & -0.325   & -0.405   & -0.434   & -0.380   & -0.421   \\
                          & (0.042)  & (0.090)  & (0.434)  & (0.233)  & (0.266)  & (0.170)  & (0.369)  & (0.379)  & (0.326)  & (0.292)  & (0.373)  \\
$\delta$ (Sentiment)     & ---      & 0.444*** & 0.818    & 0.908*   & 1.036*   & 0.772    & 0.890    & 1.050    & 1.073*   & 1.060*   & 1.055    \\
                          &          & (0.170)  & (0.886)  & (0.482)  & (0.600)  & (0.649)  & (0.758)  & (0.769)  & (0.636)  & (0.606)  & (0.746)  \\
Constant                  & 0.161*** & -0.071   & -0.301   & -0.171   & -0.408   & -0.518   & -0.108   & -0.171   & -0.375   & -0.329   & -0.175   \\
                          & (0.056)  & (0.136)  & (0.481)  & (0.203)  & (0.356)  & (0.598)  & (0.244)  & (0.295)  & (0.346)  & (0.303)  & (0.305)  \\
\hline
Obs.                      & 110 & 110    & 110    & 110   & 110   & 110   & 110   & 110   & 110    & 110   & 110   \\
First-stage $F$           & --- & 235.60 & 100.49 & 54.81 & 82.07 & 30.95 & 78.93 & 73.67 & 105.69 & 89.78 & 82.95 \\
Hansen $J$ $p$-value      & --- & 0.532  & 0.228  & 0.860 & 0.326 & 0.378 & 0.360 & 0.508 & 0.762  & 0.748 & 0.450 \\
\hline\hline
\end{tabular}
\begin{tablenotes}[para,flushleft]
\scriptsize
\item \textit{Notes:} Two-step efficient GMM with sentiment treated as endogenous. Newey--West HAC standard errors with $L=4$ in parentheses. * $p<0.10$, ** $p<0.05$, *** $p<0.01$. Dependent variable: BoC overnight policy rate; sample 1991Q1--2019Q4, $T=110$ obs. Excluded instruments (forecast-revision set): lag~2 of the policy rate, the one-quarter revision $E_{t-1}(\pi_{t+4})-E_{t-2}(\pi_{t+4})$ in the staff inflation forecast, and the analogous revision $E_{t-1}(x_{t+2})-E_{t-2}(x_{t+2})$ in the staff output-gap forecast (3 instruments, 2 over-identifying restrictions). GMM weighting and covariance use the Bartlett kernel with bandwidth $L$. The Hansen $J$ $p$-value is obtained by block wild Rademacher bootstrap ($B{=}500$, block length 4). Column headers identify the LLM family; $^z$ denotes zero-shot prompting, while few-shot prompting is the default. M-BERT abbreviates ModernFinBERT and DSeek abbreviates DeepSeek-R1.
\end{tablenotes}
\end{threeparttable}
\end{table}

\begin{table}[htbp]
\centering
\scriptsize
\setlength{\tabcolsep}{2pt}
\renewcommand{\arraystretch}{0.95}
\caption{Taylor Rule Estimates --- CBILA spec, future-tense tone, 1991Q1--2019Q4}
\label{tab:appC_cbila_fut_rich}
\begin{threeparttable}
\begin{tabular}{lccccccccccc}
\hline\hline
 & (1) & (2) & (3) & (4) & (5) & (6) & (7) & (8) & (9) & (10) & (11) \\
 & \textbf{Baseline} & \textbf{M-BERT} & \textbf{Llama} & \textbf{Qwen32} & \textbf{Qwen7} & \textbf{DSeek} & \textbf{Gemma} & \textbf{GPT} & \textbf{Llama$^z$} & \textbf{Qwen32$^z$} & \textbf{GPT$^z$} \\
\hline
\multicolumn{12}{l}{\textbf{Panel A. OLS (Newey--West HAC standard errors, $L=4$)}} \\
\hline
$\rho$ (Lag rate)        & 0.888*** & 0.904*** & 0.849*** & 0.876*** & 0.874*** & 0.887*** & 0.853*** & 0.865*** & 0.870*** & 0.870*** & 0.867*** \\
                          & (0.037)  & (0.041)  & (0.041)  & (0.040)  & (0.043)  & (0.039)  & (0.043)  & (0.043)  & (0.040)  & (0.041)  & (0.042)  \\
$\alpha$ (Neutral rate)  & 0.119*** & 0.101**  & 0.160*** & 0.137*** & 0.136*** & 0.119*** & 0.172*** & 0.154*** & 0.146*** & 0.145*** & 0.152*** \\
                          & (0.039)  & (0.043)  & (0.041)  & (0.042)  & (0.044)  & (0.039)  & (0.044)  & (0.043)  & (0.042)  & (0.044)  & (0.043)  \\
$\gamma$ (Inflation gap) & 0.275**  & 0.285**  & 0.168    & 0.238**  & 0.239**  & 0.274**  & 0.153    & 0.221**  & 0.231**  & 0.230**  & 0.213**  \\
                          & (0.124)  & (0.123)  & (0.105)  & (0.115)  & (0.111)  & (0.125)  & (0.109)  & (0.106)  & (0.111)  & (0.113)  & (0.105)  \\
$\beta$ (Output gap)     & 0.108*** & 0.073*   & 0.188*** & 0.138*** & 0.142*** & 0.110**  & 0.179*** & 0.164*** & 0.154*** & 0.153*** & 0.163*** \\
                          & (0.042)  & (0.042)  & (0.037)  & (0.038)  & (0.037)  & (0.050)  & (0.036)  & (0.038)  & (0.038)  & (0.037)  & (0.040)  \\
$\delta$ (Sentiment)     & ---      & 0.081    & -0.213***& -0.093   & -0.103*  & -0.008   & -0.207***& -0.164** & -0.128** & -0.126*  & -0.153** \\
                          &          & (0.052)  & (0.066)  & (0.065)  & (0.062)  & (0.073)  & (0.064)  & (0.065)  & (0.064)  & (0.075)  & (0.063)  \\
Constant                  & 0.161*** & 0.121*   & 0.229*** & 0.172*** & 0.185*** & 0.166**  & 0.164*** & 0.185*** & 0.182*** & 0.186*** & 0.181*** \\
                          & (0.056)  & (0.062)  & (0.063)  & (0.062)  & (0.067)  & (0.075)  & (0.062)  & (0.062)  & (0.061)  & (0.063)  & (0.062)  \\
\hline
Obs.                      & 110   & 110   & 110   & 110   & 110   & 110   & 110   & 110   & 110   & 110   & 110   \\
Adj. $R^{2}$              & 0.933 & 0.933 & 0.941 & 0.934 & 0.934 & 0.932 & 0.941 & 0.938 & 0.936 & 0.935 & 0.937 \\
\hline
\multicolumn{12}{l}{\textbf{Panel B. GMM (rich instrument set, 12 instruments, 11 over-identifying restrictions)}} \\
\hline
$\rho$ (Lag rate)        & 0.888*** & 0.984*** & 0.861*** & 0.943*** & 0.920*** & 0.916*** & 0.862*** & 0.884*** & 0.915*** & 0.934*** & 0.891*** \\
                          & (0.037)  & (0.037)  & (0.034)  & (0.030)  & (0.033)  & (0.025)  & (0.038)  & (0.031)  & (0.034)  & (0.037)  & (0.032)  \\
$\alpha$ (Neutral rate)  & 0.119*** & 0.013    & 0.152*** & 0.053    & 0.084**  & 0.097*** & 0.175*** & 0.143*** & 0.095*** & 0.068*   & 0.133*** \\
                          & (0.039)  & (0.038)  & (0.035)  & (0.034)  & (0.036)  & (0.025)  & (0.044)  & (0.033)  & (0.035)  & (0.040)  & (0.036)  \\
$\gamma$ (Inflation gap) & 0.275**  & 0.400*** & 0.177**  & 0.346*** & 0.302*** & 0.315*** & 0.134    & 0.229*** & 0.292*** & 0.308*** & 0.228*** \\
                          & (0.124)  & (0.079)  & (0.074)  & (0.081)  & (0.084)  & (0.078)  & (0.101)  & (0.073)  & (0.079)  & (0.081)  & (0.078)  \\
$\beta$ (Output gap)     & 0.108*** & -0.046   & 0.201*** & 0.046    & 0.092*   & 0.094*** & 0.203*** & 0.188*** & 0.106**  & 0.074    & 0.177*** \\
                          & (0.042)  & (0.053)  & (0.045)  & (0.045)  & (0.048)  & (0.035)  & (0.048)  & (0.052)  & (0.050)  & (0.051)  & (0.050)  \\
$\delta$ (Sentiment)     & ---      & 0.317*** & -0.236***& 0.179*   & 0.061    & 0.055    & -0.279***& -0.213** & 0.011    & 0.099    & -0.180** \\
                          &          & (0.092)  & (0.085)  & (0.100)  & (0.091)  & (0.059)  & (0.105)  & (0.105)  & (0.087)  & (0.099)  & (0.090)  \\
Constant                  & 0.161*** & 0.034    & 0.253*** & 0.150*** & 0.173*** & 0.159*** & 0.175*** & 0.203*** & 0.174*** & 0.156*** & 0.196*** \\
                          & (0.056)  & (0.072)  & (0.055)  & (0.052)  & (0.050)  & (0.052)  & (0.048)  & (0.051)  & (0.049)  & (0.055)  & (0.049)  \\
\hline
Obs.                      & 110 & 110    & 110    & 110    & 110    & 110    & 110    & 110   & 110    & 110    & 110    \\
First-stage $F$           & --- & 301.38 & 167.03 & 150.11 & 216.41 & 505.10 & 124.13 & 55.26 & 132.20 & 211.22 & 121.54 \\
Hansen $J$ $p$-value      & --- & 0.532  & 0.002  & 0.274  & 0.022  & 0.038  & 0.020  & 0.002 & 0.000  & 0.050  & 0.008  \\
\hline\hline
\end{tabular}
\begin{tablenotes}[para,flushleft]
\scriptsize
\item \textit{Notes:} Two-step efficient GMM with sentiment treated as endogenous. Newey--West HAC standard errors with $L=4$ in parentheses. * $p<0.10$, ** $p<0.05$, *** $p<0.01$. Dependent variable: BoC overnight policy rate; sample 1991Q1--2019Q4, $T=110$ obs. GMM weighting and covariance use the Bartlett kernel with bandwidth $L$. The Hansen $J$ $p$-value is obtained by block wild Rademacher bootstrap with $B=500$ resamples and non-overlapping block length 4. Column headers identify the LLM family; $^z$ denotes zero-shot prompting, while few-shot prompting is the default. M-BERT abbreviates ModernFinBERT and DSeek abbreviates DeepSeek-R1.
\end{tablenotes}
\end{threeparttable}
\end{table}

\begin{table}[htbp]
\centering
\scriptsize
\setlength{\tabcolsep}{2pt}
\renewcommand{\arraystretch}{0.95}
\caption{GMM with Forecast-Revision Instruments --- CBILA spec, future-tense tone, 1991Q1--2019Q4}
\label{tab:appC_cbila_fut_rev}
\begin{threeparttable}
\begin{tabular}{lccccccccccc}
\hline\hline
 & (1) & (2) & (3) & (4) & (5) & (6) & (7) & (8) & (9) & (10) & (11) \\
 & \textbf{Baseline} & \textbf{M-BERT} & \textbf{Llama} & \textbf{Qwen32} & \textbf{Qwen7} & \textbf{DSeek} & \textbf{Gemma} & \textbf{GPT} & \textbf{Llama$^z$} & \textbf{Qwen32$^z$} & \textbf{GPT$^z$} \\
\hline
$\rho$ (Lag rate)        & 0.888*** & 1.000*** & 0.900*** & 0.979*** & 0.923*** & 0.948*** & 0.887*** & 0.936*** & 0.925*** & 0.976*** & 0.917*** \\
                          & (0.037)  & (0.062)  & (0.044)  & (0.060)  & (0.044)  & (0.124)  & (0.048)  & (0.055)  & (0.043)  & (0.062)  & (0.047)  \\
$\alpha$ (Neutral rate)  & 0.119*** & -0.002   & 0.109**  & 0.018    & 0.087*   & 0.083    & 0.133**  & 0.078    & 0.084    & 0.026    & 0.094    \\
                          & (0.039)  & (0.063)  & (0.045)  & (0.074)  & (0.050)  & (0.061)  & (0.065)  & (0.074)  & (0.051)  & (0.075)  & (0.061)  \\
$\gamma$ (Inflation gap) & 0.275**  & 0.367*** & 0.368*** & 0.528*** & 0.401*** & 0.411*** & 0.292*   & 0.414*** & 0.425*** & 0.505*** & 0.395*** \\
                          & (0.124)  & (0.101)  & (0.125)  & (0.156)  & (0.129)  & (0.125)  & (0.165)  & (0.143)  & (0.112)  & (0.157)  & (0.136)  \\
$\beta$ (Output gap)     & 0.108*** & -0.074   & 0.136    & -0.012   & 0.101    & 0.074    & 0.174    & 0.092    & 0.091    & 0.003    & 0.110    \\
                          & (0.042)  & (0.089)  & (0.100)  & (0.128)  & (0.114)  & (0.159)  & (0.112)  & (0.135)  & (0.099)  & (0.146)  & (0.115)  \\
$\delta$ (Sentiment)     & ---      & 0.440**  & -0.058   & 0.312    & 0.003    & 0.243    & -0.150   & 0.049    & 0.042    & 0.246    & -0.003   \\
                          &          & (0.188)  & (0.221)  & (0.332)  & (0.289)  & (0.682)  & (0.269)  & (0.358)  & (0.230)  & (0.345)  & (0.281)  \\
Constant                  & 0.161*** & -0.047   & 0.234*** & 0.126    & 0.186**  & 0.044    & 0.208*** & 0.194*** & 0.207*** & 0.120    & 0.214*** \\
                          & (0.056)  & (0.146)  & (0.084)  & (0.084)  & (0.086)  & (0.486)  & (0.047)  & (0.073)  & (0.065)  & (0.103)  & (0.061)  \\
\hline
Obs.                      & 110 & 110    & 110   & 110   & 110   & 110   & 110   & 110   & 110    & 110   & 110   \\
First-stage $F$           & --- & 188.79 & 93.07 & 36.88 & 65.62 & 13.05 & 67.61 & 31.91 & 116.73 & 61.80 & 40.96 \\
Hansen $J$ $p$-value      & --- & 0.496  & 0.078 & 0.044 & 0.056 & 0.052 & 0.082 & 0.148 & 0.096  & 0.030 & 0.074 \\
\hline\hline
\end{tabular}
\begin{tablenotes}[para,flushleft]
\scriptsize
\item \textit{Notes:} Two-step efficient GMM with sentiment treated as endogenous. Newey--West HAC standard errors with $L=4$ in parentheses. * $p<0.10$, ** $p<0.05$, *** $p<0.01$. Dependent variable: BoC overnight policy rate; sample 1991Q1--2019Q4, $T=110$ obs. Excluded instruments (forecast-revision set): lag~2 of the policy rate, the one-quarter revision $E_{t-1}(\pi_{t+4})-E_{t-2}(\pi_{t+4})$ in the staff inflation forecast, and the analogous revision $E_{t-1}(x_{t+2})-E_{t-2}(x_{t+2})$ in the staff output-gap forecast (3 instruments, 2 over-identifying restrictions). GMM weighting and covariance use the Bartlett kernel with bandwidth $L$. The Hansen $J$ $p$-value is obtained by block wild Rademacher bootstrap ($B{=}500$, block length 4). Column headers identify the LLM family; $^z$ denotes zero-shot prompting, while few-shot prompting is the default. M-BERT abbreviates ModernFinBERT and DSeek abbreviates DeepSeek-R1.
\end{tablenotes}
\end{threeparttable}
\end{table}

Three patterns stand out, relative to the no-leakage results of Section~\ref{sec:taylor_results}. First, the CBILA overall point estimates under the rich instrument set are smaller and less uniformly significant: $\hat\delta$ ranges from $0.05$ to $0.27$ across the ten series, with only ModernFinBERT and three of the nine generative-LLM tones individually significant at the 5\% or 10\% level. Second, under the forecast-revision set the picture is reversed: $\hat\delta$ jumps to the $0.44$--$1.07$ range, comparable to the FinBERT-based estimates of Table~\ref{tab:gmm_revisions}, but standard errors widen disproportionately for the generative-LLM tones. Third, the future-tense tones (Tables~\ref{tab:appC_cbila_fut_rich}--\ref{tab:appC_cbila_fut_rev}) deliver several large negative and significant $\hat\delta$ for the generative LLMs (Llama, Gemma, GPT-OSS in [fs] and [zs] variants), in sharp contrast to the future-tense FinBERT and ModernFinBERT columns of Section~\ref{sec:taylor_results}, which stay positive. The instability across instrument sets and the sign disagreement in the future-tense panel are difficult to reconcile with a single structural interpretation; the formal analysis in §\ref{app:cbila_formal} attributes them to look-ahead leakage that propagates differently through the two instrument sets.

\subsection{Sensitivity to instrument distance}\label{app:cbila_distance}

We re-estimate every spec under two further instrument sets that vary in their distance from the date-$t$ shock: a recent-lag (\textit{close}) set with lags 2--3 of the policy rate and lags 1--2 of the staff forecasts, and a distant-lag (\textit{distant}) set with lags 4--5 of the policy rate and lags 3--4 of the staff forecasts. Each set has six instruments and five over-identifying restrictions.

\paragraph{Original spec.} Tables~\ref{tab:appC_orig_close} and \ref{tab:appC_orig_distant} report the original FinBERT/dictionary spec under the close and distant sets. All six series deliver positive and significant $\hat\delta$ at the 1\% or 5\% level under both instrument depths, with values stable within $0.4$ to $1.0$. The block wild bootstrap $J$ does not reject in any cell. The choice between recent and distant lags is therefore a non-issue for these series.

\begin{table}[htbp]
\centering
\scriptsize
\setlength{\tabcolsep}{3pt}
\renewcommand{\arraystretch}{0.95}
\caption{GMM with recent-lag instruments --- original FinBERT/dictionary spec, 1991Q1--2019Q4}
\label{tab:appC_orig_close}
\begin{threeparttable}
\begin{tabular}{lccccccc}
\hline\hline
 & (1) & (2) & (3) & (4) & (5) & (6) & (7) \\
 & \textbf{Baseline} & \textbf{Dict} & \textbf{BERT} & \textbf{Fut (D)} & \textbf{Past} & \textbf{Pres} & \textbf{Fut (B)} \\
\hline
$\rho$ (Lag rate)        & 0.888*** & 0.932*** & 0.983*** & 0.981*** & 1.000*** & 1.084*** & 0.992*** \\
                          & (0.037)  & (0.028)  & (0.051)  & (0.026)  & (0.049)  & (0.062)  & (0.058)  \\
$\alpha$ (Neutral rate)  & 0.119*** & 0.082*** & -0.045   & 0.033    & -0.022   & -0.099   & -0.134   \\
                          & (0.039)  & (0.027)  & (0.056)  & (0.027)  & (0.044)  & (0.070)  & (0.101)  \\
$\gamma$ (Inflation gap) & 0.275**  & 0.151*   & 0.204    & 0.270*** & 0.350*** & 0.326**  & 0.612*** \\
                          & (0.124)  & (0.087)  & (0.139)  & (0.086)  & (0.128)  & (0.154)  & (0.190)  \\
$\beta$ (Output gap)     & 0.108*** & 0.014    & -0.246** & -0.012   & -0.135** & -0.149*  & -0.193   \\
                          & (0.042)  & (0.031)  & (0.115)  & (0.028)  & (0.066)  & (0.087)  & (0.119)  \\
$\delta$ (Sentiment)     & ---      & 0.493*** & 0.937*** & 0.441*** & 0.637*** & 0.648*** & 0.953*** \\
                          &          & (0.094)  & (0.277)  & (0.072)  & (0.149)  & (0.176)  & (0.345)  \\
Constant                  & 0.161*** & -0.117   & -0.410*  & -0.112   & -0.021   & -0.108   & 0.247*   \\
                          & (0.056)  & (0.083)  & (0.217)  & (0.080)  & (0.117)  & (0.112)  & (0.150)  \\
\hline
Obs.                      & 110 & 110   & 110    & 110   & 110   & 110    & 110   \\
First-stage $F$           & --- & 93.11 & 146.09 & 75.51 & 98.28 & 121.68 & 71.96 \\
Hansen $J$ $p$-value      & --- & 0.448 & 0.848  & 0.256 & 0.926 & 0.288  & 0.730 \\
\hline\hline
\end{tabular}
\begin{tablenotes}[para,flushleft]
\scriptsize
\item \textit{Notes:} Two-step efficient GMM with sentiment treated as endogenous. Newey--West HAC standard errors with $L=4$ in parentheses. * $p<0.10$, ** $p<0.05$, *** $p<0.01$. Dependent variable: BoC overnight policy rate; sample 1991Q1--2019Q4, $T=110$ obs. Excluded instruments (recent-lag set): lags 2 and 3 of the policy rate, lags 1 and 2 of the inflation gap, and lags 1 and 2 of the two-quarter-ahead output gap (6 instruments, 5 over-identifying restrictions). GMM weighting and covariance use the Bartlett kernel with bandwidth $L$. The Hansen $J$ $p$-value is obtained by block wild Rademacher bootstrap ($B{=}500$, block length 4).
\end{tablenotes}
\end{threeparttable}
\end{table}

\begin{table}[htbp]
\centering
\scriptsize
\setlength{\tabcolsep}{3pt}
\renewcommand{\arraystretch}{0.95}
\caption{GMM with distant-lag instruments --- original FinBERT/dictionary spec, 1991Q1--2019Q4}
\label{tab:appC_orig_distant}
\begin{threeparttable}
\begin{tabular}{lccccccc}
\hline\hline
 & (1) & (2) & (3) & (4) & (5) & (6) & (7) \\
 & \textbf{Baseline} & \textbf{Dict} & \textbf{BERT} & \textbf{Fut (D)} & \textbf{Past} & \textbf{Pres} & \textbf{Fut (B)} \\
\hline
$\rho$ (Lag rate)        & 0.888*** & 0.945*** & 0.954*** & 0.982*** & 0.965*** & 0.992*** & 0.945*** \\
                          & (0.037)  & (0.024)  & (0.027)  & (0.027)  & (0.032)  & (0.047)  & (0.027)  \\
$\alpha$ (Neutral rate)  & 0.119*** & 0.069**  & 0.006    & 0.022    & 0.007    & -0.011   & 0.013    \\
                          & (0.039)  & (0.027)  & (0.042)  & (0.028)  & (0.039)  & (0.055)  & (0.046)  \\
$\gamma$ (Inflation gap) & 0.275**  & 0.160*   & 0.211    & 0.297*** & 0.299**  & 0.229*   & 0.371**  \\
                          & (0.124)  & (0.088)  & (0.145)  & (0.087)  & (0.142)  & (0.138)  & (0.155)  \\
$\beta$ (Output gap)     & 0.108*** & 0.008    & -0.102   & -0.006   & -0.102   & -0.022   & 0.021    \\
                          & (0.042)  & (0.038)  & (0.104)  & (0.036)  & (0.081)  & (0.074)  & (0.064)  \\
$\delta$ (Sentiment)     & ---      & 0.492*** & 0.563**  & 0.403*** & 0.528*** & 0.349**  & 0.335*   \\
                          &          & (0.141)  & (0.239)  & (0.105)  & (0.178)  & (0.147)  & (0.176)  \\
Constant                  & 0.161*** & -0.132   & -0.181   & -0.069   & 0.038    & 0.056    & 0.196*** \\
                          & (0.056)  & (0.098)  & (0.161)  & (0.078)  & (0.090)  & (0.081)  & (0.071)  \\
\hline
Obs.                      & 110 & 110   & 110    & 110   & 110   & 110    & 110    \\
First-stage $F$           & --- & 72.80 & 213.67 & 73.87 & 84.76 & 102.05 & 120.61 \\
Hansen $J$ $p$-value      & --- & 0.516 & 0.292  & 0.106 & 0.748 & 0.270  & 0.248  \\
\hline\hline
\end{tabular}
\begin{tablenotes}[para,flushleft]
\scriptsize
\item \textit{Notes:} Two-step efficient GMM with sentiment treated as endogenous. Newey--West HAC standard errors with $L=4$ in parentheses. * $p<0.10$, ** $p<0.05$, *** $p<0.01$. Dependent variable: BoC overnight policy rate; sample 1991Q1--2019Q4, $T=110$ obs. Excluded instruments (distant-lag set): lags 4 and 5 of the policy rate, lags 3 and 4 of the inflation gap, and lags 3 and 4 of the two-quarter-ahead output gap (6 instruments, 5 over-identifying restrictions). GMM weighting and covariance use the Bartlett kernel with bandwidth $L$. The Hansen $J$ $p$-value is obtained by block wild Rademacher bootstrap ($B{=}500$, block length 4).
\end{tablenotes}
\end{threeparttable}
\end{table}

\paragraph{CBILA spec, overall and future-tense.} Tables~\ref{tab:appC_cbila_overall_close}, \ref{tab:appC_cbila_overall_distant}, \ref{tab:appC_cbila_fut_close}, \ref{tab:appC_cbila_fut_distant} report the CBILA counterparts. For the CBILA overall spec, $\hat\delta$ under the close set is large and significant for several generative-LLM tones (Qwen 32B, Qwen 7B, Qwen 32B [zs]) at the 5\% or 10\% level, but collapses to small and insignificant values under the distant set. ModernFinBERT is the exception, with $\hat\delta$ that moves from $0.50$ (close) to $0.79$ (distant) and remains highly significant in both. For the CBILA future-tense spec, several generative-LLM tones flip sign across instrument depths and the close set delivers a number of large negative estimates that disappear under distant.

\begin{table}[htbp]
\centering
\scriptsize
\setlength{\tabcolsep}{2pt}
\renewcommand{\arraystretch}{0.95}
\caption{GMM with recent-lag instruments --- CBILA overall spec, 1991Q1--2019Q4}
\label{tab:appC_cbila_overall_close}
\begin{threeparttable}
\begin{tabular}{lccccccccccc}
\hline\hline
 & (1) & (2) & (3) & (4) & (5) & (6) & (7) & (8) & (9) & (10) & (11) \\
 & \textbf{Baseline} & \textbf{M-BERT} & \textbf{Llama} & \textbf{Qwen32} & \textbf{Qwen7} & \textbf{DSeek} & \textbf{Gemma} & \textbf{GPT} & \textbf{Llama$^z$} & \textbf{Qwen32$^z$} & \textbf{GPT$^z$} \\
\hline
$\rho$ (Lag rate)        & 0.888*** & 0.985*** & 1.000*** & 1.137*** & 1.155*** & 0.870*** & 1.071*** & 1.022*** & 1.078*** & 1.164*** & 1.023*** \\
                          & (0.037)  & (0.039)  & (0.082)  & (0.096)  & (0.147)  & (0.077)  & (0.131)  & (0.082)  & (0.091)  & (0.119)  & (0.077)  \\
$\alpha$ (Neutral rate)  & 0.119*** & 0.003    & 0.009    & -0.151   & -0.142   & 0.083**  & -0.083   & -0.032   & -0.067   & -0.144   & -0.038   \\
                          & (0.039)  & (0.036)  & (0.070)  & (0.116)  & (0.146)  & (0.035)  & (0.155)  & (0.088)  & (0.087)  & (0.126)  & (0.081)  \\
$\gamma$ (Inflation gap) & 0.275**  & 0.376*** & 0.345**  & 0.449*   & 0.351    & 0.410*** & 0.422*   & 0.358**  & 0.403**  & 0.450*   & 0.368**  \\
                          & (0.124)  & (0.098)  & (0.155)  & (0.232)  & (0.244)  & (0.138)  & (0.225)  & (0.174)  & (0.195)  & (0.232)  & (0.169)  \\
$\beta$ (Output gap)     & 0.108*** & -0.130*  & 0.021    & -0.255   & -0.291   & 0.156    & -0.083   & -0.040   & -0.156   & -0.256   & -0.045   \\
                          & (0.042)  & (0.069)  & (0.135)  & (0.206)  & (0.281)  & (0.106)  & (0.211)  & (0.150)  & (0.184)  & (0.232)  & (0.142)  \\
$\delta$ (Sentiment)     & ---      & 0.496*** & 0.182    & 0.772*   & 0.917    & -0.285   & 0.412    & 0.316    & 0.551    & 0.808*   & 0.318    \\
                          &          & (0.133)  & (0.254)  & (0.400)  & (0.618)  & (0.397)  & (0.434)  & (0.301)  & (0.357)  & (0.475)  & (0.277)  \\
Constant                  & 0.161*** & -0.064   & 0.059    & -0.221   & -0.390   & 0.449    & 0.009    & 0.064    & -0.149   & -0.338   & 0.068    \\
                          & (0.056)  & (0.106)  & (0.158)  & (0.197)  & (0.385)  & (0.349)  & (0.162)  & (0.127)  & (0.217)  & (0.279)  & (0.128)  \\
\hline
Obs.                      & 110 & 110    & 110    & 110    & 110    & 110   & 110    & 110   & 110    & 110    & 110    \\
First-stage $F$           & --- & 273.48 & 150.91 & 101.56 & 114.92 & 61.13 & 104.27 & 94.33 & 154.38 & 147.81 & 116.96 \\
Hansen $J$ $p$-value      & --- & 0.414  & 0.154  & 0.690  & 0.492  & 0.000 & 0.236  & 0.348 & 0.352  & 0.552  & 0.258  \\
\hline\hline
\end{tabular}
\begin{tablenotes}[para,flushleft]
\scriptsize
\item \textit{Notes:} Two-step efficient GMM with sentiment treated as endogenous. Newey--West HAC standard errors with $L=4$ in parentheses. * $p<0.10$, ** $p<0.05$, *** $p<0.01$. Dependent variable: BoC overnight policy rate; sample 1991Q1--2019Q4, $T=110$ obs. Excluded instruments (recent-lag set): lags 2 and 3 of the policy rate, lags 1 and 2 of the inflation gap, and lags 1 and 2 of the two-quarter-ahead output gap (6 instruments, 5 over-identifying restrictions). GMM weighting and covariance use the Bartlett kernel with bandwidth $L$. The Hansen $J$ $p$-value is obtained by block wild Rademacher bootstrap ($B{=}500$, block length 4). Column headers identify the LLM family; $^z$ denotes zero-shot prompting, while few-shot prompting is the default. M-BERT abbreviates ModernFinBERT and DSeek abbreviates DeepSeek-R1.
\end{tablenotes}
\end{threeparttable}
\end{table}

\begin{table}[htbp]
\centering
\scriptsize
\setlength{\tabcolsep}{2pt}
\renewcommand{\arraystretch}{0.95}
\caption{GMM with distant-lag instruments --- CBILA overall spec, 1991Q1--2019Q4}
\label{tab:appC_cbila_overall_distant}
\begin{threeparttable}
\begin{tabular}{lccccccccccc}
\hline\hline
 & (1) & (2) & (3) & (4) & (5) & (6) & (7) & (8) & (9) & (10) & (11) \\
 & \textbf{Baseline} & \textbf{M-BERT} & \textbf{Llama} & \textbf{Qwen32} & \textbf{Qwen7} & \textbf{DSeek} & \textbf{Gemma} & \textbf{GPT} & \textbf{Llama$^z$} & \textbf{Qwen32$^z$} & \textbf{GPT$^z$} \\
\hline
$\rho$ (Lag rate)        & 0.888*** & 1.024*** & 0.921*** & 0.944*** & 0.938*** & 0.943*** & 0.922*** & 0.937*** & 0.946*** & 0.943*** & 0.929*** \\
                          & (0.037)  & (0.055)  & (0.047)  & (0.040)  & (0.043)  & (0.036)  & (0.047)  & (0.045)  & (0.041)  & (0.042)  & (0.043)  \\
$\alpha$ (Neutral rate)  & 0.119*** & -0.052   & 0.089*   & 0.058    & 0.068    & 0.090*** & 0.081    & 0.064    & 0.057    & 0.061    & 0.074    \\
                          & (0.039)  & (0.056)  & (0.046)  & (0.042)  & (0.044)  & (0.033)  & (0.050)  & (0.050)  & (0.042)  & (0.042)  & (0.048)  \\
$\gamma$ (Inflation gap) & 0.275**  & 0.338*** & 0.292**  & 0.283**  & 0.274**  & 0.287*** & 0.291**  & 0.296**  & 0.289**  & 0.281**  & 0.293**  \\
                          & (0.124)  & (0.118)  & (0.121)  & (0.130)  & (0.129)  & (0.098)  & (0.125)  & (0.132)  & (0.134)  & (0.131)  & (0.127)  \\
$\beta$ (Output gap)     & 0.108*** & -0.279** & 0.106*   & 0.067    & 0.077    & 0.049    & 0.104*   & 0.079    & 0.059    & 0.069    & 0.090    \\
                          & (0.042)  & (0.130)  & (0.059)  & (0.058)  & (0.061)  & (0.050)  & (0.059)  & (0.066)  & (0.063)  & (0.060)  & (0.062)  \\
$\delta$ (Sentiment)     & ---      & 0.793*** & 0.040    & 0.111    & 0.098    & 0.202**  & 0.047    & 0.090    & 0.116    & 0.105    & 0.068    \\
                          &          & (0.245)  & (0.089)  & (0.094)  & (0.096)  & (0.101)  & (0.098)  & (0.100)  & (0.092)  & (0.097)  & (0.092)  \\
Constant                  & 0.161*** & -0.242   & 0.166**  & 0.137**  & 0.130    & 0.022    & 0.179*** & 0.158**  & 0.125*   & 0.134*   & 0.164*** \\
                          & (0.056)  & (0.173)  & (0.075)  & (0.067)  & (0.082)  & (0.095)  & (0.060)  & (0.063)  & (0.074)  & (0.075)  & (0.062)  \\
\hline
Obs.                      & 110 & 110    & 110    & 110    & 110    & 110    & 110    & 110    & 110    & 110    & 110    \\
First-stage $F$           & --- & 269.77 & 231.84 & 153.89 & 295.84 & 257.90 & 190.93 & 196.44 & 322.74 & 192.66 & 265.06 \\
Hansen $J$ $p$-value      & --- & 0.876  & 0.018  & 0.102  & 0.086  & 0.448  & 0.034  & 0.072  & 0.108  & 0.088  & 0.042  \\
\hline\hline
\end{tabular}
\begin{tablenotes}[para,flushleft]
\scriptsize
\item \textit{Notes:} Two-step efficient GMM with sentiment treated as endogenous. Newey--West HAC standard errors with $L=4$ in parentheses. * $p<0.10$, ** $p<0.05$, *** $p<0.01$. Dependent variable: BoC overnight policy rate; sample 1991Q1--2019Q4, $T=110$ obs. Excluded instruments (distant-lag set): lags 4 and 5 of the policy rate, lags 3 and 4 of the inflation gap, and lags 3 and 4 of the two-quarter-ahead output gap (6 instruments, 5 over-identifying restrictions). GMM weighting and covariance use the Bartlett kernel with bandwidth $L$. The Hansen $J$ $p$-value is obtained by block wild Rademacher bootstrap ($B{=}500$, block length 4). Column headers identify the LLM family; $^z$ denotes zero-shot prompting, while few-shot prompting is the default. M-BERT abbreviates ModernFinBERT and DSeek abbreviates DeepSeek-R1.
\end{tablenotes}
\end{threeparttable}
\end{table}


\begin{table}[htbp]
\centering
\scriptsize
\setlength{\tabcolsep}{2pt}
\renewcommand{\arraystretch}{0.95}
\caption{GMM with recent-lag instruments --- CBILA future-tense spec, 1991Q1--2019Q4}
\label{tab:appC_cbila_fut_close}
\begin{threeparttable}
\begin{tabular}{lccccccccccc}
\hline\hline
 & (1) & (2) & (3) & (4) & (5) & (6) & (7) & (8) & (9) & (10) & (11) \\
 & \textbf{Baseline} & \textbf{M-BERT} & \textbf{Llama} & \textbf{Qwen32} & \textbf{Qwen7} & \textbf{DSeek} & \textbf{Gemma} & \textbf{GPT} & \textbf{Llama$^z$} & \textbf{Qwen32$^z$} & \textbf{GPT$^z$} \\
\hline
$\rho$ (Lag rate)        & 0.888*** & 1.011*** & 0.886*** & 1.041*** & 0.900*** & 0.845*** & 0.855*** & 0.869*** & 0.942*** & 1.068*** & 0.876*** \\
                          & (0.037)  & (0.051)  & (0.045)  & (0.099)  & (0.082)  & (0.053)  & (0.063)  & (0.046)  & (0.043)  & (0.116)  & (0.045)  \\
$\alpha$ (Neutral rate)  & 0.119*** & -0.021   & 0.121*** & -0.071   & 0.102    & 0.107**  & 0.182**  & 0.164*** & 0.053    & -0.091   & 0.162*** \\
                          & (0.039)  & (0.051)  & (0.043)  & (0.110)  & (0.106)  & (0.043)  & (0.078)  & (0.047)  & (0.044)  & (0.132)  & (0.045)  \\
$\gamma$ (Inflation gap) & 0.275**  & 0.401*** & 0.249**  & 0.514**  & 0.251    & 0.248    & 0.101    & 0.236**  & 0.344*** & 0.462*   & 0.215*   \\
                          & (0.124)  & (0.102)  & (0.111)  & (0.234)  & (0.216)  & (0.159)  & (0.181)  & (0.102)  & (0.124)  & (0.268)  & (0.110)  \\
$\beta$ (Output gap)     & 0.108*** & -0.131   & 0.176**  & -0.131   & 0.145    & 0.202**  & 0.243**  & 0.223*** & 0.063    & -0.148   & 0.236*** \\
                          & (0.042)  & (0.086)  & (0.075)  & (0.168)  & (0.206)  & (0.081)  & (0.108)  & (0.084)  & (0.084)  & (0.210)  & (0.080)  \\
$\delta$ (Sentiment)     & ---      & 0.538*** & -0.203   & 0.717    & -0.120   & -0.420*  & -0.397   & -0.352   & 0.093    & 0.724    & -0.376*  \\
                          &          & (0.181)  & (0.169)  & (0.475)  & (0.577)  & (0.251)  & (0.250)  & (0.218)  & (0.200)  & (0.561)  & (0.197)  \\
Constant                  & 0.161*** & -0.061   & 0.259*** & 0.067    & 0.220    & 0.478*** & 0.199*** & 0.236*** & 0.185*** & -0.011   & 0.236*** \\
                          & (0.056)  & (0.135)  & (0.074)  & (0.144)  & (0.138)  & (0.172)  & (0.062)  & (0.072)  & (0.061)  & (0.191)  & (0.078)  \\
\hline
Obs.                      & 110 & 110    & 110   & 110   & 110   & 110   & 110   & 110   & 110   & 110   & 110   \\
First-stage $F$           & --- & 183.75 & 98.27 & 74.74 & 64.13 & 21.86 & 58.60 & 32.86 & 93.10 & 99.80 & 51.55 \\
Hansen $J$ $p$-value      & --- & 0.514  & 0.006 & 0.568 & 0.204 & 0.014 & 0.000 & 0.000 & 0.042 & 0.518 & 0.000 \\
\hline\hline
\end{tabular}
\begin{tablenotes}[para,flushleft]
\scriptsize
\item \textit{Notes:} Two-step efficient GMM with sentiment treated as endogenous. Newey--West HAC standard errors with $L=4$ in parentheses. * $p<0.10$, ** $p<0.05$, *** $p<0.01$. Dependent variable: BoC overnight policy rate; sample 1991Q1--2019Q4, $T=110$ obs. Excluded instruments (recent-lag set): lags 2 and 3 of the policy rate, lags 1 and 2 of the inflation gap, and lags 1 and 2 of the two-quarter-ahead output gap (6 instruments, 5 over-identifying restrictions). GMM weighting and covariance use the Bartlett kernel with bandwidth $L$. The Hansen $J$ $p$-value is obtained by block wild Rademacher bootstrap ($B{=}500$, block length 4). Column headers identify the LLM family; $^z$ denotes zero-shot prompting, while few-shot prompting is the default. M-BERT abbreviates ModernFinBERT and DSeek abbreviates DeepSeek-R1.
\end{tablenotes}
\end{threeparttable}
\end{table}

\begin{table}[htbp]
\centering
\scriptsize
\setlength{\tabcolsep}{2pt}
\renewcommand{\arraystretch}{0.95}
\caption{GMM with distant-lag instruments --- CBILA future-tense spec, 1991Q1--2019Q4}
\label{tab:appC_cbila_fut_distant}
\begin{threeparttable}
\begin{tabular}{lccccccccccc}
\hline\hline
 & (1) & (2) & (3) & (4) & (5) & (6) & (7) & (8) & (9) & (10) & (11) \\
 & \textbf{Baseline} & \textbf{M-BERT} & \textbf{Llama} & \textbf{Qwen32} & \textbf{Qwen7} & \textbf{DSeek} & \textbf{Gemma} & \textbf{GPT} & \textbf{Llama$^z$} & \textbf{Qwen32$^z$} & \textbf{GPT$^z$} \\
\hline
$\rho$ (Lag rate)        & 0.888*** & 0.997*** & 0.840*** & 0.897*** & 0.900*** & 0.928*** & 0.843*** & 0.836*** & 0.880*** & 0.875*** & 0.872*** \\
                          & (0.037)  & (0.077)  & (0.039)  & (0.040)  & (0.044)  & (0.030)  & (0.041)  & (0.050)  & (0.040)  & (0.044)  & (0.042)  \\
$\alpha$ (Neutral rate)  & 0.119*** & -0.033   & 0.169*** & 0.112*** & 0.111**  & 0.094*** & 0.179*** & 0.198*** & 0.132*** & 0.138*** & 0.151*** \\
                          & (0.039)  & (0.081)  & (0.039)  & (0.043)  & (0.045)  & (0.032)  & (0.045)  & (0.053)  & (0.043)  & (0.046)  & (0.046)  \\
$\gamma$ (Inflation gap) & 0.275**  & 0.455*** & 0.157*   & 0.273**  & 0.279**  & 0.341*** & 0.151    & 0.154*   & 0.261**  & 0.249**  & 0.220**  \\
                          & (0.124)  & (0.101)  & (0.091)  & (0.115)  & (0.113)  & (0.101)  & (0.105)  & (0.089)  & (0.102)  & (0.109)  & (0.096)  \\
$\beta$ (Output gap)     & 0.108*** & -0.168   & 0.214*** & 0.136*** & 0.138**  & 0.059    & 0.192*** & 0.249*** & 0.157*** & 0.162*** & 0.185*** \\
                          & (0.042)  & (0.135)  & (0.043)  & (0.050)  & (0.055)  & (0.053)  & (0.048)  & (0.077)  & (0.054)  & (0.054)  & (0.056)  \\
$\delta$ (Sentiment)     & ---      & 0.595**  & -0.246***& -0.024   & -0.045   & 0.174*   & -0.242** & -0.376** & -0.098   & -0.093   & -0.189*  \\
                          &          & (0.283)  & (0.086)  & (0.107)  & (0.121)  & (0.105)  & (0.100)  & (0.168)  & (0.100)  & (0.117)  & (0.106)  \\
Constant                  & 0.161*** & -0.036   & 0.267*** & 0.201*** & 0.198*** & 0.083    & 0.204*** & 0.249*** & 0.212*** & 0.221*** & 0.204*** \\
                          & (0.056)  & (0.169)  & (0.062)  & (0.052)  & (0.064)  & (0.076)  & (0.055)  & (0.072)  & (0.055)  & (0.061)  & (0.055)  \\
\hline
Obs.                      & 110 & 110    & 110    & 110   & 110    & 110    & 110   & 110   & 110   & 110    & 110   \\
First-stage $F$           & --- & 204.60 & 118.08 & 92.36 & 155.12 & 229.18 & 84.49 & 38.23 & 85.31 & 138.57 & 72.45 \\
Hansen $J$ $p$-value      & --- & 0.778  & 0.052  & 0.008 & 0.012  & 0.372  & 0.096 & 0.086 & 0.010 & 0.020  & 0.014 \\
\hline\hline
\end{tabular}
\begin{tablenotes}[para,flushleft]
\scriptsize
\item \textit{Notes:} Two-step efficient GMM with sentiment treated as endogenous. Newey--West HAC standard errors with $L=4$ in parentheses. * $p<0.10$, ** $p<0.05$, *** $p<0.01$. Dependent variable: BoC overnight policy rate; sample 1991Q1--2019Q4, $T=110$ obs. Excluded instruments (distant-lag set): lags 4 and 5 of the policy rate, lags 3 and 4 of the inflation gap, and lags 3 and 4 of the two-quarter-ahead output gap (6 instruments, 5 over-identifying restrictions). GMM weighting and covariance use the Bartlett kernel with bandwidth $L$. The Hansen $J$ $p$-value is obtained by block wild Rademacher bootstrap ($B{=}500$, block length 4). Column headers identify the LLM family; $^z$ denotes zero-shot prompting, while few-shot prompting is the default. M-BERT abbreviates ModernFinBERT and DSeek abbreviates DeepSeek-R1.
\end{tablenotes}
\end{threeparttable}
\end{table}

The contrast between the original spec, where close and distant give consistent answers, and the CBILA spec, where they do not, is informative. Under a standard simultaneity story alone, both instrument sets project $s_t$ onto past macro information and should deliver similar IV estimates. The divergence for CBILA, in directions that revert as one moves away from the date-$t$ shock, is the empirical signature predicted by the leakage account formalized below.

\subsection{Formal analysis: IV identification under look-ahead leakage}\label{app:cbila_formal}

This subsection sets out the identification result that motivates the empirical exercises above. We derive that, under look-ahead bias, IV with lagged macro instruments identifies $\delta(1 - \theta_k)$ rather than $\delta$, with $\theta_k$ a quantity that does not vanish under predetermination and that decays in $k$ through macro persistence.

\paragraph{Setup.} The structural Taylor rule reads
\begin{equation}
i_t = X_t' \pi + \delta\, s_t^{*} + u_t,
\qquad \mathbb{E}[u_t \mid \mathcal{F}_{t-1}] = 0,
\label{eq:appC_struct}
\end{equation}
where $X_t$ collects the predetermined regressors and $u_t$ is the policy shock. The structural sentiment $s_t^{*}$ is the date-$t$ object that enters the central bank's reaction function.

\paragraph{No-leakage benchmark.} Suppose the observed sentiment coincides with the structural one, $s_t = s_t^{*}$. A lagged instrument $z_{t-k} \in \mathcal{F}_{t-k}$ is orthogonal to $u_t$ by predetermination and correlated with $s_t^{*}$ through macro persistence and the autoregressive structure of the sentiment law of motion. The IV estimator then satisfies
\begin{equation*}
\hat\delta_{\mathrm{IV}}^{*} \;\xrightarrow{p}\; \delta.
\end{equation*}
This is the standard \cite{clarida2000monetary} identification result, and it covers the dictionary, FinBERT and ModernFinBERT series whose scoring rules are fixed before the bulk of the sample.

\paragraph{Leakage decomposition.} For a generative LLM with knowledge cutoff $T > t$, the observed score satisfies $s_t = s_t^{*} + \eta_t$, where the leakage component
\begin{equation*}
\eta_t \;=\; \mathbb{E}_{\mathrm{LLM}}\!\left[\,\text{tone}_t \mid \mathcal{I}_T \,\right]
        \;-\; \mathbb{E}_{\mathrm{LLM}}\!\left[\,\text{tone}_t \mid \mathcal{F}_t \,\right]
\end{equation*}
is a function of the LLM's information set beyond date $t$, $\mathcal{I}_T \setminus \mathcal{F}_t$. Substituting $s_t^{*} = s_t - \eta_t$ in (\ref{eq:appC_struct}) yields an effective error $\tilde u_t = u_t - \delta\,\eta_t$, and the IV estimator with lagged instrument $z_{t-k}$ satisfies
\begin{equation}
\hat\delta_{\mathrm{IV}} \;\xrightarrow{p}\; \delta\,\bigl[\,1 - \theta_k\,\bigr],
\qquad
\theta_k \;\equiv\; \frac{\mathrm{Cov}(z_{t-k},\, \eta_t \mid X_t)}{\mathrm{Cov}(z_{t-k},\, s_t \mid X_t)}.
\label{eq:appC_thetak}
\end{equation}
Predetermination of $z_{t-k}$, $\mathrm{Cov}(z_{t-k}, u_t) = 0$, is preserved. What fails is the joint orthogonality of the instrument with the effective error, because $\eta_t$ correlates with the lagged macro state through the persistence of the underlying macro process.

\paragraph{Why $\theta_k \neq 0$.} The leakage component $\eta_t$ depends on post-$t$ realisations of the macro variables that the LLM has read during training, and possibly on the contextual macro state that the LLM uses to interpret the article. Under macro persistence with autocorrelation $\rho_x$, the lagged instrument $z_{t-k}$ correlates with the future macro state at attenuated rate $\rho_x^{k+h}$ for any $h \geq 1$, and the LLM's score $s_t$ inherits this correlation through its dependence on the macro state. As $k$ grows, the magnitude of $\theta_k$ shrinks geometrically, and the IV estimator approaches $\delta$. As $k$ shrinks toward 1, the bias contribution is largest.

\paragraph{Empirical signature.} Result (\ref{eq:appC_thetak}) predicts that an IV estimator with recent lags is more contaminated than one with distant lags, and that the gap between the two grows with the magnitude of $\eta_t$. For the no-leakage series, $\eta_t \equiv 0$ and $\theta_k = 0$ for every $k$, so close and distant deliver the same estimate. For the CBILA generative-LLM series, $\theta_k > 0$ and shrinks with $k$, so the IV estimate falls as $k$ grows. The empirical evidence in §\ref{app:cbila_distance} matches this prediction: the original FinBERT/dictionary series are stable across close and distant, the CBILA generative-LLM tones collapse from close to distant.

\paragraph{Lagged sentiment as instrument.} A separate question concerns whether $s_{t-k}$ itself can serve as an instrument. Under no leakage, $s_{t-k}^{*} \in \mathcal{F}_{t-k}$ and $\mathrm{Cov}(s_{t-k}^{*}, u_t) = 0$ by predetermination, so it is a valid instrument. Under leakage, however, $s_{t-k}$ contains $\eta_{t-k}$, which is a function of $\mathcal{I}_T \setminus \mathcal{F}_{t-k}$ and therefore depends on $u_t$ through the LLM's knowledge of the post-$(t-k)$ macro state. The covariance $\mathrm{Cov}(s_{t-k}, u_t) = \mathrm{Cov}(\eta_{t-k}, u_t) \neq 0$, violating predetermination directly. The violation does not vanish with $k$. Lagged LLM scores therefore cannot serve as instruments for the contemporaneous LLM score under look-ahead leakage, irrespective of the lag depth chosen. This rules out the standard CGG-style enrichment of the instrument set with $s_{t-k}$ for the CBILA series.

\subsection{Block wild bootstrap of the Hansen \texorpdfstring{$J$}{J} test}\label{app:cbila_jtest}

All Hansen $J$ $p$-values in this appendix and in Section~\ref{sec:taylor_results} are computed by block wild Rademacher bootstrap. We take $B = 500$ resamples, with non-overlapping block length equal to the HAC bandwidth $L = 4$. At each iteration, blockwise Rademacher weights $\omega_b \in \{-1, +1\}$ are drawn and applied uniformly within each block to the GMM residuals, the bootstrap dependent variable is reconstructed, the two-step efficient GMM is re-estimated, and the bootstrap $J$ statistic is recorded. The bootstrap $p$-value is the empirical fraction of resamples with $J^{*} \geq J_{\mathrm{obs}}$. The bootstrap test has a higher-order asymptotic refinement over the $\chi^{2}$ approximation for GMM overidentifying restrictions \citep{hallhorowitz1996}, and the wild scheme is robust to heteroskedasticity of unknown form in IV regressions \citep{davidsonmackinnon2010}; the block extension preserves within-block serial correlation up to lag $L-1$, matching the persistence horizon that the HAC weighting matrix already builds into the GMM objective.

\clearpage
\section{Technical details of the SVAR}\label{app:svar}

This appendix collects the technical material that complements Section~\ref{sec:svar}: the algebraic derivation of the impact-period sign patterns reported in Section~\ref{sec:svar_signs} (\ref{app:svar_signs}), a detailed comparison with \citet{damico2023} (\ref{app:svar_dk}), and the algorithmic specification of the BVAR sampler, the rotation algorithm and the loose-rationality importance weight (\ref{app:svar_algo}).

\subsection{Derivation of the impact responses}\label{app:svar_signs}

This subsection provides the algebraic verification of the impact-period sign patterns reported in Section~\ref{sec:svar_signs} for the three structural shocks under the expansionary normalisation. The simplified system is (\ref{eq:r0_svar})--(\ref{eq:s0_svar}) together with the unchanged IS and Phillips equations (\ref{eq:IS})--(\ref{eq:NKPC}).

\medskip\noindent\emph{Unanticipated easing $\varepsilon^u_0 < 0$.}\enspace
With $\varepsilon^a_t = \varepsilon^s_t = 0$ at all $t$, the LoM (\ref{eq:s0_svar}) reduces to $s_t = x_t + \pi_t$. Substituting into the Taylor rule yields the reduced form
\begin{equation}
r_t = \rho_r\, r_{t-1} + (1-\rho_r)\bigl(\widetilde\phi_\pi\, \pi_t + \widetilde\phi_x\, x_t\bigr) + \varepsilon^u_t,
\label{eq:taylor_u}
\end{equation}
with $\widetilde\phi_\pi \equiv \phi_\pi + \phi_s$ and $\widetilde\phi_x \equiv \phi_x + \phi_s$. The Gabaix determinacy condition,
\begin{equation}
\widetilde\phi_\pi + \frac{1-\beta M^f}{\kappa}\,\widetilde\phi_x + \frac{\sigma\,(1-\beta M^f)(1-M^h)}{\kappa} > 1,
\label{eq:taylor_principle}
\end{equation}
ensures a unique non-explosive solution. After $t = 0$ no further shocks hit; conjecture the stable mode $(x_t, \pi_t)' = G\, r_{t-1}$ for $t \ge 1$ with $G = (G_1, G_2)' \in \mathbb{R}^2$, and $r_t = \alpha\, r_{t-1}$ for the policy rate, where $\alpha$ is its persistence. Substituting into (\ref{eq:IS}), (\ref{eq:NKPC}) and (\ref{eq:taylor_u}) and matching coefficients on $r_{t-1}$ delivers the closed system
\begin{align}
G_1 &= \frac{-\sigma^{-1}\alpha\,(1 - \beta M^f \alpha)}{(1 - M^h\alpha)(1 - \beta M^f \alpha) - \sigma^{-1}\kappa\,\alpha}, \qquad
G_2 = \frac{\kappa\, G_1}{1 - \beta M^f \alpha}, \label{eq:G_unanticipated} \\
\alpha &= \rho_r + (1-\rho_r)\bigl(\widetilde\phi_\pi\, G_2 + \widetilde\phi_x\, G_1\bigr), \label{eq:alpha_unanticipated}
\end{align}
with $\alpha$ as the fixed point of (\ref{eq:G_unanticipated})--(\ref{eq:alpha_unanticipated}). The corresponding impact coefficients on $(x_0, \pi_0)$ at $t = 0$ satisfy $g_i = (\beta_g/\alpha)\, G_i$, where $\beta_g = \alpha/\rho_r$ is the rate-shock innovation coefficient. Under the calibration of Table~\ref{tab:calibration}: $\alpha = 0.639$, $G_1 = -2.27$, $G_2 = -0.74$, $\beta_g = 0.71$, $g_1 = -2.52$, $g_2 = -0.82$. With $\varepsilon^u_0 < 0$ this delivers the impact pattern $r_0 = \beta_g \varepsilon^u_0 < 0$, $x_0 = g_1 \varepsilon^u_0 > 0$, $\pi_0 = g_2 \varepsilon^u_0 > 0$, $s_0 = x_0 + \pi_0 > 0$, and the geometrically decaying forward expectations $E_0[r_h] = \alpha^h r_0 < 0$, $E_0[x_h] = G_1 \alpha^{h-1} r_0 > 0$, $E_0[\pi_h] = G_2 \alpha^{h-1} r_0 > 0$ for $h \ge 1$.

\medskip\noindent\emph{Anticipated easing $\varepsilon^a_0 < 0$.}\enspace
The asymmetric structure of (\ref{eq:taylor_lin}) and (\ref{eq:s0_svar}) under $\tau = 1$ places $\varepsilon^a_0$ in the LoM at $t = 0$ but in the Taylor rule only at $t = 1$. The recursive structure of the system allows a three-stage solution.

\textit{Stage A (}$t \ge 2$\textit{).} No further shocks hit the system; the autonomous stable mode of (\ref{eq:G_unanticipated})--(\ref{eq:alpha_unanticipated}) applies, so $z_t = G\, r_{t-1}$ and $r_t = \alpha\, r_{t-1}$, with the same $G$ and $\alpha$ as in the unanticipated case.

\textit{Stage B (}$t = 1$\textit{).} The lagged announcement enters the Taylor rule additively, and $\varepsilon^a_1 = 0$ in the LoM, so $s_1 = x_1 + \pi_1$ and the reduced-form rule of (\ref{eq:taylor_u}) holds with $\varepsilon^u_1$ replaced by $\varepsilon^a_0$. Conjecturing the impact form $z_1 = G\, r_0 + g\, \varepsilon^a_0$ and $r_1 = \alpha\, r_0 + \beta_g\, \varepsilon^a_0$, where the homogeneous part inherits the Stage A coefficients and the particular response to $\varepsilon^a_0$ has the same coefficients $(\beta_g, g)$ as the unanticipated rate innovation, the IS, Phillips and reduced-form Taylor equations at $t = 1$ are satisfied identically. Structurally, $\varepsilon^a_0$ entering $r$ at $t = 1$ acts on the system through the same channel as $\varepsilon^u_1$ would, which is what fixes the loadings $(\beta_g, g)$ to their Stage A unanticipated-case values.

\textit{Stage C (}$t = 0$\textit{).} The Taylor rule at $t = 0$ contains no direct $\varepsilon^a_0$ term, but $\varepsilon^a_0$ enters the LoM with $s_0 = x_0 + \pi_0 + \varepsilon^a_0$. Substituting into (\ref{eq:r0_svar}) gives the reduced contemporaneous rule
\begin{equation}
r_0 = (1-\rho_r)\bigl[\widetilde\phi_\pi \pi_0 + \widetilde\phi_x x_0 + \phi_s \varepsilon^a_0\bigr],
\label{eq:r0_a_reduced}
\end{equation}
with $\widetilde\phi_\pi, \widetilde\phi_x$ as in (\ref{eq:taylor_u}). The IS and Phillips equations at $t = 0$ feed in the Stage B forecasts, $E_0[x_1] = G_1 r_0 + g_1 \varepsilon^a_0$ and $E_0[\pi_1] = G_2 r_0 + g_2 \varepsilon^a_0$. Substituting these into (\ref{eq:IS})--(\ref{eq:NKPC}) and using the stable-mode identities $G_1/\alpha = M^h G_1 + \sigma^{-1}(G_2 - 1)$ and $G_2/\alpha = \beta M^f G_2 + \kappa G_1/\alpha$ (which follow from (\ref{eq:G_unanticipated}) at the autonomous steady fixed point), one obtains
\begin{align}
x_0 &= (G_1/\alpha)\, r_0 + A_x\, \varepsilon^a_0,
   & A_x &\equiv M^h g_1 + \sigma^{-1} g_2, \label{eq:x0_a}\\
\pi_0 &= (G_2/\alpha)\, r_0 + A_\pi\, \varepsilon^a_0,
   & A_\pi &\equiv \beta M^f g_2 + \kappa A_x. \label{eq:pi0_a}
\end{align}
Inserting (\ref{eq:x0_a})--(\ref{eq:pi0_a}) into (\ref{eq:r0_a_reduced}) and using $\alpha = \rho_r + (1-\rho_r)(\widetilde\phi_\pi G_2 + \widetilde\phi_x G_1)$ from (\ref{eq:alpha_unanticipated}) to simplify the homogeneous coefficient on $r_0$,
\begin{equation}
r_0 = \beta_g (1-\rho_r)\bigl[\widetilde\phi_\pi A_\pi + \widetilde\phi_x A_x + \phi_s\bigr]\, \varepsilon^a_0,
\qquad \beta_g \equiv \alpha/\rho_r.
\label{eq:r0_a_closed}
\end{equation}
Under the calibration of Table~\ref{tab:calibration} (with $\phi_s = 0.5$ from the GMM estimates of Section~\ref{sec:taylor_results} and $M^h = 0.8$): $g_1 < 0$, $g_2 < 0$, hence $A_x < 0$ and $A_\pi < 0$, so $\widetilde\phi_\pi A_\pi + \widetilde\phi_x A_x + \phi_s < 0$ in the calibration since the negative fundamentals contribution dominates the positive direct loading $\phi_s$. With $\varepsilon^a_0 < 0$, this delivers $r_0 > 0$, the contemporaneous lean against the wind of (\ref{eq:r0_a}) in Section~\ref{sec:svar_signs}. Back-substitution into (\ref{eq:x0_a})--(\ref{eq:pi0_a}) gives $x_0 > 0$ and $\pi_0 > 0$, since both the coefficient on $r_0$ (negative because $G_i < 0$) and the direct loading on $\varepsilon^a_0$ (negative because $A_x, A_\pi < 0$) contribute positively when multiplied by the equilibrium signs of $r_0$ and $\varepsilon^a_0$. The impact response of sentiment is
\begin{equation}
s_0 = \underbrace{(x_0 + \pi_0)}_{>0} + \underbrace{\varepsilon^a_0}_{<0},
\label{eq:s0_a}
\end{equation}
whose sign depends on whether the fundamentals block of the LoM dominates the direct loading of the announcement on $s_t$. We adopt the dominance condition (\ref{eq:fundamentals_dominance}) as an identifying assumption; under the alternative case $\lambda_a < 0$ of Section~\ref{sec:model_narratives}, the sign restriction $s_0 > 0$ holds unconditionally. Numerical evaluation under the calibration of Table~\ref{tab:calibration} confirms $(x_0 + \pi_0) > |\varepsilon^a_0|$ for one-standard-deviation announcements. The forward expectations follow from Stages A and B: $E_0[r_1] = \alpha r_0 + \beta_g \varepsilon^a_0$, which is negative under the calibration because $\beta_g |\varepsilon^a_0|$ dominates $\alpha r_0$ (the announcement materialises at $t = 1$); $E_0[x_1] = G_1 r_0 + g_1 \varepsilon^a_0 > 0$ and $E_0[\pi_1] = G_2 r_0 + g_2 \varepsilon^a_0 > 0$ since both $G_i$ and $g_i$ are negative and the two contributions reinforce each other. For $h \ge 2$ the Stage A geometric decay applies, with the impact-period signs preserved.

\medskip\noindent\emph{Positive narrative shock $\varepsilon^s_0 > 0$.}\enspace
The narrative innovation enters only the LoM (\ref{eq:s0_svar}) at $t = 0$ and does not appear in the IS, Phillips or Taylor equations at any date. The system has only two stages: the autonomous Stage~A from $t \ge 1$ and the contemporaneous Stage~C at $t = 0$. There is no analogue to Stage~B of the anticipated case.

\textit{Stage A (}$t \ge 1$\textit{).} No shocks hit the system; the autonomous stable mode of (\ref{eq:G_unanticipated})--(\ref{eq:alpha_unanticipated}) applies, so $z_t = G\, r_{t-1}$ and $r_t = \alpha\, r_{t-1}$, with the same $G$ and $\alpha$ as in the unanticipated case.

\textit{Stage C (}$t = 0$\textit{).} The Taylor rule at $t = 0$ contains no direct $\varepsilon^s_0$ term, but $\varepsilon^s_0$ enters the LoM with $s_0 = x_0 + \pi_0 + \varepsilon^s_0$. Substituting into (\ref{eq:r0_svar}) gives the reduced contemporaneous rule
\begin{equation}
r_0 = (1-\rho_r)\bigl[\widetilde\phi_\pi \pi_0 + \widetilde\phi_x x_0 + \phi_s \varepsilon^s_0\bigr],
\label{eq:r0_eps_reduced}
\end{equation}
with $\widetilde\phi_\pi, \widetilde\phi_x$ as in (\ref{eq:taylor_u}). Unlike the anticipated case, the Stage~A forecasts at $t = 0$ are clean of any $\varepsilon^s_0$ term: $E_0[x_1] = G_1 r_0$ and $E_0[\pi_1] = G_2 r_0$, since $\varepsilon^s_0$ does not enter the Stage~A dynamics. The IS and Phillips equations at $t = 0$ then yield
\begin{equation}
x_0 = (G_1/\alpha)\, r_0, \qquad \pi_0 = (G_2/\alpha)\, r_0,
\label{eq:x0_pi0_eps}
\end{equation}
using the same stable-mode identities $G_1/\alpha = M^h G_1 + \sigma^{-1}(G_2 - 1)$ and $G_2/\alpha = \beta M^f G_2 + \kappa G_1/\alpha$ as in the anticipated case. Inserting (\ref{eq:x0_pi0_eps}) into (\ref{eq:r0_eps_reduced}) and using $\alpha = \rho_r + (1-\rho_r)(\widetilde\phi_\pi G_2 + \widetilde\phi_x G_1)$ from (\ref{eq:alpha_unanticipated}) collapses the result to
\begin{equation}
r_0 = \beta_g\,(1-\rho_r)\,\phi_s\, \varepsilon^s_0,
\qquad \beta_g \equiv \alpha/\rho_r.
\label{eq:r0_eps_closed}
\end{equation}
All factors in $\beta_g (1-\rho_r)\phi_s$ are positive under the calibration of Table~\ref{tab:calibration} and the GMM-supported $\phi_s > 0$, so $r_0 > 0$ for $\varepsilon^s_0 > 0$. This is the lean against the wind in the narrative case: the central bank reads the autonomous rise in sentiment as an incipient overheating signal and tightens, with no contribution from the standard Taylor inputs since $\pi_0, x_0$ are themselves driven by $r_0$ rather than by the shock directly. Back-substitution into (\ref{eq:x0_pi0_eps}) gives $x_0 < 0$ and $\pi_0 < 0$ since $G_1, G_2 < 0$ and $r_0 > 0$. The impact response of sentiment is
\begin{equation}
s_0 = \underbrace{(x_0 + \pi_0)}_{<\,0} + \underbrace{\varepsilon^s_0}_{>\,0}
    = \left[1 - |G_1 + G_2|\,\frac{1-\rho_r}{\rho_r}\,\phi_s\right] \varepsilon^s_0,
\label{eq:s0_eps}
\end{equation}
where the closed form uses (\ref{eq:r0_eps_closed}) and $\beta_g/\alpha = 1/\rho_r$. Under the calibration of Table~\ref{tab:calibration}, $|G_1 + G_2|\cdot (1-\rho_r)/\rho_r \cdot \phi_s = 3.01 \times 0.111 \times 0.5 \approx 0.17 \ll 1$, so the bracket is positive and $s_0 > 0$: the direct innovation dominates the equilibrium contraction in fundamentals by a factor of roughly six. The forward expectations follow from Stage~A: $E_0[r_h] = \alpha^h r_0 > 0$, $E_0[x_h] = G_1 \alpha^{h-1} r_0 < 0$, and $E_0[\pi_h] = G_2 \alpha^{h-1} r_0 < 0$ for $h \ge 1$, with the impact-period signs preserved and damped geometrically at rate $\alpha$.

\subsection{Detailed comparison with \texorpdfstring{\citet{damico2023}}{D'Amico and King (2023)}}\label{app:svar_dk}

\citet{damico2023} (D\&K hereafter) identify $\varepsilon^a$ and $\varepsilon^u$ in a survey-augmented VAR with the same sign pattern on $(E_t[r_{t+h}], E_t[x_{t+h}], E_t[\pi_{t+h}])$ as the first two rows of Table~\ref{tab:svar_signs}. The substantive differences between the two schemes lie in three places: the source of the contemporaneous lean on $r_t$, the nature of the friction that delivers a less-than-one-for-one forecast response to anticipated shocks, and the addition of a third structural shock on $s_t$.

The lean against the wind on $r_t$ in D\&K relies on the standard NK model with no role for $s_t$. Setting $\phi_s = 0$ in (\ref{eq:r0_svar}) and using (\ref{eq:s0_svar}) only as an unobserved auxiliary, the impact policy rate under an anticipated easing reduces to $r_0 = \phi_\pi \pi_0 + \phi_x x_0$, whose sign is positive when $\phi_\pi \pi_0 + \phi_x x_0 > 0$. With $\pi_0, x_0 > 0$ in the expansionary normalisation, the rule does deliver the lean against the wind in sign, but its magnitude is small: the contemporaneous moves of $\pi_0, x_0$ to a four-period-ahead announcement are damped by forward-looking algebra under any sensible set of parameters. D\&K acknowledge this and consider a second, more realistic scenario in which the systematic response of $r_0$ to current $\pi_0, x_0$ is shut down. In their identifying restriction the lean is imposed as a sign restriction whose structural magnitude is small. In our model the lean acquires an additional source through $\phi_s s_0$: equation (\ref{eq:r0_a}) shows that under $\phi_s > 0$ the announcement raises $s_0$ (for an easing) and the policy rule responds through both $\phi_\pi \pi_0 + \phi_x x_0$ and $\phi_s s_0$. This additional $\phi_s$ channel makes the lean a structural force that can be estimated and, as Section~\ref{sec:taylor} shows, is empirically supported.

The friction that produces a damped forecast response to anticipated shocks differs in the same way. D\&K assume agents observe a noisy signal of the announced change, $\widetilde{\varepsilon}^a_t = \varepsilon^a_t + \nu_t$, with signal-to-noise ratio $K = \sigma^2_\varepsilon / (\sigma^2_\varepsilon + \sigma^2_\nu)$. Posterior expectations of the future rate satisfy $E_0[r_\tau] = K\, \widetilde{\varepsilon}^a_0 + \text{endogenous terms}$, so that when $K < 1$ the forecast response is damped relative to the rational-expectations benchmark by a factor that they estimate at $\widehat{K} \approx 0.27$. Our agents observe the announcement exactly but discount its forward-looking content by $(M^h)^\tau$ along the perceived path. The two reduced-form predictions for $E_0[r_\tau]/\varepsilon^a_0$ are observationally close, but the underlying interpretations differ: $K$ captures credibility (the announcement may be a noisy signal of the truth), while $(M^h)^\tau$ captures salience (the announcement is correctly understood but receives reduced cognitive weight).

The third difference is structural. D\&K's seven-variable system contains macro variables and survey forecasts, with the survey-implied expectation $E^S_t[\cdot]$ entering as an imperfect proxy for the rational expectation, $E_t[\cdot] = \alpha E^S_t[\cdot] + (1-\alpha) E^{VAR}_t[\cdot]$, and a loose-rationality importance weight on $\|E^S - E^{VAR}\|^2$ disciplining the rotations. We adopt the same loose-rationality weighting (see \ref{app:svar_algo}), but we add a ninth variable, the FinBERT sentiment $s_t$ used in Section~\ref{sec:taylor}, and we identify a third structural shock $\varepsilon^s$ that operates on $s_t$ and feeds back into the policy rate through $\phi_s$. The narrative shock has no counterpart in D\&K's setup. It is the empirical counterpart of the residual $\varepsilon^s_t$ in the sentiment law of motion (\ref{eq:sentiment_lom}), and it allows the SVAR to separate the variation in $s_t$ that is orthogonal to fundamentals and to communication from the variation that follows mechanically from MP shocks.

\subsection{Algorithm and diagnostics}\label{app:svar_algo}

The VAR is estimated under a diffuse Normal-Inverse-Wishart conjugate prior. Conditional on $\Sigma$, the prior on $\text{vec}(\Phi)$ is centred at the OLS estimate and has covariance $\lambda^{-1}_{\text{shr}} (X'X)^{-1} \otimes \Sigma$ with $\lambda_{\text{shr}} = 4$. This is an empirical-Bayes $g$-prior in the spirit of Zellner: it shrinks the posterior toward the OLS coefficients rather than toward zero or a random walk, and is therefore a numerical regulariser rather than a Litterman-style Minnesota prior. Posterior draws of $(\Phi, \Sigma)$ are obtained by direct sampling from the conjugate Normal-Inverse-Wishart, retaining only draws for which the companion matrix has all eigenvalues strictly inside the unit circle. We sample $D_{\text{post}} = 500$ such posterior draws.

Given a draw $(\Phi^d, \Sigma^d)$, structural identification proceeds as follows. The Cholesky factor $P^d = \text{chol}(\Sigma^d)$ provides one orthogonal decomposition, $\Sigma^d = P^d (P^d)'$, and any rotation $B^d = P^d Q$ with $Q$ orthogonal yields another. We draw $N_{\text{rot}} = 8000$ random orthogonal matrices $Q$ per posterior draw using the QR algorithm of \citet{rubio2010structural}. For each rotation we test all $n = 9$ columns of $B^d$ and both signs ($\pm$) and ask whether one column can be labelled $\varepsilon^a$ (anticipated easing), another $\varepsilon^u$ (unanticipated easing) and a third $\varepsilon^s$ (positive narrative) such that the resulting impact responses satisfy the sign restrictions of Table~\ref{tab:svar_signs} and additionally the corresponding signs of $E_0[r_h], E_0[x_h], E_0[\pi_h]$ at the four-period-ahead horizon (computed from $\Phi^d$ for $h = 4$, and from the average of the four impulse responses at $h = 1, 2, 3, 4$ for the policy-rate expectation). The remaining six columns of $B^d$ are unrestricted and represent unmodelled shocks.

The acceptance criterion is augmented with a loose-rationality importance weight in the spirit of \citet[eqn 14]{damico2023},
\begin{equation}
w_j(B^d) = \exp\!\Bigl(- \tfrac{1}{2\delta} \bigl[\bigl(B^d_{E[r],j} - \bar r^d_j\bigr)^2 + \bigl(B^d_{E[x],j} - x^d_j(4)\bigr)^2 + \bigl(B^d_{E[\pi],j} - \pi^d_j(4)\bigr)^2\bigr]\Bigr),
\label{eq:loose_rationality_app}
\end{equation}
where $j \in \{\varepsilon^a, \varepsilon^u, \varepsilon^s\}$ and the three terms compare the impact response of the three staff expectations under shock $j$ to the corresponding four-period-ahead trajectory of the underlying realised variable from the same rotation. Specifically, $B^d_{E[r],j}, B^d_{E[x],j}, B^d_{E[\pi],j}$ are the impact responses (i.e.\ entries of $B^d$ in the rows for $E^S_t[r_{t+4}], E^S_t[\log Y_{t+4}], E^S_t[\log P_{t+4}]$ and the column assigned to shock $j$). On the other side, $\bar r^d_j = \tfrac{1}{4}\sum_{h=1}^{4} r^d_j(h)$ is the average over $h=1, 2, 3, 4$ of the realised policy-rate IRFs computed from $\Phi^d$ and the same column of $B^d$ (this average is the natural counterpart of an annualised four-quarter-ahead forecast of the policy rate, which is the convention used by the BoC staff for the policy-rate path), while $x^d_j(4)$ and $\pi^d_j(4)$ are the IRFs of realised log GDP and log CPI at horizon $h = 4$ from the same draw and rotation. The weight $w_j$ penalises rotations in which the survey-implied impact responses diverge sharply from the VAR-implied trajectories of the underlying variables four periods ahead. We set $\delta = 0.5$, the central calibration of D\&K. The three weights $(w_{\varepsilon^a}, w_{\varepsilon^u}, w_{\varepsilon^s})$ enter the posterior moments \emph{shock by shock}: weighted medians and 68\% credible bands of objects associated with shock $j$ are computed using $w_j$ alone (rather than a product or sum across the three shocks), so that the discipline imposed by the loose-rationality criterion on each shock is independent of the other two.

\begin{figure}[H]
\centering
\caption{Sensitivity of IRFs to the loose-rationality penalty $\delta$.\label{fig:svar_delta_robust}}
\includegraphics[width=0.99\linewidth]{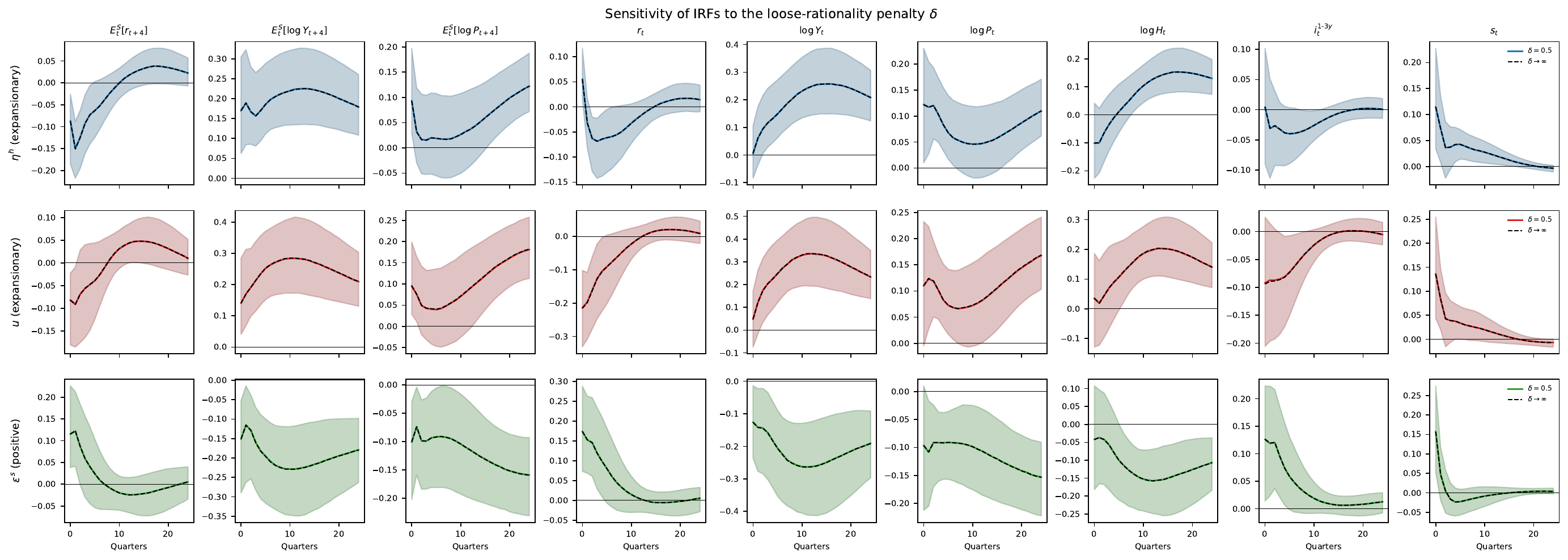}
\par\smallskip
\begin{minipage}{0.95\linewidth}\footnotesize\textit{Note}: Each panel shows the impulse response of one of the nine variables (columns) to one of the three identified shocks (rows). Coloured bands and solid medians: baseline $\delta = 0.5$, weighted by the loose-rationality importance weight (\ref{eq:loose_rationality_app}). Grey bands and dashed black medians: uniform weights on the same accepted rotation set, equivalent to $\delta \to \infty$. The two sets of medians are visually indistinguishable; the maximum absolute difference across all horizons, variables and shocks is $0.0019$ percentage points. The loose-rationality weight tightens the dispersion of admissible rotations without moving the median impulse responses materially, in line with \citet[Fig.~E.5]{damico2023}.\end{minipage}
\end{figure}

The procedure delivers $3{,}179$ accepted draws out of $4{,}000{,}000$ rotations tested, an acceptance rate of $\approx 8\times 10^{-4}$, with effective sample sizes of $3{,}177$, $3{,}176$ and $3{,}177$ for the anticipated, unanticipated and narrative shocks respectively.\footnote{Total run time on a single laptop core is approximately seven minutes for the BVAR sampling and rotation testing, and a further three minutes for the FEVD, historical decomposition and BGW counterfactual.} The high effective sample sizes are consistent with the sensitivity exercise reported by \citet[Appendix~E, Fig.~E.5]{damico2023}, who overlay IRFs under $\delta = 0$ (strict rationality), $\delta = 0.5$ (baseline) and $\delta = \infty$ (flat prior) in their seven-variable US system and conclude that the median responses are essentially unchanged across the three values of $\delta$, while the credible bands widen and become upside-skewed when the prior is removed. We replicate that exercise on our 9-variable Canadian system in Figure~\ref{fig:svar_delta_robust}, comparing the baseline $\delta = 0.5$ summaries (coloured bands and solid medians) to the uniform-weight summaries on the same accepted rotation set (i.e.\ $\delta \to \infty$, grey bands and dashed black medians). The two sets of medians are visually indistinguishable for all three shocks and all nine variables. Quantitatively, the maximum absolute difference between the two sets of medians across all $H = 24$ horizons, all nine variables and all three shocks is $0.0019$ percentage points (in $\log Y_t$ at the unanticipated shock); all other entries are smaller. The loose-rationality weight in our setting is therefore best interpreted, in the sense of D\&K, as a tightening of the dispersion of admissible rotations rather than as a driver of the median impulse responses; the high effective sample sizes are the local empirical signature of this property. The orthogonality of the three identified shock series is essentially exact at the posterior level: pairwise within-draw correlations have weighted medians of $-0.036$, $-0.027$ and $-0.029$ for $(\varepsilon^a, \varepsilon^u)$, $(\varepsilon^a, \varepsilon^s)$, and $(\varepsilon^u, \varepsilon^s)$ respectively, with 68\% credible bands inside $[-0.13, +0.07]$ in all three cases.\footnote{The $99\%$ tails of the within-draw correlations remain inside $[-0.27, +0.21]$ across all three pairs, with no evidence of quasi-collinear draws.} First-order autocorrelations of the identified shock series are below $0.04$ in absolute value, and the posterior shock standard deviations cluster around $0.94$, close to the unit-variance normalisation.

\end{document}